\pdfoutput=1
\documentclass[cernpreprint,txtfonts,english,USenglish]{na61doc}
\usepackage{multicol}
\usepackage{xspace}
\usepackage{multirow}
\usepackage{enumerate}
\usepackage{flushend}
\usepackage{cite}
\usepackage{upgreek}

\usepackage{colortbl}
\definecolor{darkred}{rgb}{0.5,0,0}
\definecolor{darkblue}{rgb}{0,0,0.5}
\definecolor{firebrick}{rgb}{0.75,0.125,0.125}
\definecolor{darkgreen}{rgb}{0,0.5,0}

\usepackage[colorlinks=true,linkcolor=firebrick,citecolor=darkgreen,urlcolor=darkblue]{hyperref}

\usepackage{subfigure}
\usepackage{url}
\usepackage{wrapfig}

\graphicspath{{./figures/}}

\newcommand{\VTPCOne}{{VTPC-1}\xspace}

\newcommand{\eV}{\ensuremath{\mbox{e\kern-0.1em V}}\xspace}
\newcommand{\GeV}{\ensuremath{\mbox{Ge\kern-0.1em V}}\xspace}
\newcommand{\MeV}{\ensuremath{\mbox{Me\kern-0.1em V}}\xspace}
\newcommand{\GeVc}{\ensuremath{\mbox{Ge\kern-0.1em V}\kern-0.1em/\kern-0.05em c}\xspace}
\newcommand{\GeVcc}{\ensuremath{\mbox{Ge\kern-0.1em V}\kern-0.1em/\kern-0.05em c^2}\xspace}
\newcommand{\AGeV}{\ensuremath{A\,\mbox{Ge\kern-0.1em V}}\xspace}
\newcommand{\AGeVc}{\ensuremath{A\,\mbox{Ge\kern-0.1em V}\kern-0.1em/\kern-0.05em c}\xspace}
\newcommand{\MeVc}{\ensuremath{\mbox{Me\kern-0.1em V}\kern-0.1em/\kern-0.05em c}\xspace}
\newcommand{\MeVcc}{\ensuremath{\mbox{Me\kern-0.1em V}\kern-0.1em/\kern-0.05em c^2}\xspace}

\newcommand{\mb}{\ensuremath{\mbox{mb}}\xspace}

\newcommand{\pT}{\ensuremath{p_\text{T}}\xspace}
\newcommand{\pL}{\ensuremath{p_\text{L}}\xspace}
\newcommand{\xF}{\ensuremath{x_\text{F}}\xspace}
\newcommand{\xpF}{\ensuremath{x'_\text{F}}\xspace}
\newcommand{\minv}{\ensuremath{m_\text{inv}}\xspace}

\newcommand{\Geant}{{\scshape Geant}\xspace}

\newcommand{\Epos}{{\scshape Epos}\xspace}
\newcommand{\EposLong}{{\scshape Epos\,1.99}\xspace}
\newcommand{\QGSJetLong}{{\scshape QGSJet\,II-04}\xspace}

\newcommand{\DPMJetLong}{{\scshape DPMJet\,3.06}\xspace}
\newcommand{\SibyllLong}{{\scshape Sibyll\,2.1}\xspace}
\newcommand{\SibyllNewLong}{{\scshape Sibyll\,2.3}\xspace}
\newcommand{\EposLHCLong}{{\scshape Epos\,LHC}\xspace}

\def\avg#1{\langle{#1}\rangle}

\newcommand{\CernVM}{\textsc{Cern\-\kern-0.05emVM}\xspace}

\ShineTitle{Measurement of Meson Resonance Production in
         $\pi^{-} + $C Interactions at SPS energies
    }

\PreprintIdNumber{CERN-EP-2017-105}

\ShineJournal{\mbox{Eur.\ Phys.\ J.\ C}}

\ShineAbstract{
 We present measurements of $\rho^0$, $\omega$ and K$^{*0}$ spectra in
  $\pi^{-} + $C production interactions at 158\,\GeVc and $\rho^0$
  spectra at 350\,\GeVc using
  the \NASixtyOne spectrometer at the CERN SPS. Spectra are presented
  as a function of the Feynman's variable $x_\text{F}$ in the range $0
  < x_\text{F} < 1$ and $0 < x_\text{F} < 0.5$ for 158\,\GeVc and
  350\,\GeVc respectively. Furthermore, we show comparisons with
  previous measurements and predictions of several hadronic
  interaction models. These measurements are essential for a better
  understanding of hadronic shower development and for improving the
  modeling of cosmic ray air showers.
}
\begin{document}
\maketitle
\sloppy



%


  \noindent
  \section{Introduction}
\label{sec:intro}

When cosmic rays of high energy collide with the nuclei of the
atmosphere, they initiate extensive air showers (EAS). Earth's
atmosphere then acts as a medium in which the particle shower
evolves. It proceeds mainly through the production and interaction of
secondary pions and kaons. Depending on the particle energy and
density of the medium in which the shower evolves, secondary particles
either decay or re-interact, producing further secondaries. Neutral
pions have a special role.  Instead of interacting hadronically, they
immediately decay ($c\bar\tau = 25$\,nm) into two photons with a
branching ratio of 99.9\%, giving rise to an electromagnetic shower
component. When only the primary particle energy is of interest, and
all shower components are sampled, a detailed understanding of the
energy transfer from the hadronic particles to the electromagnetic
shower component is not needed.  However, for other measurements of
air shower properties this understanding is of central importance.

A complete measurement of an air shower is not possible and particles
are typically sampled only in select positions at the ground level or
the ionization energy deposited in the atmosphere is measured.
Therefore, the interpretation of EAS data, and in particular the
determination of the composition of cosmic rays, relies to a large
extent on a correct modelling of hadron-air interactions that occur
during the shower development (see
e.g.~\cite{Engel:2011zzb}). Experiments such as the Pierre Auger
Observatory~\cite{Abraham:2004dt}, IceTop~\cite{IceCube:2012nn},
KASCADE-Grande~\cite{Navarra:2004hi} or the Telescope
Array~\cite{AbuZayyad:2012kk} use models for the interpretation of
measurements.  However, there is mounting evidence that current
hadronic interaction models do not provide a satisfactory description
of the muon production in air showers and that there is a deficit in
the  number of muons predicted at the ground level by the models when
compared to the air shower measurements (see
Refs.~\cite{abu2000evidence, Arteaga-Velazquez:2013ira,Aab:2014pza,Aab:2014dua,Aab:2016hkv}).

To understand the possible cause of this deficit it is instructive to
study the air shower development in a very simplified
model~\cite{Matthews:2005sd} in which mesons are produced in
subsequent interactions of the air cascade until the average meson
energy is low enough such that its decay length is smaller than its
interaction length. In each interaction a fraction $f_\mathrm{em}$ of
the shower energy is transferred to the electromagnetic shower
component via the production and decay of neutral mesons.  After $n$
interactions the energy available in the hadronic part of the shower
to produce muons is therefore $E_\mathrm{had} = E_0 \,
(1-f_\mathrm{em})^n$, where $E_0$ denotes the primary energy of the
cosmic ray initiating the air shower.  In the standard simplified
picture, one third of the interactions products of charged pions with
air are neutral mesons. Assuming a typical value of $n=7$ for the
number of interactions needed to reach particle energies low enough
that the charged mesons decay to muons rather than interact again, the
simplistic model gives $E_\mathrm{had} / E_0 \simeq 6\%$.  One way to
increase this number is to account for the production of baryons and
antibaryons to decrease $f_\mathrm{em}$~\cite{Pierog:2006qv}.  Another
possibilty has been recently
identified~\cite{Drescher:2007hc,Ostapchenko:2013pia} by noting that
accelerator data on $\pi^+ + \text{p}$
interactions~\cite{Adamus:1986ta,Azhinenko:1990hx,Agababyan:1990df}
indicate that most of the neutral mesons produced in the forward
direction are not $\pi^0$s but $\rho^0$ mesons.  With $\rho^0$
decaying into $\pi^+\,\pi^-$ this would imply that the energy of the
leading particle is not transferred to the electromagnetic shower
component as it would be in the case of neutral pions and
corresponingly $f_\mathrm{em}$ is decreased leading to more muons at
ground level.

Given these considerations it is evident that the modeling of air
showers depends crucially on our knowledge of pion interactions with
air. It can be shown (see e.g.~\cite{Drescher:2002vp,ICRC2009}) that the
relevant energies for the interactions in the last stage of the air
shower development are in the range from 10 to $10^3$\,\GeV.  This
range is accessible to fixed-target experiments with charged pion
beams.

A large body of data is available at these energies for
proton-nucleus interactions (e.g.\ \cite{Eichten:1972nw,Abbott:1991en,Ambrosini:1998th,Alt:2006fr,Apollonio:2009en}),
but only a very limited amount of data exists for pion or kaon beams.  A
number of dedicated measurements for air-shower simulations have been
performed by studying particle production on light nuclei at beam
momenta up to $12$\,\GeVc (see,
e.g.\ Ref.~\cite{Catanesi:2008rf,Catanesi:2008ui}).  Unfortunately, at
higher energies, there are no comprehensive and precise particle
production measurements of $\pi$ interactions with light nuclei of
masses similar to air. Earlier measurements were either limited to a
small acceptance in momentum space (e.g.\ Ref.~\cite{Barton:1982dg})
or protons as
target~\cite{Adamus:1986ta,Azhinenko:1990hx,Agababyan:1990df,AguilarBenitez:1989fn},
or did not discriminate between the different
secondaries~\cite{Elias:1979cp}.

To address the lack of suitable data for the tuning of hadronic
interaction models used in air shower simulations,
\mbox{\NASixtyOne}~\cite{Abgrall:2014fa} collected new data with
negatively charged pion beams at 158 and 350\,\GeVc on a thin carbon
target. Preliminary spectra of unidentified hadrons and
identified pions were previously derived from this data
set~\cite{ISVHECRI12_MU, ICRC13_HD, ICRC2015} and in this paper, we
present the results of the measurement of $\rho^0$, $\omega$ and
K$^{*0}$ spectra in $\pi^{-}$+C interactions at 158 and 350\,\GeVc.

It is worthwhile noting that the measurements presented in this paper
will not only be useful for interpretation of cosmic-ray calorimetry
in air, but can also be beneficial for the understanding of hadronic
calorimeters used in high-energy laboratory experiments.  Hadronic
interaction models used for calorimeter simulations are mostly tuned
to and validated with the overall calorimeter response from test-beam
data (see e.g.~\cite{Kiryunin:2006cm, Damgov:2006ef,
  Adloff:2013kio}). A tuning of these models to the data presented
here will improve the description of the energy transfer from the
hadronic to the electromagnetic shower component for individual
interactions inside the calorimeter and thus increase the predictive
power of the calorimeter simulation.

The paper is organized as follows: A brief description of the
experimental setup, the collected data, data reconstruction and
simulation is presented in Sec.~\ref{sec:setup}. The analysis
technique used to measure meson resonance production in $\pi$+C
interactions is described in Sec.~\ref{sec:analysis}. The final
results, with comparison to model predictions, and other experimental
data are presented in Sec.~\ref{sec:results}. A summary in
Sec.~\ref{sec:Summary} closes the paper.

  \section{Experimental setup, data processing and simulation}
\label{sec:setup}

\begin{figure*}[t]
\centering
\subfigure[Beam and trigger configuration]{
\label{fig:beam_setup}
\includegraphics[width=0.8\textwidth]{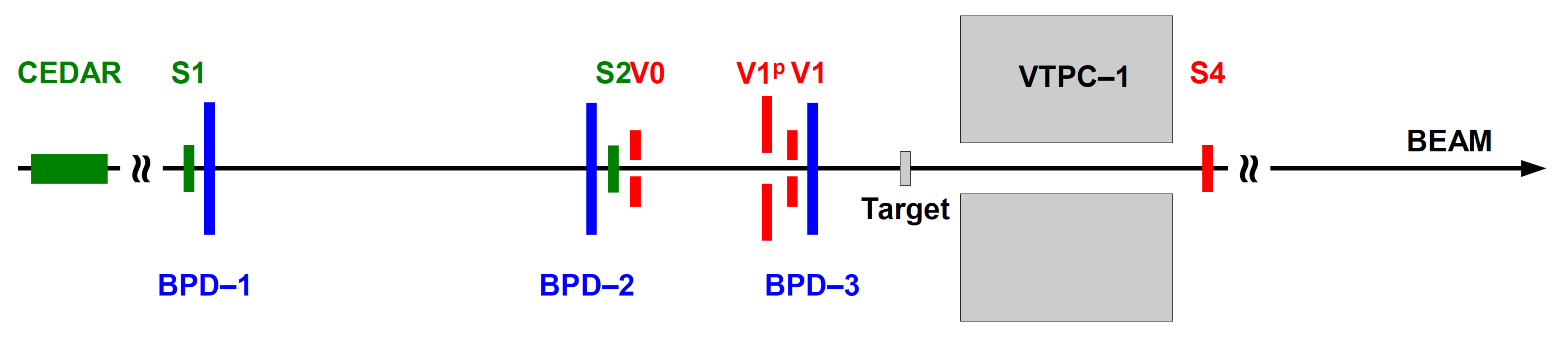}
}
\subfigure[Schematic layout in the beam plane]{
\label{fig:na61}
\includegraphics[width=\textwidth]{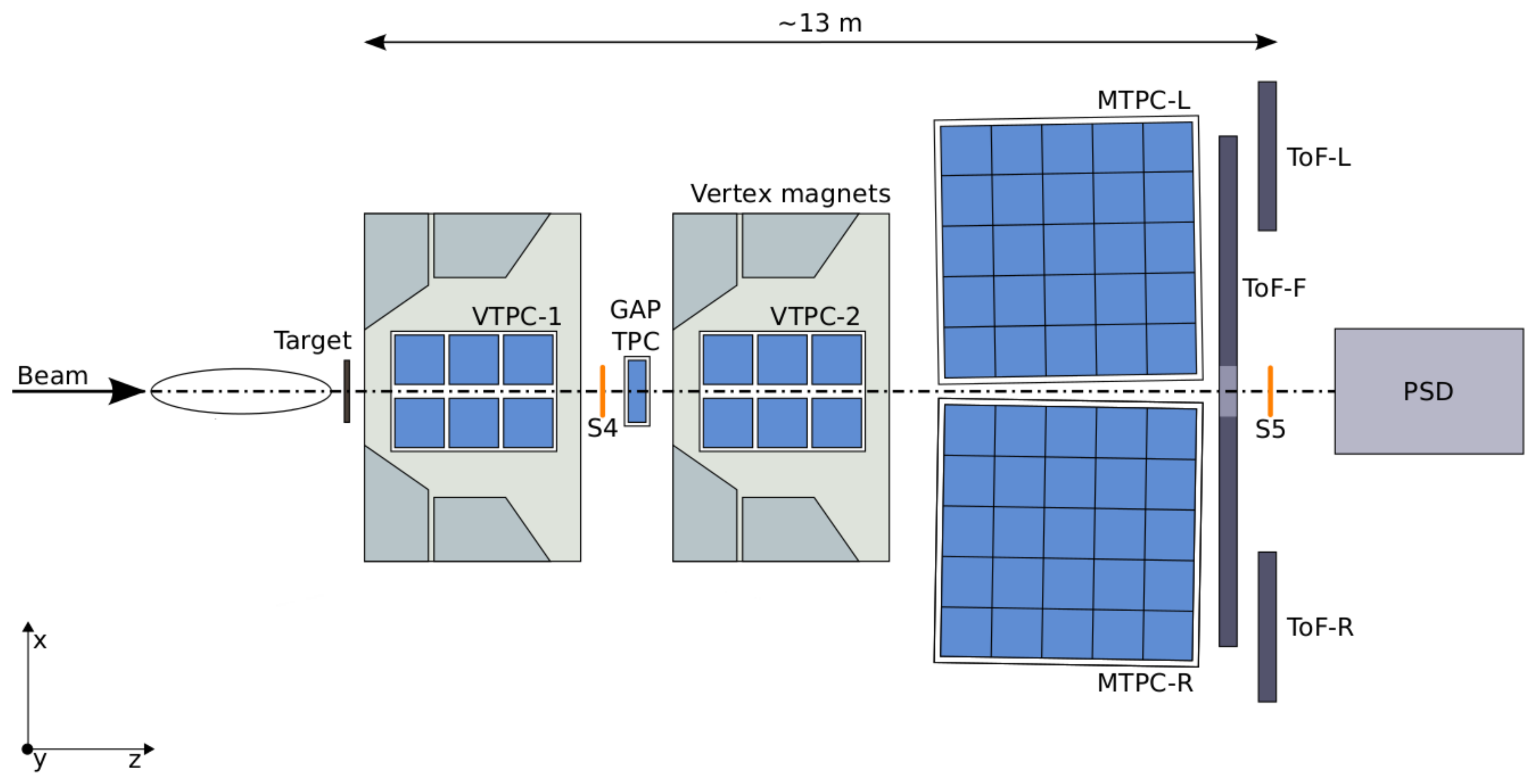}
}
\caption{Experimental Setup of the NA61/SHINE
  experiment~\cite{Abgrall:2014fa} (configuration for the $\pi^-$+C
  data taking).  The coordinate system used in this paper is indicated
  on the lower left.  The incoming beam direction is along the $z$
  axis.  The magnetic field bends charged particle trajectories in the
  $x{-}z$ (horizontal) plane. The drift direction in the TPCs is along
  the $y$ (vertical) axis. The beam and trigger instrumentation is indicated
  as an ellipse in the lower panel and detailed in the upper panel.}
\label{fig:na61Layout}
\end{figure*}

The NA61/SHINE apparatus is a wide-acceptance hadron spectrometer
at the CERN SPS on the H2 beam line of the CERN North Area.
A detailed description of the experiment is presented
in Ref.~\cite{Abgrall:2014fa}. Only features relevant for the $\pi^-$+C
data are briefly mentioned here.
Numerous components of the NA61/SHINE setup were inherited from its
predecessor, the NA49 experiment~\cite{Afanasev:1999iu}.
An overview of the setup used for data taking on $\pi^-$+C interactions in 2009
is shown in Fig.~\ref{fig:na61Layout}.

The detector is built around five
Time Projection Chambers (TPCs), as shown in Fig.~\ref{fig:na61}. Two Vertex TPCs
(\mbox{VTPC-1} and \mbox{VTPC-2}) are placed in the magnetic field produced by two
superconducting dipole magnets and two Main-TPCs (\mbox{MTPC-L} and
\mbox{MTPC-R}) are located
downstream symmetrically with respect to the beamline.
An additional small TPC is placed between VTPC-1 and VTPC-2,
covering the very-forward region, and is referred to as the GAP TPC (GTPC).

The magnet current setting for data taking at 158 and 350\,\GeVc corresponds
to 1.5\,T in the first and 1.1\,T, in the second magnet. It results in
a precise measurement of the particle momenta $p$ with a resolution
of $\sigma(p)/p^2\approx (0.3{-}7){\times}10^{-4}\,\mathrm{(GeV/c)}^{-1}$.

Two scintillation counters, S1 and S2,
together with the three veto counters V0, V1 and V1$^\text{p}$,
define the beam upstream of the target. The setup of these
counters can be seen in Fig.~\ref{fig:beam_setup} for the 158\,\GeVc
run. The S1 counter also provides the start time for all timing measurements.

\begin{figure}[t]
\centering
\subfigure{
\includegraphics[width=0.485\textwidth]{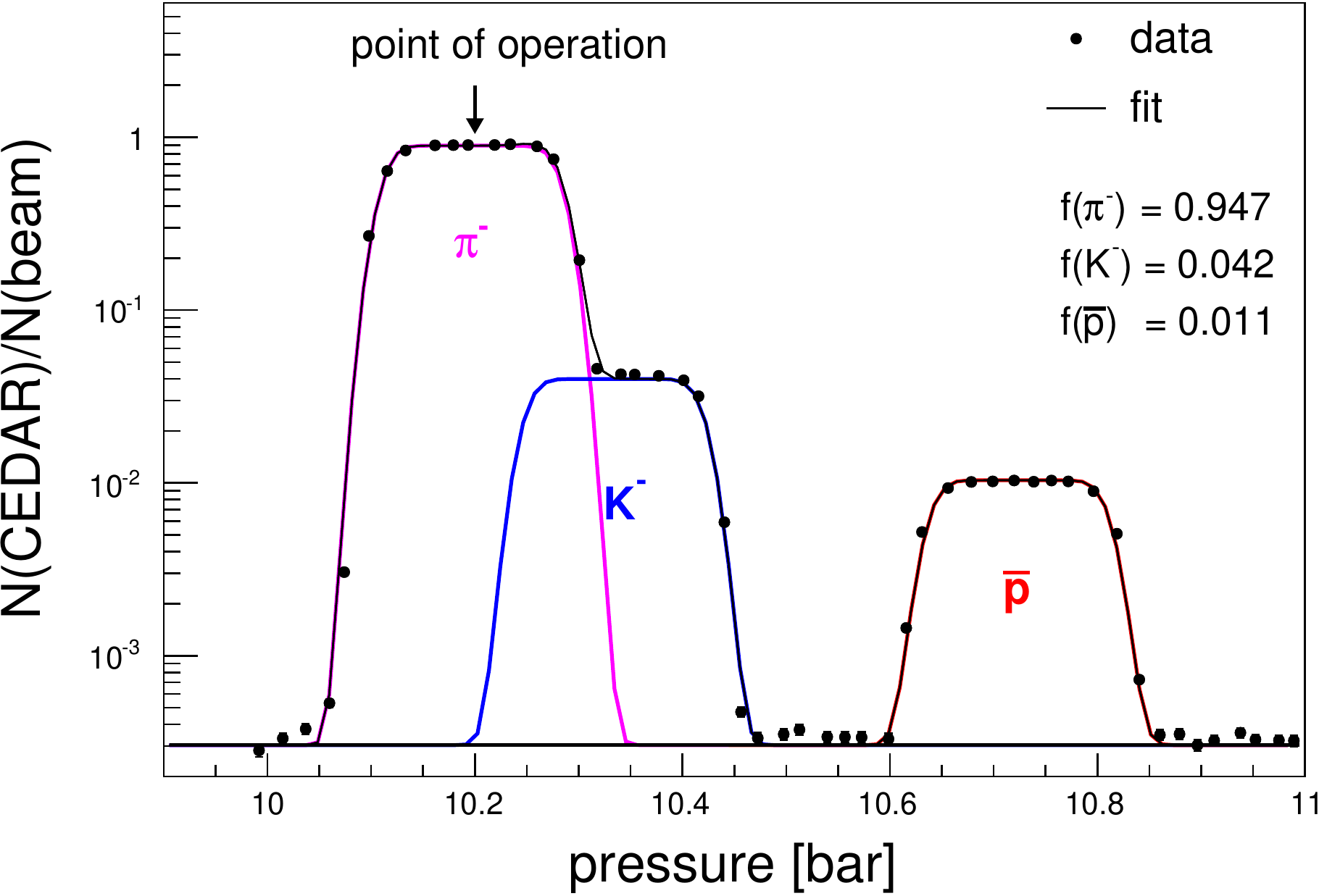}
}
\subfigure{
\includegraphics[width=0.485\textwidth]{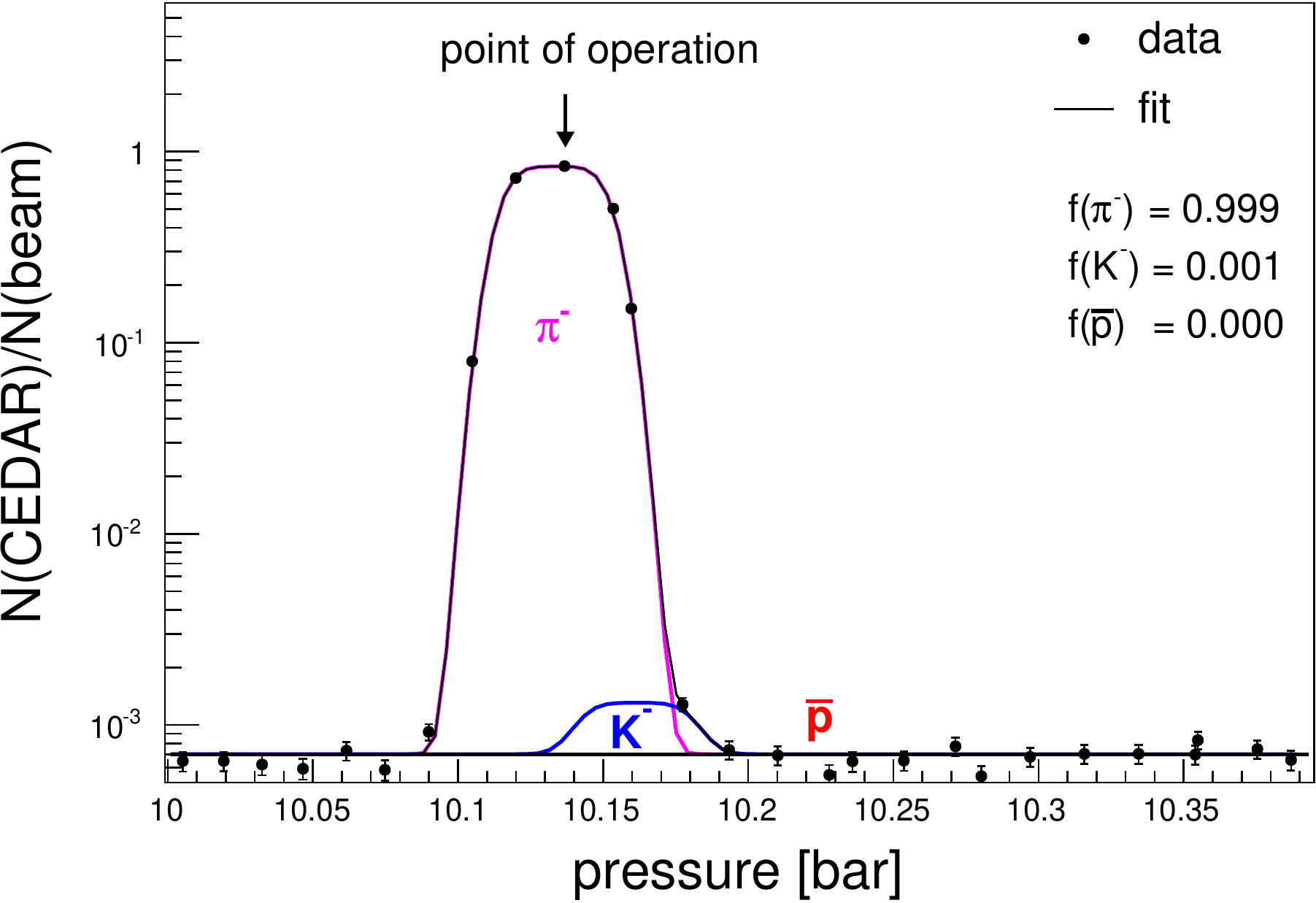}
}
\caption{The fraction of CEDAR triggers as a function of its gas
  pressure for beam momenta of 158 \emph{(left)} and 350
  \emph{(right)} \GeVc. The fitted fractions $f$ of pions, kaons and
  anti-protons are quoted within the figures and the
  point of operation during data taking is indicated by an arrow.}
\label{fig:CEDAR}
\end{figure}

The 158 and 350\,\GeVc secondary hadron beam was produced by
400\,\GeVc primary protons impinging on a 10\,cm long beryllium target.
Negatively charged hadrons ($\text{h}^-$) produced at the
target are transported downstream to the \NASixtyOne experiment by the
H2 beamline, in which collimation and momentum selection occur.
The beam particles, mostly $\pi^-$ mesons, are identified by
a differential ring-imaging Cherenkov detector CEDAR~\cite{Bovet:1975bx}.
The fraction of pions is ${\approx}95\%$
for 158\,\GeVc and ${\approx}100\%$ for 350\,\GeVc (see Fig.~\ref{fig:CEDAR}).
The CEDAR signal is recorded during data taking and then used as
an offline selection cut (see Sec.~\ref{subsec:selection}).
The beam particles are selected by the beam trigger,
T$_\text{beam}$, then defined by the coincidence
$\text{S1}\land\text{S2}\land\overline{\text{V0}}
\land\overline{\text{V1}}\land\overline{\text{V1}^\text{p}}$.
The interaction trigger ($\text{T}_\text{int} = \text{T}_\text{beam} \land \overline{\text{S4}}$)
is given by the anti-coincidence of the incoming beam particle and S4,
a scintillation counter, with a diameter of 2\,cm, placed
between the VTPC-1 and VTPC-2 detectors along the  beam trajectory
at about 3.7\,m from the target,
see Figs.~\ref{fig:na61} and \ref{fig:beam_setup}.
Almost all beam particles that interact inelastically in the target do not reach S4.
The interaction and beam triggers were recorded in parallel.
The beam trigger events were recorded with a frequency
by a factor of about 10
lower than the frequency of interaction trigger events.

The incoming beam trajectory is measured by a set
of three Beam Position Detectors (BPDs), placed along the beamline
upstream of the target, as shown in Fig.~\ref{fig:beam_setup}.
These detectors are $4.8 \times 4.8$\,cm$^2$ proportional chambers.
Each BPD measures the position of the beam particle on the transverse plane
with respect to the beam direction with a resolution of
${\sim}100\,\upmu$m (see Ref.~\cite{Abgrall:2014fa} for more details).

For data taking on $\pi^-$+C interactions, the target was an isotropic graphite plate
with a thickness along the beam axis of 2\,cm with a density of $\rho=1.84$\,g/cm$^3$, equivalent to about 4\% of a
nuclear interaction length. During the data taking
the target was placed 80~cm upstream of \VTPCOne.
90\% of data was recorded with the target inserted and 10\% with the removed target.
The latter set was used to estimate the bias due to interactions with
the material upstream and downstream of the target.

Detector parameters were optimised using a data-based calibration
procedure which also took into account their time dependences. Minor
adjustments were determined in consecutive steps for:
\begin{enumerate}[(i)]
\item detector geometry and TPC drift velocities and
\item magnetic field map.
\end{enumerate}

Each step involved reconstruction of the data required to optimise a given
set of calibration constants and time dependent corrections
followed by verification procedures.
Details of the procedure and quality assessment are presented in
Ref.~\cite{Abgrall:2008zz}.

\begin{figure*}[t]
\centering
\includegraphics[width=0.88\textwidth]{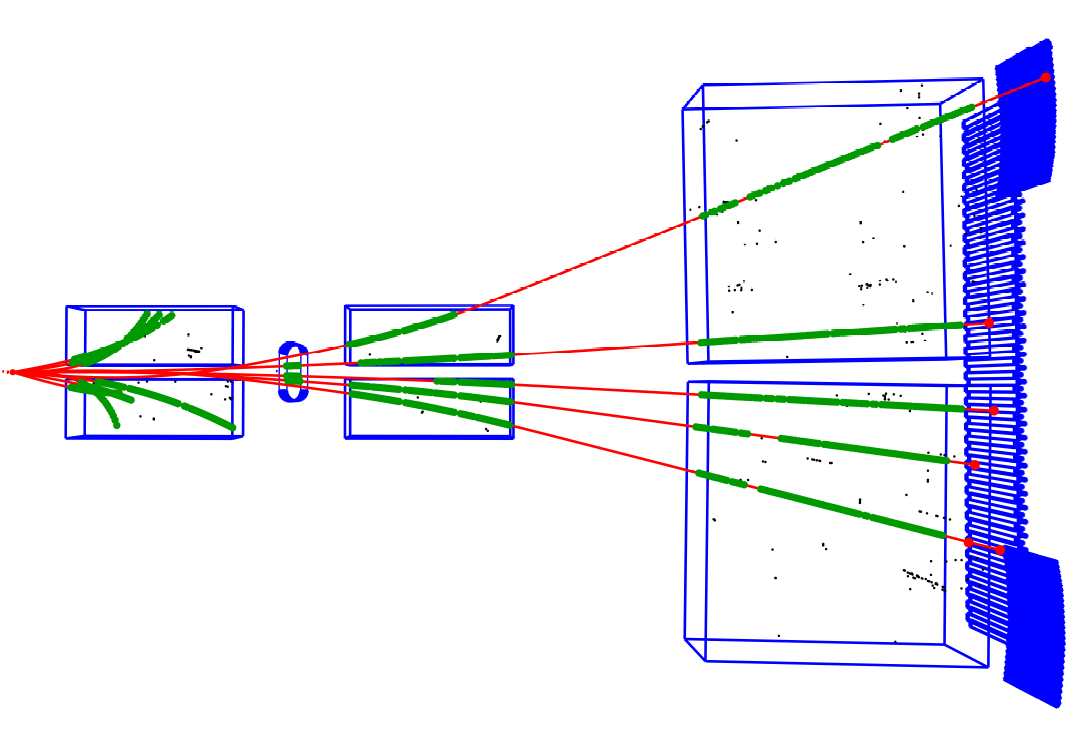}
\caption{An example of a $\pi^-$+C interaction at 158\,\GeVc measured in the
\NASixtyOne detector (top view). The measured points (\textit{green dots}) are used to fit tracks
(\textit{red lines}) to the interaction point. The \textit{black dots} show the
noise clusters and the \textit{red dots} show matched Time of Flight hits
(not used in this analysis).}
\label{fig:real_data_event}
\end{figure*}

The main steps of the data reconstruction procedure are:
\begin{enumerate}[(i)]
 \item finding of clusters in the TPC raw data, calculation of the cluster
 centre-of-gravity and total charge,
 \item reconstruction of local track segments in each TPC separately,
 \item matching of track segments into global tracks,
 \item fitting of the track through the magnetic field and determination of track
 parameters at the first measured TPC cluster,
 \item determination of the interaction vertex using the beam trajectory
 fitted in the BPDs and the trajectories of tracks
 reconstructed in the TPCs (the final data analysis uses the middle of
 the target as the $z$-position, $z=-580\,$cm) and
 \item
 refitting of the particle trajectory using the interaction vertex as an
 additional point and determining the particle momentum at the interaction
 vertex.
\end{enumerate}
An example of a reconstructed $\pi^-$+C interaction at 158\,\GeVc
is shown in Fig.~\ref{fig:real_data_event}.
Amongst the many tracks visible are five long tracks of three negatively
charged and two positively charged particles, with momentum ranging $5{-}50$\,\GeVc.

A simulation of the \NASixtyOne detector response is used to correct the measured
raw yields of resonances. For the purposes of this analysis, the \EposLong model
was used for the simulation and calculation of correction factors.
\DPMJetLong~\cite{Roesler:2000he} was used as a comparison for estimation
of systematic uncertainties. The choice of \Epos was made due to
both the number of resonances included in the model, as well as the ability
to include the intrinsic width of these resonances in the simulation. \EposLong
rather than \EposLHCLong was used as it is better tuned
to the measurements at SPS energies \cite{Tanguy_Epos}.

The simulation consists of the following steps:
\begin{enumerate}[(i)]
 \item generation of inelastic $\pi^-$+C interactions using the \EposLong
 model,
 \item propagation of outgoing particles through the detector material
 using the \Geant 3.21 package~\cite{Geant3} which takes into account the
 magnetic field as well as relevant physics processes, such as particle
 interactions and decays,
 \item
 simulation of the detector response using dedicated \NASixtyOne packages
 which also introduce distortions corresponding to all corrections applied to
 the real data,
 \item simulation of the interaction trigger selection by checking whether
 a charged particle hits the S4 counter,
 \item storage of the simulated events in a file which has the same format
 as the raw data,
 \item reconstruction of the simulated events with the same reconstruction
 chain as used for the real data and
 \item matching of the reconstructed to the simulated tracks
 based on the cluster positions.
\end{enumerate}

For more details on the reconstruction and calibration algorithms applied to the raw
data, as well as the simulation of the NA61/SHINE detector response, used to correct
the raw data, see Ref.~\cite{Abgrall:2011ae}.

  \section{Analysis}
\label{sec:analysis}

In this section we present the analysis technique developed for the
measurement of the $\rho^0$, $\omega$ and K$^{*0}$ spectra in
$\pi^-$+C production interactions. Production interactions are
interactions with at least one new particle produced,
i.e.\ interactions where only elastic or quasi-elastic scattering
occurred are excluded. The procedure used for the data analysis
consists of the following steps:
\begin{enumerate}[(i)]

\item application of event and track selection criteria,

\item combination of oppositely charged tracks,

\item accumulating the combinations in bins of Feynman-$x$, \xF, calculated by using the
mass of the $\rho^0$ meson for the boost between the lab and centre of mass frames,

\item calculation of the invariant mass of each combination, assuming pion masses for the particles,

\item fitting of the invariant mass distributions with templates of resonance decays to obtain raw yields and

\item application of corrections to the raw yields calculated from simulations.

\end{enumerate}
These steps are described in the following subsections.

\subsection{Event and track selection}
\label{subsec:selection}

A total of $5.49{\times}10^6$ events were recorded at 158\,\GeVc and
$4.48{\times}10^6$ events were recorded at 350\,\GeVc.
All events used in the analysis are required to pass cuts to ensure
both an interaction event and events of good quality. These cuts are:
\begin{enumerate}[(i)]

\item Well-contained measurements of the beam with the BPDs and a
     successful reconstruction of the beam direction.

\item Pion identification with the CEDAR (only for 158\,\GeVc as the
impurity of the 350\,\GeVc beam is below 0.1\%).

\item No extra (off-time) beam particles detected within $\pm2\,\upmu$s
of the triggered beam particle.

\item All events must have an interaction trigger as defined in Sec.~\ref{sec:setup}.

\item The main vertex point is properly reconstructed.

\item The $z$-position of the interaction vertex must be between $-597$\,cm and $-563$\,cm.

\end{enumerate}

\begin{figure}[t]
\centering
\includegraphics[width=0.6\linewidth]{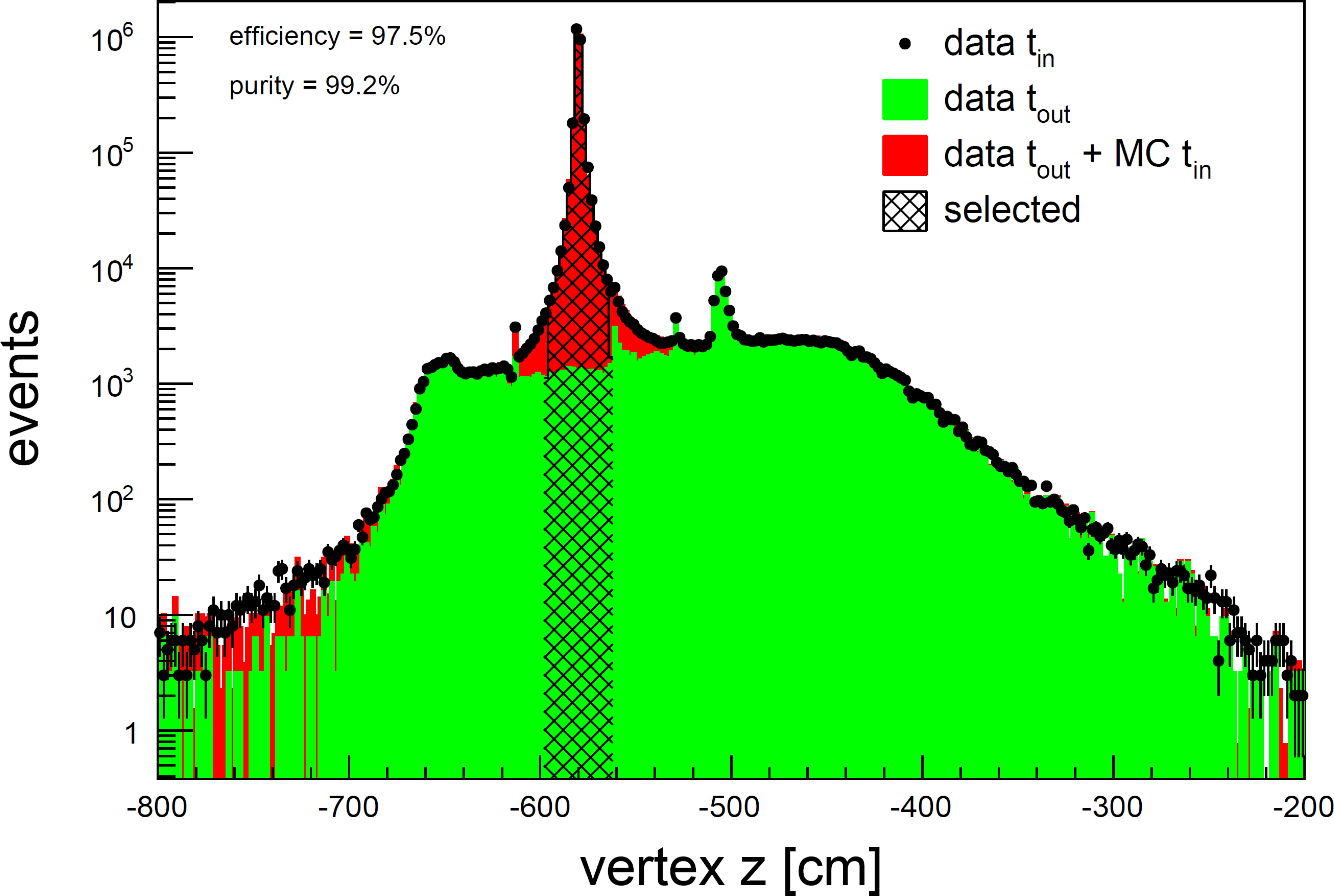}
\caption{Distribution of fitted vertex $z$ positions for $\pi^-$+C interactions at
 158\,\GeVc. The filled green area shows the distribution for events recorded with the
 target removed, while the filled red area shows the distribution for the reconstructed
 Monte Carlo simulation. The dashed area indicates events selected
 for this analysis.}
\label{fig:Vertex_Z}
\end{figure}

The cut (vi) is illustrated in Fig.~\ref{fig:Vertex_Z} and its purpose
is to remove the majority of interactions that do not occur in the
target.  This cut will increase the Monte Carlo correction because
some in-target events are removed due to the vertex $z$
resolution. However, as there is good agreement between the data and
Monte Carlo vertex $z$ distributions, this will only have a minor
impact on systematic uncertainties. An alternative method to correct
for out-of-target interactions would be to measure the resonance
yields in the target-removed data, but the template-fitting method
used in this paper can not be applied to data sets with small statistics
such as the target-removed data.

The range of this cut, ($-597,-563$)\,cm, was selected to maximise
the event number, while minimising the contamination due to off-target
events. The residual contribution of non-target interactions after
applying this cut is 0.8\%.

\begin{table}[!htb]
\caption{
Number of events after each event selection cut and
selection efficiency with respect to the previous cut for the
target inserted data set for 158\,\GeVc and 350\,\GeVc beam momentum.
}
\begin{center}
\begin{tabular}{cc|cc|cc}
\cline{2-6}
 & $p_\text{beam}$ & \multicolumn{2}{c|}{$158\,\GeVc$} & \multicolumn{2}{c}{$350\,\GeVc$}
\\
 & & & & &
\\[-1em]
 & Cut & $N_\text{events}$ & Efficiency (\%) & $N_\text{events}$ & Efficiency (\%)
\\
\cline{2-6}
 & & & & &
\\[-1em]
 & Total & $5.49{\times}10^6$ & 100 & $4.48{\times}10^6$ & 100
\\
(i) & BPD & $4.96{\times}10^6$ & 90.3 & $4.08{\times}10^6$ & 91.1
\\
(ii) & CEDAR & $4.26{\times}10^6$ & 85.9 & $4.08{\times}10^6$ & 100
\\
(iii) & Off-time & $4.03{\times}10^6$ & 94.5 & $3.94{\times}10^6$ & 96.5
\\
(iv) & Trigger & $3.34{\times}10^6$ & 83.0 & $2.97{\times}10^6$ & 75.3
\\
(v) & Vertex fit & $3.29{\times}10^6$ & 98.5 & $2.95{\times}10^6$ & 99.5
\\
(vi) & $z$-position & $2.78{\times}10^6$ & 84.6 & $2.59{\times}10^6$ & 87.9
\\
\cline{2-6}
\end{tabular}
\end{center}
\label{tbl:event-cuts-target-in}
\end{table}

\begin{table}[!htb]
\caption{
Number of tracks after each track selection cut and
selection efficiency with respect to the previous cut for the
target inserted data set for 350\,\GeVc beam momentum.
}
\begin{center}
\begin{tabular}{cc|cc|cc}
\cline{2-6}
 & $p_\text{beam}$ & \multicolumn{2}{c|}{$158\,\GeVc$} & \multicolumn{2}{c}{$350\,\GeVc$}
\\
 & & & & &
\\[-1em]
 & Cut & $N_\text{tracks}$ & Efficiency (\%) & $N_\text{tracks}$ & Efficiency (\%)
\\
\cline{2-6}
 & & & & &
\\[-1em]
 & Total & $3.85{\times}10^7$ & 100 & $4.41{\times}10^7$ & 100
\\
(i) & Track quality & $2.27{\times}10^7$ & 59.0 & $2.77{\times}10^7$ & 62.8
\\
(ii) & Acceptance & $1.57{\times}10^7$ & 69.0 & $1.99{\times}10^7$ & 72.0
\\
(iii) & Total clusters & $1.54{\times}10^7$ & 98.1 & $1.95{\times}10^7$ & 98.2
\\
(iv) & TPC clusters & $1.51{\times}10^7$ & 98.0 & $1.91{\times}10^7$ & 97.8
\\
(v) & Impact parameters & $1.42{\times}10^7$ & 94.4 & $1.80{\times}10^7$ & 94.1
\\
\cline{2-6}
\end{tabular}
\end{center}
\label{tbl:track-cuts-target-in}
\end{table}

The number of events after these cuts is $2.78{\times}10^6$ for 158\,\GeVc
and $2.59{\times}10^6$ for 350\,\GeVc. The efficiency of these cuts is
shown in Table \ref{tbl:event-cuts-target-in} for 158\,\GeVc and 350\,\GeVc
beam momentum.

After the event cuts were applied, a further set of quality cuts were
applied to the individual tracks. These were used to ensure a high
reconstruction efficiency as well as reducing contamination by tracks from
secondary interactions. These cuts are:
\begin{enumerate}[(i)]
\item The track is well reconstructed at the interaction vertex.
\item The fitted track is inside the geometrical acceptance of the detector.
\item The total number of clusters on the track should be greater than or equal to 30.
\item The sum of clusters on the track in VTPC-1 and VTPC-2 should
be greater than or equal to 15, or the total number of clusters on the track in GTPC should
be greater than or equal to 6.
\item The distance of closest approach of the fitted track to the
interaction point (impact parameter) is required to be less than 2\,cm
in the $x$-plane and 0.4\,cm in the $y$-plane.
\end{enumerate}

For the acceptance cut, (ii), we studied the selection efficiency with
simulations as a function of azimuthal angle $\phi$ for bins in total
momentum $p$ and transverse momentum $\pT$.  This leads to a
three-dimensional lookup table that defines the regions in $(\phi, p,
\pT)$ for which the selection efficiency is larger than 90\%.  Within
this region, the detector is close to fully efficient and the
corresponding correction factor is purely \emph{geometric}, since the
production of resonances is uniform in $\phi$ for an unpolarised beam and target.

The efficiency
of each track-selection cut is shown in Table~\ref{tbl:track-cuts-target-in} for
the data collected at 158\,\GeVc and 350\,\GeVc.

No particle identification was used in this analysis. This increases the background but
simplifies the analysis and increases the longitudinal momentum range of the results.
The longitudinal momentum fraction, $x'_\text{F}$, was calculated as
\begin{equation}
\label{eq:xf}
x'_\text{F} = \frac{2p_\text{L}}{\sqrt{s}} \quad \left(\approx \frac{p_\text{L}}{p_\text{L}(\mathrm{max})}\right),
\end{equation}
where $p_\text{L}$ is the longitudinal momentum of the $\rho^0$-candidate in the centre of
mass frame in the pion-nucleon interaction and $\sqrt{s}$ is the centre of
mass energy of the interaction. $p_\text{L}$ is calculated using the mean mass of the $\rho^0$ meson
($m_{\rho^0} = 0.775\,\GeVcc$) when boosting between the lab frame and the centre of mass frame. The mass of the nucleon used
in the calculations is taken to be the average of the proton and neutron masses. There is no
difference between $x'_\text{F}$ and the Feynman-$x$, $\xF=\pL/\pL(\text{max})$,
for a particle pair originating from a $\rho^0$ meson decay. For $\omega$ or K$^{*0}$ decays
the difference is less than 0.01\,in the \xpF range covered by the results presented
here. This difference approaches zero with
increasing \xpF. For simplicity, in the following, \xpF is denoted as \xF.

\subsection{Signal extraction}

\begin{figure}[t]
\centering
\def\figw{0.485}
\includegraphics[width=\figw\textwidth]{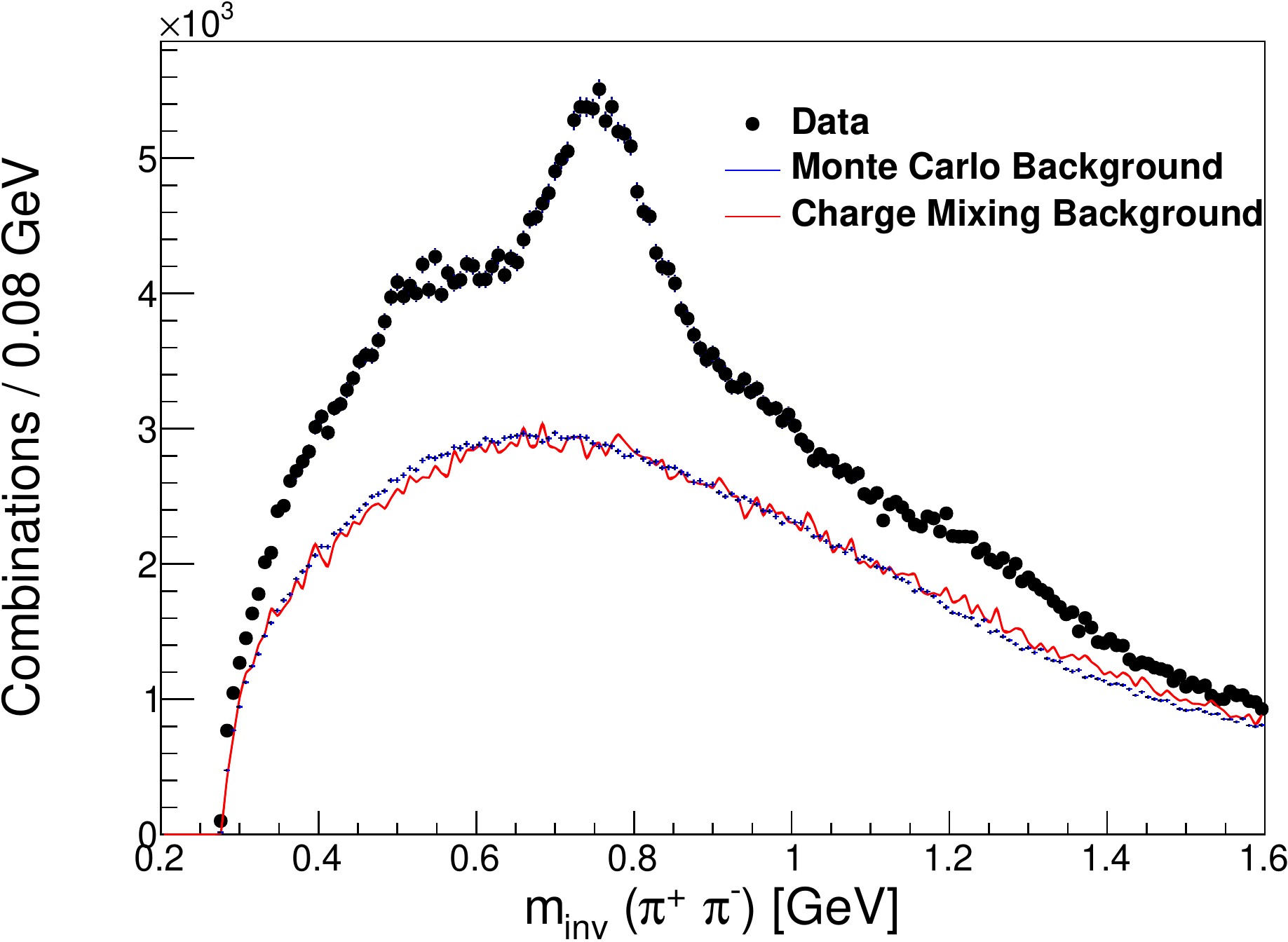}\hfill
\includegraphics[width=\figw\textwidth]{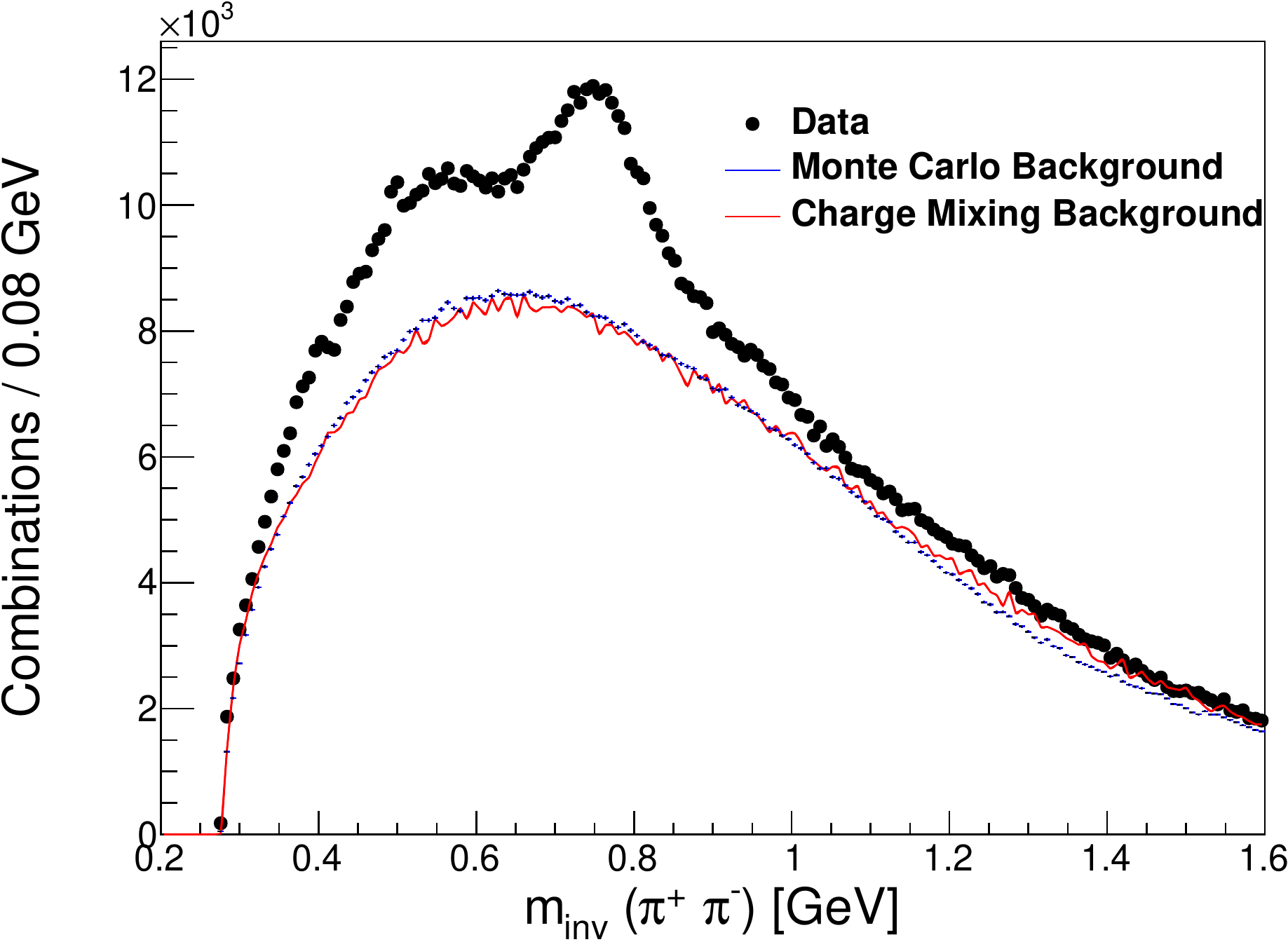}
\caption{Invariant mass distribution of opposite charge particles, calculated assuming
pion masses, in $\pi^-$+C production interactions
at 158\,\GeVc in the range $0.4 < \xF < 0.5$ (\emph{left}) and $0.3 < \xF < 0.4$ (\emph{right}). The background estimated
through the charge mixing method is shown in red and the background from the
simulation is shown in blue.}
\label{fig:Example_Data}
\end{figure}

The raw yields of $\rho^0$, $\omega$ and K$^{*0}$ mesons were obtained by performing a fit of
inclusive invariant mass spectra. These were calculated by
assuming every track that passes the cuts is a charged $\pi$. Then, for all pairs
of positively and negatively charged particles, the invariant mass
was calculated assuming pion masses for both particles. Examples of invariant mass
spectra at 158\,\GeVc and 350\,\GeVc are shown in Fig.~\ref{fig:Example_Data}.

In the inclusive invariant mass spectra, there is a large
combinatorial background, especially at low \xF. The method used to estimate
the background is the so-called charge mixing method, which uses the
invariant mass spectra calculated exactly as explained above, but using
same-charge instead of opposite-charge tracks. The resulting charge
mixing background spectra are shown in Fig.~\ref{fig:Example_Data}.
As the normalisation of these spectra will
differ from the true background, the normalisation of the charge-mixed spectra is included
as a parameter in the fit to the data. The uncertainty introduced by choosing
this method of calculating the background is estimated by comparing it with a
background found from simulations. This Monte Carlo background is defined as the sum
of:
\begin{itemize}
\item combinations of tracks that come from decays of different resonances,
i.e.\ one track from a $\rho^0$ and one from an $\omega$ (this can be done
as the parent particles of tracks are known in the simulation),
\item combinations of tracks coming directly from the interaction vertex and
\item combinations of tracks coming from resonances (both meson and baryon) that are
not included in the individual fitting-templates listed below.
\end{itemize}

As can be seen in Fig.~\ref{fig:Example_Data}, there is a good overall
agreement between the two background estimation methods and the
residual differences are used to estimate the systematics due to
background subtraction.
The boundaries of the default fit range are chosen
to include all resonances of interest and to select the invariant mass
region for which there is good agreement between the two background
estimates, and hence the results have small systematic biases.  This
leads to the fit range in $\minv(\pi^+\pi^-)$ of
$0.475{-}1.35\,\GeVcc$.

Event mixing was also investigated as an alternative way to estimate
the background by taking particles from different events to make
invariant mass spectra of $\pi^+\pi^-$ candidates, but this method was
found to not describe the shape of the background in simulations over
the mass range of the $\rho^0$, $\omega$ and K$^{*0}$ distributions
needed to obtain reliable fit results.  Refining the event mixing
method by splitting the data into multiplicity
classes did not improve the quality of this method.

As there is a large number of resonances in the $\minv(\pi^+\pi^-)$ region around
the mass of the $\rho^0$, such as the $\omega$ and K$^{*0}$ mesons,
they all have to be taken into account. This has previously
been shown in Ref.~\cite{jancso1977evidence}, where only taking into account
$\rho^0$ and $\omega$ mesons resulted in an inadequate fit, with a spurious
peak at 0.6\,\GeVcc in the $\pi^+\pi^-$ invariant mass spectra,
due to decays of K$^{*0}$ mesons, where the kaon is assigned the mass of a pion.
As there is no particle identification
used in this analysis, the effect due to K$^{*0}$ meson production is expected to be strong
and it must be included in the fitting procedure. Other contributions that are not represented by an
individual template, such as $\Lambda$ decay products, are included in the Monte Carlo background.

The fitting procedure uses templates of the invariant mass distribution
for each resonance of importance. This method of template fitting
is similar to ideas used by many other experiments such as ALICE~\cite{kohler2014dielectron},
ATLAS~\cite{ATLAS:2012vma}, CDF~\cite{Abulencia:2005aj} and CMS~\cite{CMS:2012sta},
where it is also known as the cocktail fit method. The
templates are constructed by passing simulated $\pi^-$+C production interactions,
generated with the \EposLong~\cite{Pierog:2006qv} hadronic interaction
model using \textsc{Crmc}\,1.5.3~\cite{crmc}, through the full \NASixtyOne
detector Monte Carlo chain and then through the same reconstruction
routines as the data. \textsc{Crmc} is an event generator package with access
to a variety of different event generators, such as \DPMJetLong~\cite{Roesler:2000he}
and \EposLHCLong~\cite{Pierog:2013ria}.

The template method also allows for the fitting of resonances
with dominant three body decays, such as $\omega$, as well as resonances
with two-body non-$\pi^+\pi^-$ decays, such as K$^{*0}$. A list of all decays
with a branching ratio of over $1\%$ that are used in the templates is
shown in Table~\ref{tbl:template_decays}.
The templates and the data are split into bins of \xF, calculated as in Eq.~\ref{eq:xf}.

\begin{table}
\caption{Decays of resonances for which $\minv(\pi^+\pi^-)$ templates
were calculated and fitted. Only decays
with a branching ratio greater than 1\% into at least one positively and
one negatively charged particle are considered. Branching ratios were taken from \cite{Olive:2016xmw}}.
\label{tbl:template_decays}
\begin{center}
\begin{tabular}{ccc}
\hline
Resonance & Decay & Branching ratio
\\
\hline
\multirow{1}{*}{$\rho^0$} & $\pi^+\pi^-$ & 100.0
\\
\hline
\multirow{2}{*}{$\omega$} & $\pi^+\pi^-\pi^0$ & 89.1
\\
 & $\pi^+\pi^-$ & 1.53
\\
\hline
\multirow{1}{*}{K$^{*0}$} & K\,$\pi$ & 100.0
\\
\hline
\multirow{4}{*}{f$_2$} & $\pi^+\pi^-$ & 57.0
\\
 & $\pi^+\pi^-\,2\pi^0$ & 7.7
\\
 & K$^+$K$^-$ & 4.6
\\
 & $2\pi^+\,2\pi^-$ & 2.8
\\
\hline
\multirow{2}{*}{$\eta$}	& $\pi^+\pi^-\pi^0$ & 22.7
\\
 & $\pi^+\pi^-\gamma$ & 4.6
\\
\hline
\multirow{2}{*}{f$_0$ (980)} & $\pi^+\pi^-$ & 50.0
\\
 & K$^+$K$^-$ & 12.5
\\
\hline
\multirow{4}{*}{a$_2$} & $3\pi$ & 70.1
\\
 & $\eta\,\pi$ & 14.5
\\
 & $\omega\,\pi\,\pi$ & 10.6
\\
 & K\,$\bar{\text{K}}$ & 4.9
\\
\hline
\multirow{4}{*}{$\rho_3$} & $4\pi$ & 71.1
\\
 & $\pi\,\pi$ & 23.6
\\
 & K\,K\,$\pi$ & 3.8
\\
 & K\,$\bar{\text{K}}$ & 1.58
\\
\hline
\multirow{1}{*}{K$^0_\text{S}$} & $\pi^+\pi^-$ & 69.20
\\
\hline
\end{tabular}
\end{center}
\end{table}

The templates in the fit are the charge mixing background and
the following resonances: $\rho^0$, K$^{*0}$, $\omega$, f$_2$, f$_0$ (980),
a$_2$, $\rho_3$, $\eta$ and K$^0_\text{S}$. The templates were generated
from reconstructed simulations that have all the standard reconstruction cuts
applied; they include effects due to the resolution of the detector and the
fiducial acceptance. The templates used in the fits are presented in App.~\ref{app:templates}. As can be seen, the a$_2$ and $\rho_3$ templates are broad and featureless
similar to the background template. For this reason, these resonances cannot be
fitted reliably and will be subtracted together with the background from figures
displaying the result of the template fitting in the following.

The fit to the $\pi^+\pi^-$ mass
spectrum is performed between masses of 0.475\,\GeVcc and 1.35\,\GeVcc using
the expression
\begin{equation}
\label{eq:Template_fit}
\mu(\minv) = \sum_i f_i \, T_i(\minv),
\end{equation}
where $f_i$ is the contribution for particle $i$, $T_i$ is the associated invariant
mass template and \minv is the invariant mass. $f_i$ is constrained to be between 0 and 1. The templates are normalised
to the same number of combinations as the data over the range of the fit.
The fit uses a standard Poissonian likelihood function
\begin{equation}
\mathcal{L} = \prod_j \frac{\mu_j^{k_j} e^{-\mu_j}}{k_j!},
\end{equation}
where $k_j$ is the actual number of combinations in the invariant
mass bin $j$ and $\mu_j$ is the expected number of combinations,
taken from Eq.~\eqref{eq:Template_fit}.

\begin{figure}[t]
\centering
\def\figw{0.49}
\includegraphics[width=\figw\textwidth]{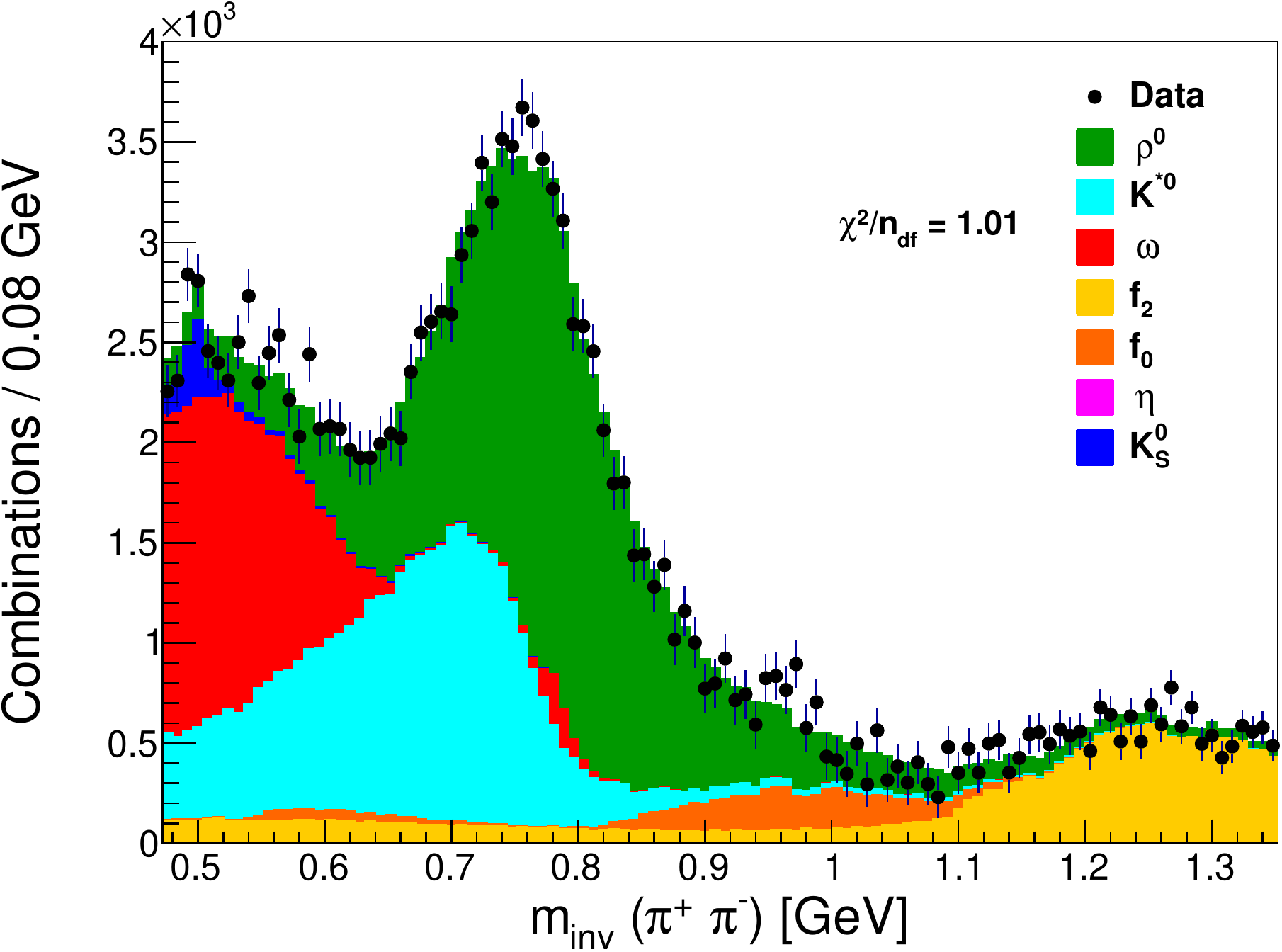}\hfill
\includegraphics[width=\figw\textwidth]{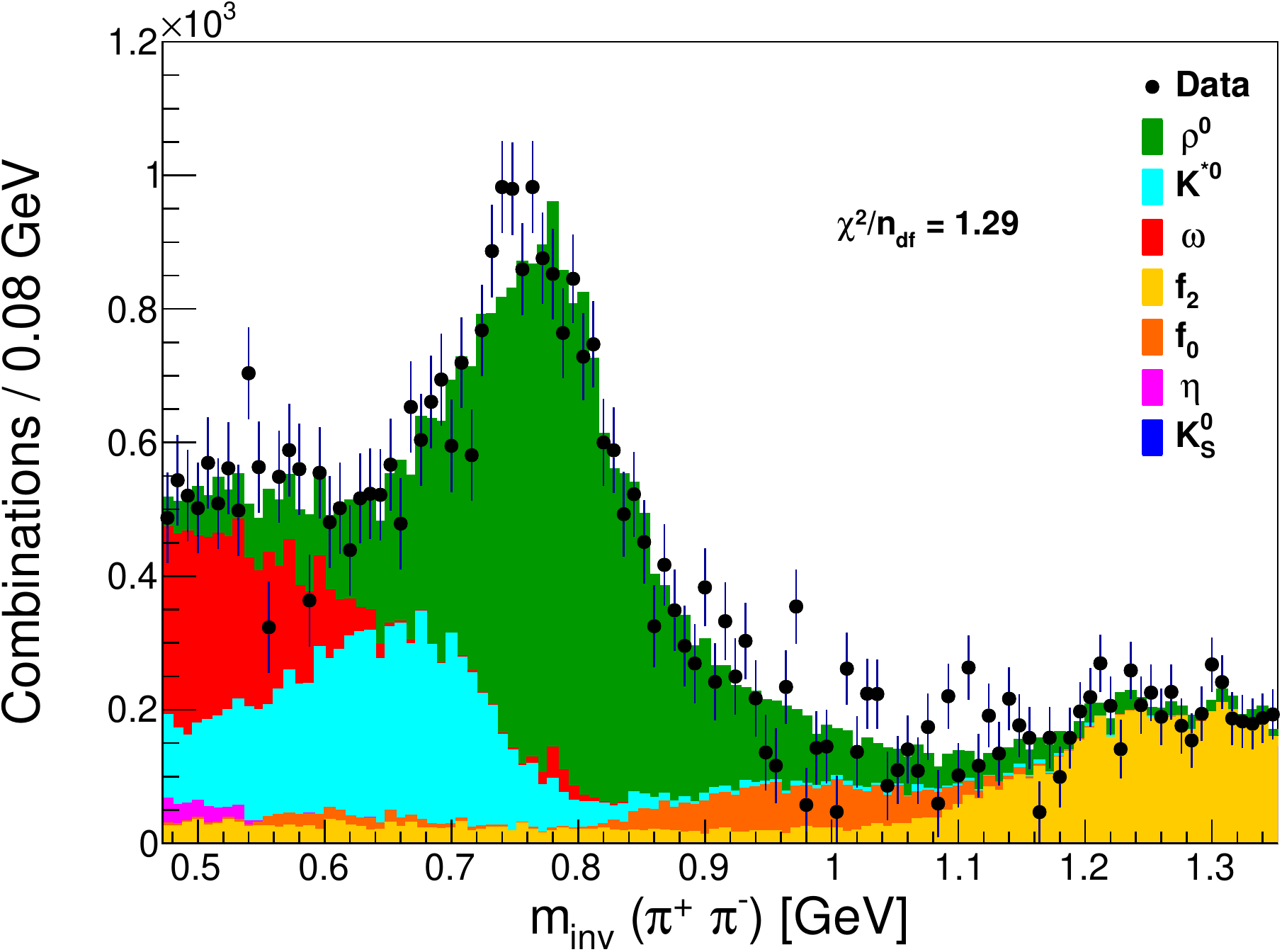}
\caption{Invariant mass distribution of oppositely charged particles, calculated assuming
pion masses, in $\pi^-$+C production interactions in the range
$0.3 < \xF < 0.4$  at 158\,\GeVc (\emph{left}) and at 350\,\GeVc
(\emph{right}). Dots with error bars denote the
data and the fitted resonance templates are shown as filled histograms.
The fitted background and featureless resonances have been subtracted.}
\label{fig:Example_Fit}
\end{figure}

Two examples of the template-fitting are shown in
Fig.~\ref{fig:Example_Fit} for 158\,\GeVc and 350\,\GeVc. The fitted
charge-mixing background as well as the contribution of the
featureless a$_2$ and $\rho_3$ resonances are subtracted to highlight
the different resonances. The full set of template fits are displayed in
App.~\ref{app:fits} for all \xF-bins and the two beam energies.

After the fractions of each templates have been determined in the fit,
the raw mean multiplicity $n_i$ of meson $i$ per event in a given \xF
bin is determined from
\begin{equation}
\label{eq:rawyield}
n_i(\xF) = \frac{1}{N_\text{acc}} \sum_j f_i \, T_i(j),
\end{equation}
where $N_\text{acc}$ is the number of events after selection cuts, $f_i$ is the result of the fit and $T_i$ is the template of the
meson of interest $i$, e.g.\ $\rho^0$.

\subsection{Correction factors}
\label{subsec:correction}

In order to obtain the true number of $\rho^0$, $\omega$ and K$^{*0}$ mesons produced in
$\pi^-$+C production interactions, three different corrections were applied to the raw yields.
These corrections were calculated using 20 million events
generated by the \EposLong model using the \textsc{Crmc} package.

\begin{enumerate}[(i)]
\item The Monte Carlo simulations that were used to obtain the templates
for the fitting procedure were used to calculate corrections due to
geometrical acceptance, reconstruction efficiency, losses due to trigger
bias, quality cuts and bin migration effects. For each \xF bin, the
correction factor $C(\xF)$ is given by
\begin{equation}
C(\xF) = \frac{n_\text{MC}^\text{gen}(\xF)}{n_\text{MC}^\text{acc}(\xF)},
\end{equation}
where
\begin{enumerate}
\item $n_\text{MC}^\text{gen}(\xF)$ is the mean multiplicity per event of $\rho^0$ ($\omega$, K$^{*0}$) mesons produced in a given
\xF bin in $\pi^-$+C production interactions at a given beam momentum, including $\rho^0$ ($\omega$, K$^{*0}$) mesons from higher
mass resonance decays and
\item $n_\text{MC}^\text{acc}(\xF)$ is the mean multiplicity per event of reconstructed $\rho^0$ ($\omega$, K$^{*0}$) mesons that
are accepted after applying all event and track cuts.
\end{enumerate}

The statistical uncertainties of the corrections factors were calculated assuming binomial distributions for
the number of events and resonances.

\item The contribution from $\rho^0$ mesons produced by re-interactions
in the target. This was estimated from the simulations. This
contribution is less than 1\% for all bins apart from $\xF<0.15$,
where the contribution is 1.7\%.

\item The fitting method was validated by applying the same procedure to
the simulated data set, using the background  estimated from either the charge
mixing method or the true background obtained from the simulation. This difference is then applied
as a multiplicative correction to the raw yield, $f_i^\text{true} / f_i^\text{fit}$,
where $f_i^\text{true}$ is the true yield of resonance $i$ and $f_i^\text{fit}$
is the yield that comes from the fit to the simulations. This correction is
calculated separately for both background estimations and applied to the fits
to the data that used the same estimation.

\end{enumerate}

\begin{figure}[t]
\centering
\def\figw{0.495}
\includegraphics[width=\figw\textwidth]{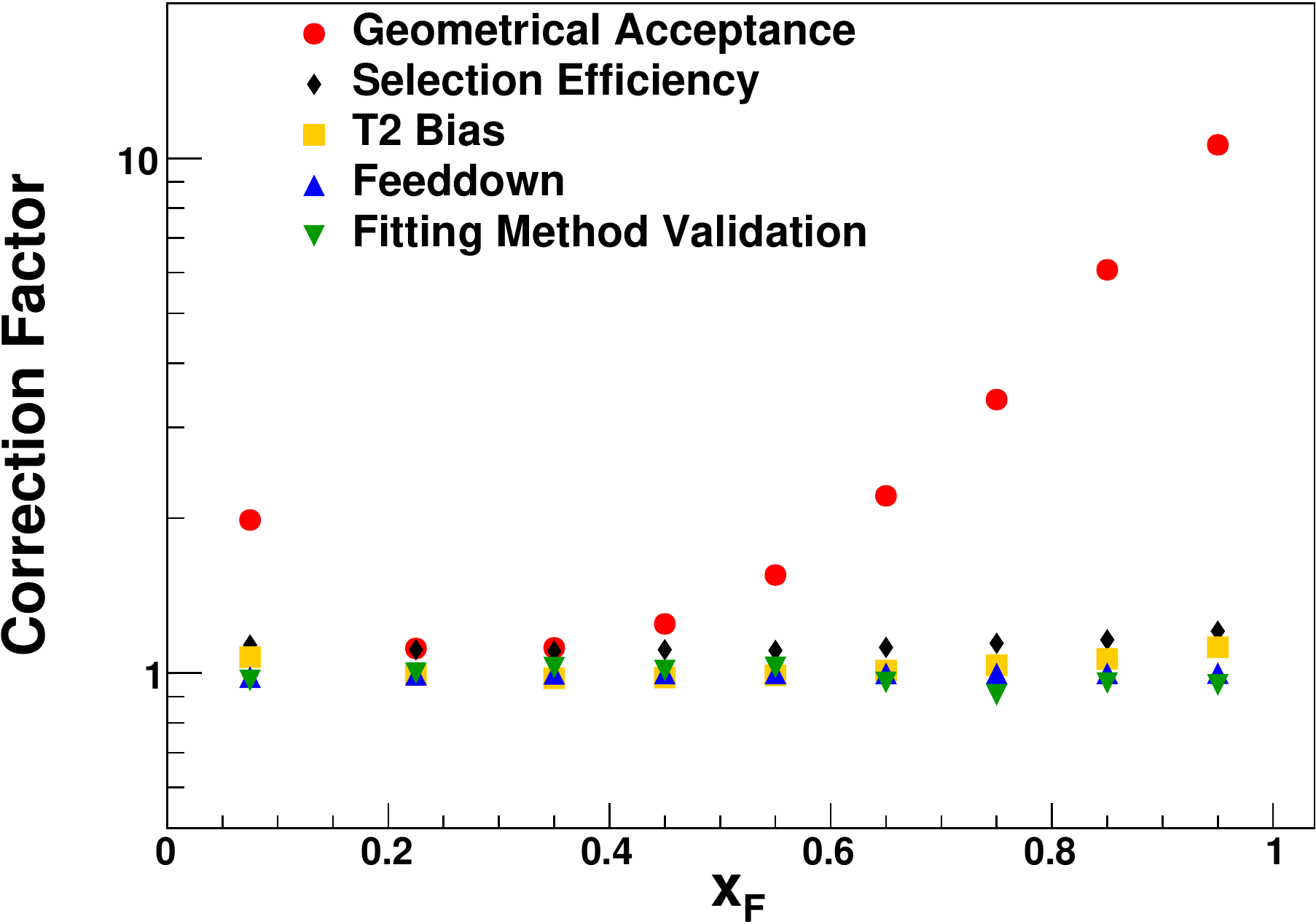}
\includegraphics[width=\figw\textwidth]{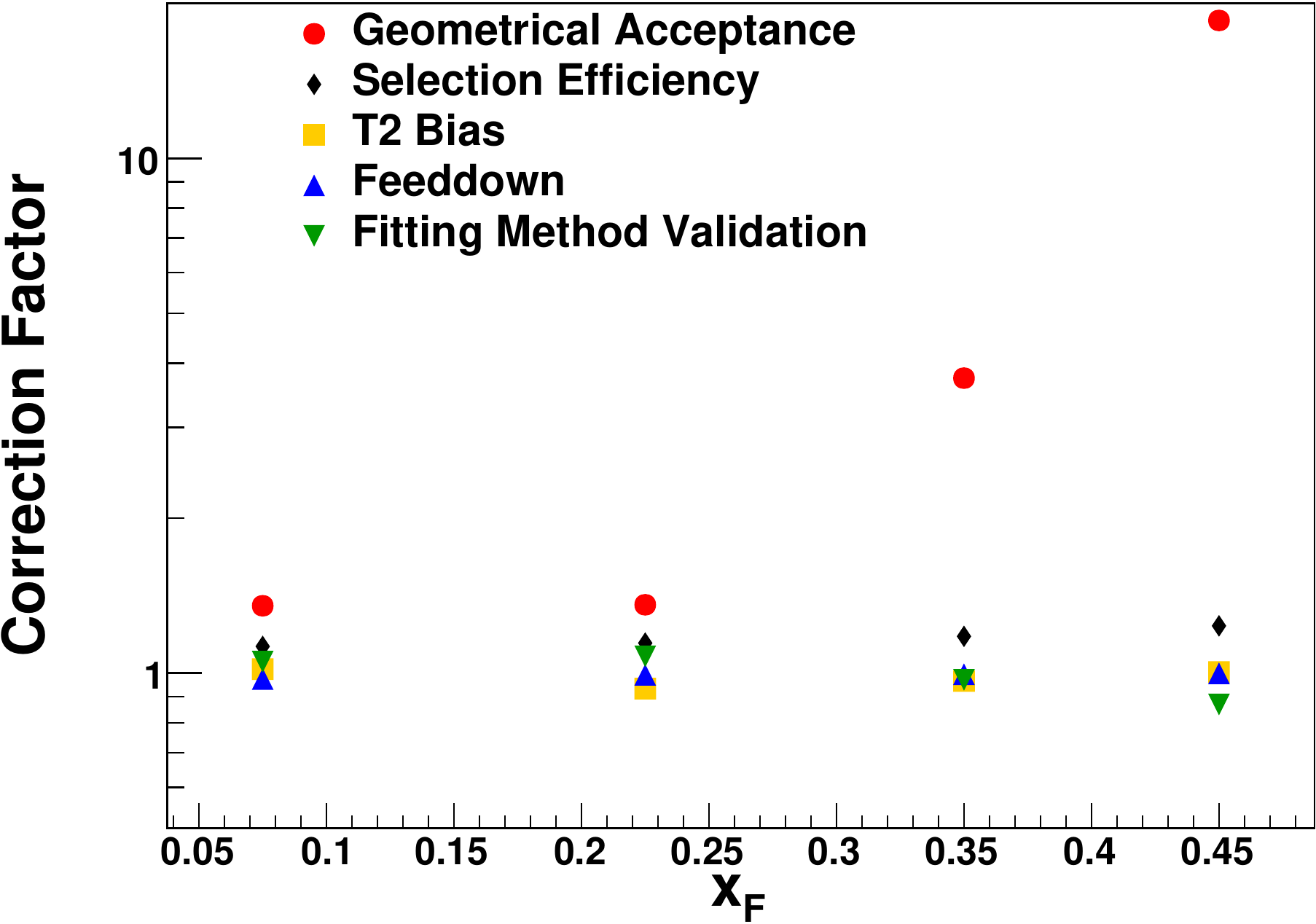}
\caption{Correction factors for the $\rho^0$ spectra in $\pi^-$+C production interactions
at 158\,\GeVc (\emph{left}) and 350\,\GeVc (\emph{right}).
It can easily be seen that the correction for geometrical acceptance dominates in
almost every bin.}
\label{fig:Corr_Factors}
\end{figure}

The breakdown of these correction factors can be seen, for the $\rho^0$ spectra
at $p_\text{beam}=158$ and $350\,\GeVc$, in Fig.~\ref{fig:Corr_Factors}.
The correction factor $C(\xF)$ is broken down into three contributions:
bias from the interaction trigger (T2), geometrical acceptance, and selection efficiency. The geometrical acceptance
dominates for large \xF values.

The correction derived from Monte Carlo simulations could introduce a bias in the
result if the \pT spectrum of the model differed from the true shape. This is
because the extrapolation to full \pT phase space is based on the model spectrum.
To investigate this effect another hadronic interaction model was used,
\DPMJetLong~\cite{Roesler:2000he}. This model also provides \pT spectra for each resonance
measured in this analysis, and the difference between the correction factors found
for \DPMJetLong and \EposLong is less than 4\%. This suggests that any bias introduced by the
extrapolation to full \pT phase space is small. The difference between the correction
factors is used in the estimate of the systematic uncertainties.

The final measurement is calculated by taking the average of the result using the
two different background description methods, charge mixing and Monte Carlo background,
with all the correction factors that change calculated separately for the two methods.
The difference between these two methods is taken to be a systematic uncertainty.

\subsection{Uncertainties and Cross Checks}
\label{subsec:uncertainties}

The statistical uncertainties in the $i$th \xF-bin are given by
\begin{equation}
\sigma_i^2 = \left(\Delta C_i \, n_i\right)^2 + \left(\frac{\sigma(n_i)}{C_i}\right)^2,
\end{equation}
where $n_i$ and $\sigma(n_i)$ are the raw meson mean multiplicity per
event and the uncertainty on this multiplicity that comes from the
template fit. The contribution due to the uncertainty of the meson
multiplicity dominates as the uncertainty $\Delta C_i$ of the
corrections factors is only from the statistics of the simulation (20
million events) which is much larger than that of the data.

The main contributors to the systematic uncertainties are
\begin{enumerate}[(i)]
\item The fitting method used for estimating the background shape and the fit procedure.
The systematic uncertainty is taken to be half the difference between the two methods,
using either charge mixing or Monte Carlo background, after the respective validation
corrections have been applied. This estimate therefore
combines the systematic uncertainty due to both the fitting method validation correction and
the background estimation used and this is the dominant systematic uncertainty.

\item Correction factors. The correction factors calculated above were compared
with factors found using a different hadronic interaction model, \DPMJetLong.

\item Track cuts. The effect of the event and track selection cuts were checked
by performing the analysis with the following cuts changed, compared to the values
shown in Sec.~\ref{subsec:selection}.
\begin{enumerate}
\item The cut on the $z$-position of the interaction vertex was changed to be between
$-590$\,cm and $-570$\,cm.
\item The window in which off-time beam particles were not allowed was decreased from
2\,$\upmu$s to 1.5\,$\upmu$s.
\item The minimum number of clusters on the track was decreased to 25.
\item The sum of clusters on the track in VTPC-1 and VTPC-2 was
decreased to 12 or increased to 18.
\item The impact parameter cuts were increased to less than 4\,cm in the $x$-plane
and 2\,cm in the $y$-plane.
\end{enumerate}

\end{enumerate}

The systematic uncertainties were estimated from the differences between the results
obtained using the standard analysis and ones obtained when adjusting the method as
listed above. The individual systematic uncertainties were added in quadrature to obtain
the total systematic uncertainties.
They are dominated by
the correction factor contribution, up to 15\%, whereas the other contributions are
less than 4\%. Other sources of uncertainty, such as using templates
from a different model, are found to be much smaller.

The fraction of target removed tracks is less than $0.15\%$ in all \xF bins.
The shape of the target removed distributions, after applying all the track and event cuts, is
consistent with the background description so there is no additional correction or systematic
uncertainty considered.

Several cross checks were performed to validate the results and check their stability.
These include extending the range of the $\minv(\pi^+\pi^-)$ fit,
using the Breit-Wigner function to describe the $\rho^0$ instead of a template as
well as a few other more simple checks.

\subsubsection{Fit range}

The default fit range used in this analysis was restricted to the mass
ranges of the resonances of interest. We tested an extended fit range
by including all data down to the kinematic threshold of $\minv(\pi^+\pi^-) = 2m_\pi$. For this
purpose additional templates needed to be taken into account including
electrons and positrons pair-produced in the target by photons from
$\pi^0$ decays. The sum of all resonances produced by the \EposLong model can however
not describe the low $\minv(\pi^+\pi^-)$ region satisfactorily.  In
particular, a significant bump at a mass of ${\approx}0.4\,\GeVcc$
appears to be in the data that does not have a counterpart in the
templates.  No resonance, meson or baryon, could be found in \EposLong that
could describe this bump. To avoid any bias the region of $0.35\,\GeVcc < \minv(\pi^+\pi^-)
< 0.4\,\GeVcc$ was excluded from the fit. Further discussions about the
study of this bump are given in App.~\ref{app:discussion}.

\begin{figure}[t]
\centering
\def\figw{0.485}
\includegraphics[width=\figw\textwidth]{Fits_158-03}\hfill
\includegraphics[width=\figw\textwidth]{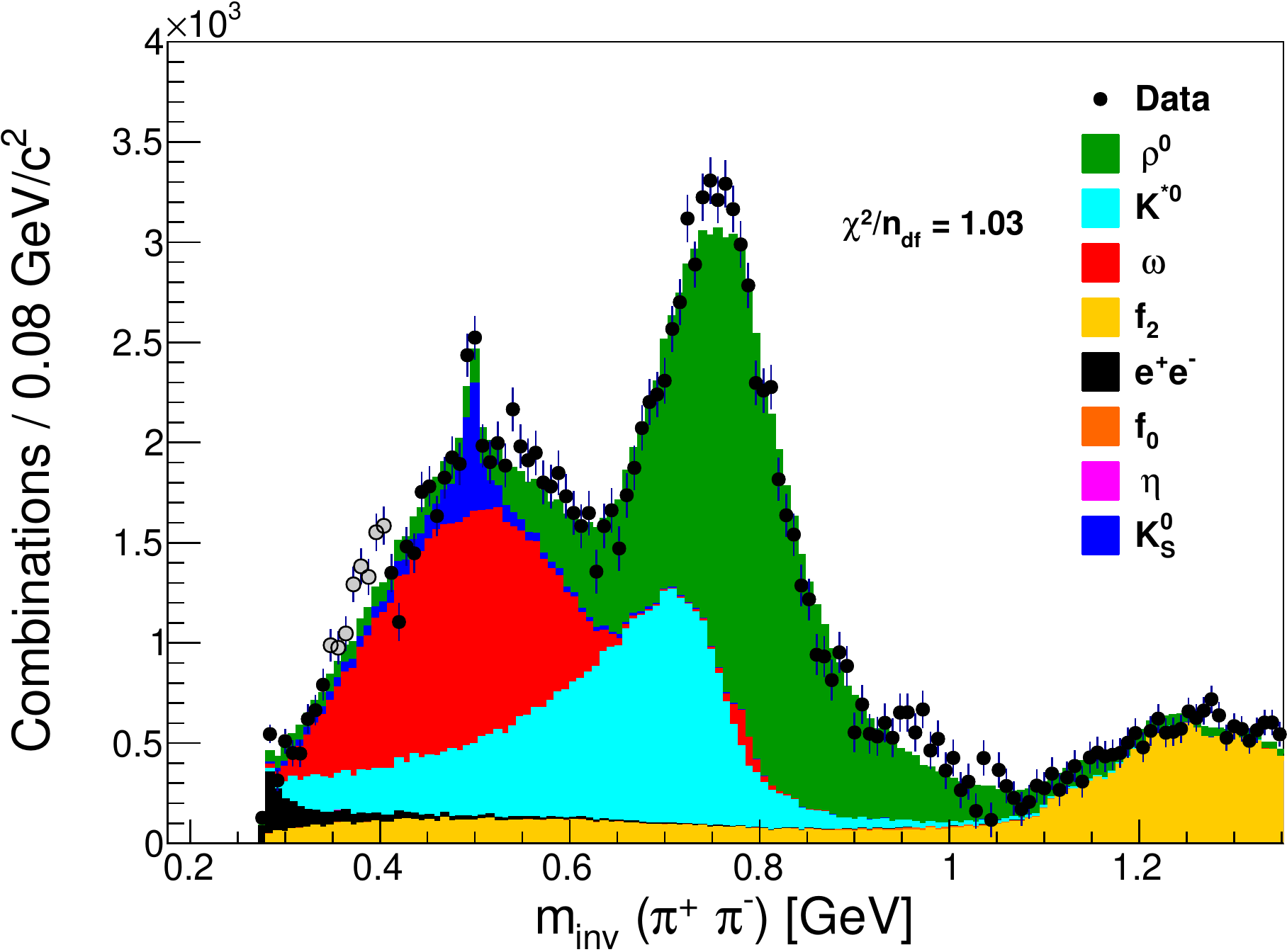}
\caption{An example of the template fit to 158\,\GeVc data in the range $0.3 < \xF < 0.4$
using the nominal fit range (\emph{left}) and the extended fit range (\emph{right}) including
a template for $e^+e^-$ pair production.}
\label{fig:ExtendedRange}
\end{figure}

Once this region is excluded from the fit a reasonable description
of the \minv distribution down to the kinematic limit can be achieved,
as shown in Fig.~\ref{fig:ExtendedRange}. However, the fit
quality is worse and the agreement between the two background estimates is weaker.
The poorer fit quality is most likely a combination
of poorer performance of the estimate of the combinatorial background close to the
kinematic threshold and the missing template to describe the bump at ${\approx}0.375\,\GeVcc$.

The yields obtained with the extended range differ by less than the systematic
uncertainties from the yields with the original range, with the exception of one bin,
and, to be conservative, the corresponding differences, which are of the order
of 10\%, are included in the systematic uncertainty.

\subsubsection[rho0 mass]{$\rho^0$ mass}

\begin{figure}[t]
\centering
\includegraphics[width=0.74\linewidth]{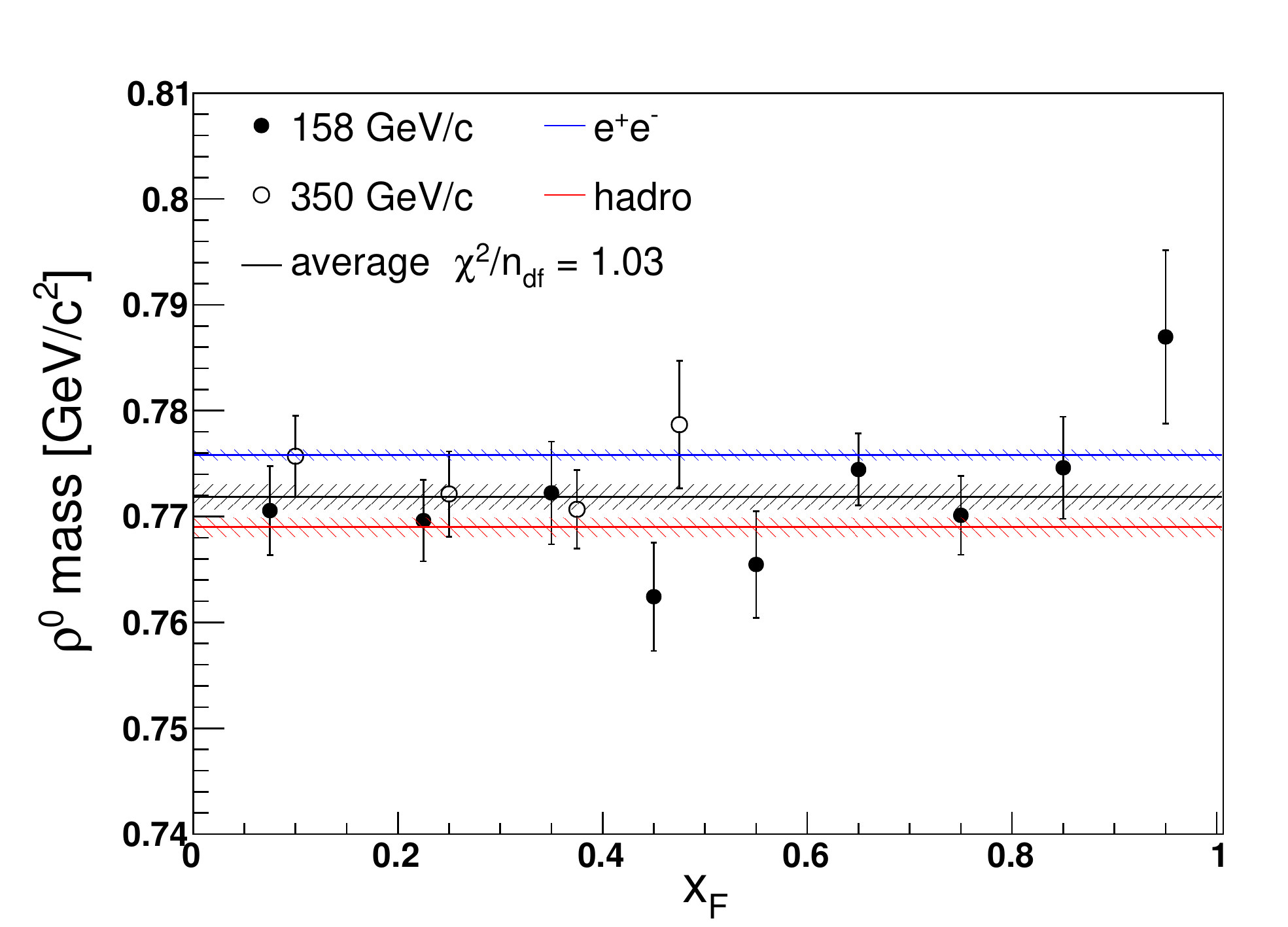}
\caption{Fitted $\rho^0$ mass for $\pi^-$+C production interactions as a function of \xF. The blue line indicates the average mass
from $e^+e^-$ annihilation and the red line indicates the average mass from other reactions, which is
dominated by hadroproduction measurements~\cite{Olive:2016xmw}. The black line is the weighted mean
of all measurements, combining both 158 and 350\,\GeVc data.}
\label{fig:RhoMass}
\end{figure}

We checked for possible nuclear effects on the $\rho^0$ mass~\cite{Hayano:2008vn, Jin:1995qg}
by removing the $\rho^0$ template from the fit and replacing
it with a Breit-Wigner function. The function used is the one used in Ref.~\cite{Adler:2002sc} with a modification to
the decay width following Refs.~\cite{Teis:1996kx} and \cite{Koch:1983tf}, where
the decay width is a function of mass \minv,
\begin{equation}
\operatorname{BW}(\minv) =
  \frac{\minv \, m_\text{R} \, \Gamma}
       {(m_\text{R}^2 - \minv^2)^2 + m_\text{R}^2 \, \Gamma^2},
\label{eq:BW}
\end{equation}
where $m_\text{R}$ is the mean mass of the fitted resonance and $\Gamma$ is given by
\begin{equation}
\Gamma(\minv) =
  \Gamma_0
  \left(
    \frac{m_\text{R}}
         {\minv}
  \right)
  \left(
    \frac{q}{q_\text{R}}
  \right)^{3/2}
  \left(
    \frac{q_\text{R}^2 + \delta^2}
         {q^2 + \delta^2}
  \right),
\end{equation}
where $q$ and $q_\text{R}$ are the pion three-momenta in the rest frame of the resonance, calculated
with mass \minv and $m_\text{R}$, respectively. The parameter $\delta$ in the cutoff function
has a value $\delta = 0.3\,\GeVc$.

We considered the mass as a free parameter and fixed
the width value to the one provided by the particle data group~\cite{Olive:2016xmw}.
The obtained mass values are consistent with the values quoted
by the particle data group as shown in Fig.~\ref{fig:RhoMass}. The weighted average of the fitted masses is
$0.772\pm0.001\,\GeVcc$, with no significant difference between the 158 and 350\,\GeVc data.

A simpler Breit-Wigner function was also tested,
\begin{equation}
\label{eq:sBW}
\operatorname{BW}(M) =
  \frac{\Gamma^2}
       {(M - m_\text{R})^2 + \Gamma^2}
\end{equation}
It is the function used to both sample resonances and generate their
widths in \EposLong. Even though this function does not directly take into account effects which are
considered in the event generator, such as decay products approaching the lower kinematic limit, or
energy conservation for decay products at higher mass, the resulting fitted masses are compatible with
the results from the more complicated Breit-Wigner function, Eq.~\eqref{eq:BW}.

The yields of the $\rho^0$ when fitting with this Breit-Wigner function differ slightly from the yields
calculated using the standard analysis method. These small differences of the order of 3\% are included in the
systematic uncertainties.

\begin{figure}[t]
\centering
\def\figw{0.485}
\includegraphics[width=\figw\textwidth]{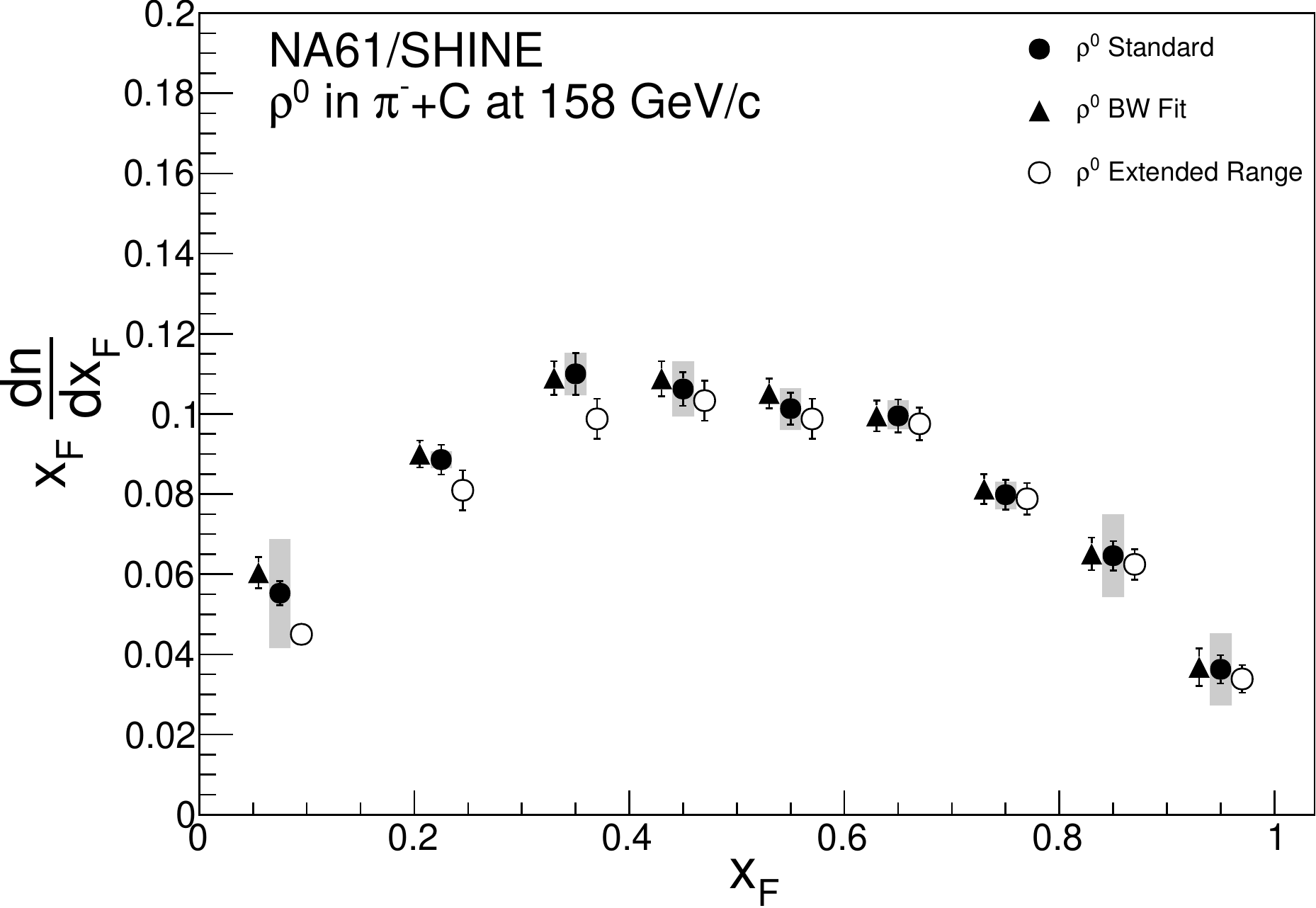}\hfill
\includegraphics[width=\figw\textwidth]{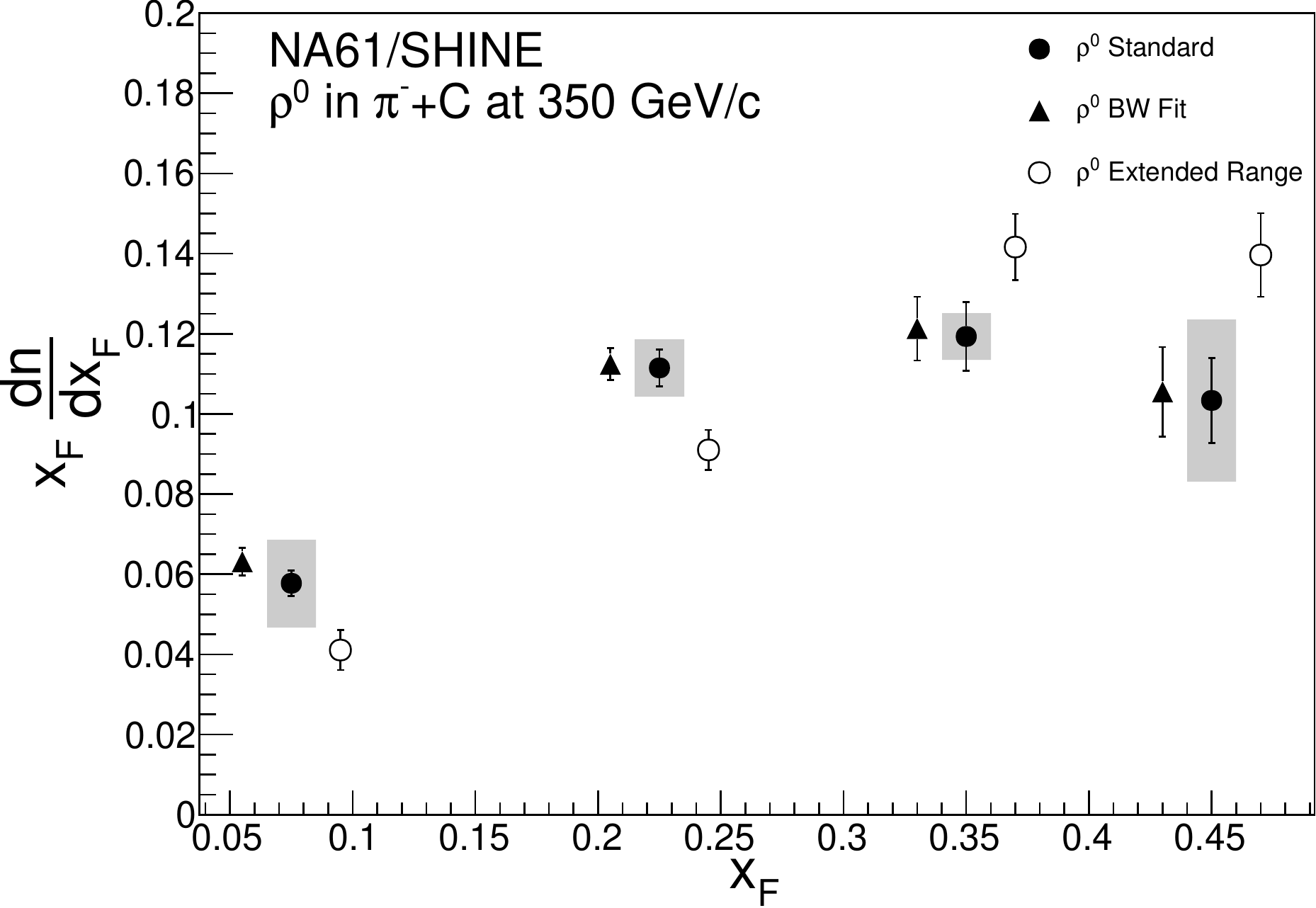}
\caption{Comparison of $\rho^0$ \xF spectra at 158\,\GeVc (\emph{left}) and 350\,\GeVc (\emph{right}) from the standard template analysis method, the extended fit range and when using
a Breit-Wigner function to parameterise the $\rho^0$. The systematic uncertainties shown are before adding contributions from the
differences to the extended fit range and Breit-Wigner function fits.}
\label{fig:CrossCheck}
\end{figure}

A comparison of the yields from the standard template analysis method, the extended fit range and when fitting
a Breit-Wigner function (Eq.~\eqref{eq:BW}) is shown in Fig.~\ref{fig:CrossCheck}. As can be seen the differences are
within the systematic uncertainties of the standard analysis. These small differences, of the order of 3\% for the fits with
a Breit-Wigner function and 10\% for the extended fit range, are added in
quadrature to the systematic uncertainties.

\subsubsection{Further checks}

Further cross checks were performed to probe the
stability of the fit and yield result. These include
\begin{enumerate}[(i)]
\item The data, along with the templates, were split into two equally sized regions of polar
angle. If there was any polar-angle dependence of the result introduced
by insufficient modeling of different parts of the detector, this would appear in a
difference between the spectra from these independent data sets.
The resulting multiplicity spectra were consistent within statistical uncertainties.
\item The data set was split according to different time ranges, both a night and day split
as well as a first half and second half split in run taking. Any possible systematic
differences in the detector which depend on time would result in discrepancies in the
spectra from the different time ranges. Both resulting \xF
spectra were again consistent within statistical uncertainties.
\item Instead of assuming the pion mass for both tracks, one track was allocated the
kaon mass. This means that the number of combinations used has to double, as both
combinations of masses have to be taken into account for any given pair of tracks to
allow for the kaon to be either of the two charges.
This also then increases the background even further and because of the different
shape of the background under the $\pi$\,K invariant mass distribution, the systematic
uncertainty for this method is larger than for the $\pi\,\pi$ method. The multiplicity spectra
from this method were consistent within statistical and systematic uncertainties of the
standard analysis method.
\end{enumerate}

All these performed cross checks gave results consistent within the total uncertainties of the
standard analysis.

  \section{Results}
\label{sec:results}

The yields of $\rho^0$, $\omega$, and K$^{*0}$ mesons in $\pi^{-}$+C
production interactions at 158\,\GeVc and 350\,\GeVc
were calculated in bins of $x_\text{F}$ as follows
\begin{equation}
\frac{\mathrm{d}n}{\mathrm{d}\xF} =
  \frac{1}
       {N_\text{prod}}
  \frac{\mathrm{d}N_\text{part}}
       {\mathrm{d}\xF} =
  \frac{C(\xF) \, n(\xF)}
       {\Delta\xF},
\end{equation}
where $N_\text{prod}$ is the number of interaction events minus
the events with elastic and quasi-elastic scattering (which are not
included), $N_\text{part}$ is the true number of produced
resonances, $n(\xF)$ is the raw mean multiplicity per event of
the meson from Eq.~\eqref{eq:rawyield}, $\Delta\xF$ is the
width of the $\xF$ bin and $C(\xF)$ is the total
correction factor for losses of event and multiplicity, as detailed
above. Measured points with large statistical or systematic
uncertainties (greater than 50\%) are not shown. This cut removes
three data points at large $\xF$ for the $\omega$ spectrum and
one data point at large $\xF$ for the K$^{*0}$ spectrum at
158\,\GeVc. In case of the data taken at $350\,\GeVc$ only a limited
$\xF$-range between 0 and 0.5 is accessible within the
acceptance of \NASixtyOne. Only one data point of the $\omega$
spectrum survived the cut on the maximum uncertainty and none for the
K$^{*0}$ spectrum. Therefore we present only $\rho^0$ spectra for
the $350\,\GeVc$ data.

\begin{figure*}[t]
\centering
\includegraphics[width=0.9\textwidth]{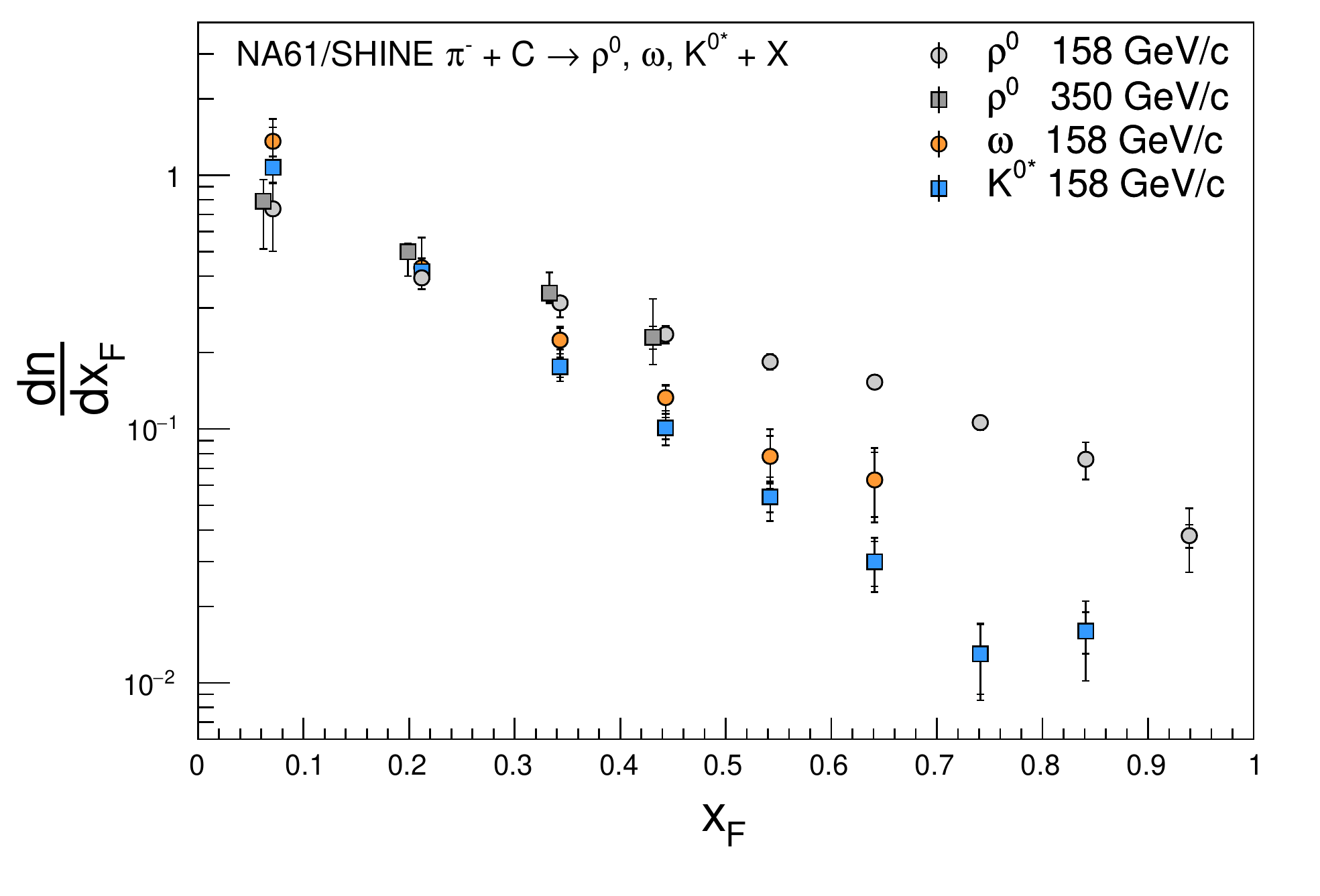}
\caption{Average multiplicity of meson resonances in $\pi^-$+C
  collisions.  The results for $\rho^0$ mesons are shown for $p_\text{beam}=158$
  and 350\,\GeVc and the spectra of $\omega$ and K$^{*0}$
  mesons were measured at $p_\text{beam}=158$\,\GeVc. The inner error
  bars show the statistical uncertainty and the outer error bar
  denotes the total uncertainty obtained by adding statistical and
  systematic uncertainties in quadrature.}
\label{fig:mesonData}
\end{figure*}

The spectra of $\rho^0$, $\omega$, and K$^{*0}$ mesons produced in
production $\pi^-$+C interactions are shown in
Fig.~\ref{fig:mesonData}.
The average $\xF$ in each bin is used
to display the data points in this and in the following figures. It is
worthwhile noting that this average is not corrected for the detector
acceptance within the bin and is calculated from all oppositely charge
combinations including combinatorial background. For a detailed
comparison of this data with model predictions it is therefore
recommended to compare to model predictions binned in the same way as
the data rather than comparing them at the average $\xF$.

As can be seen in Fig.~\ref{fig:mesonData}, no dependence of the
$\rho^0$ multiplicities on beam energy was found within the
uncertainties of the measurement. Out of the three resonances studied
here, the multiplicity of $\rho^0$ mesons is the largest at large
$\xF$, i.e.\ the region most relevant for the development of
cosmic-ray air showers.  Numerical results, including statistical and
systematic uncertainties, are given in
Tables~\ref{tbl:rho0_multiplicity}, \ref{tbl:omega_multiplicity},
and~\ref{tbl:kstar_multiplicity}. It is worthwhile noting that due to
improvements in the analysis procedure the final $\rho^0$
multiplicities at 158~\GeVc listed in Tab.~\ref{tbl:rho0_multiplicity}
are about 25\% smaller than the preliminary results presented
in~\cite{ICRC2015}.

\begin{figure}[t]
\centering
\def\figw{0.485}
\includegraphics[width=\figw\textwidth]{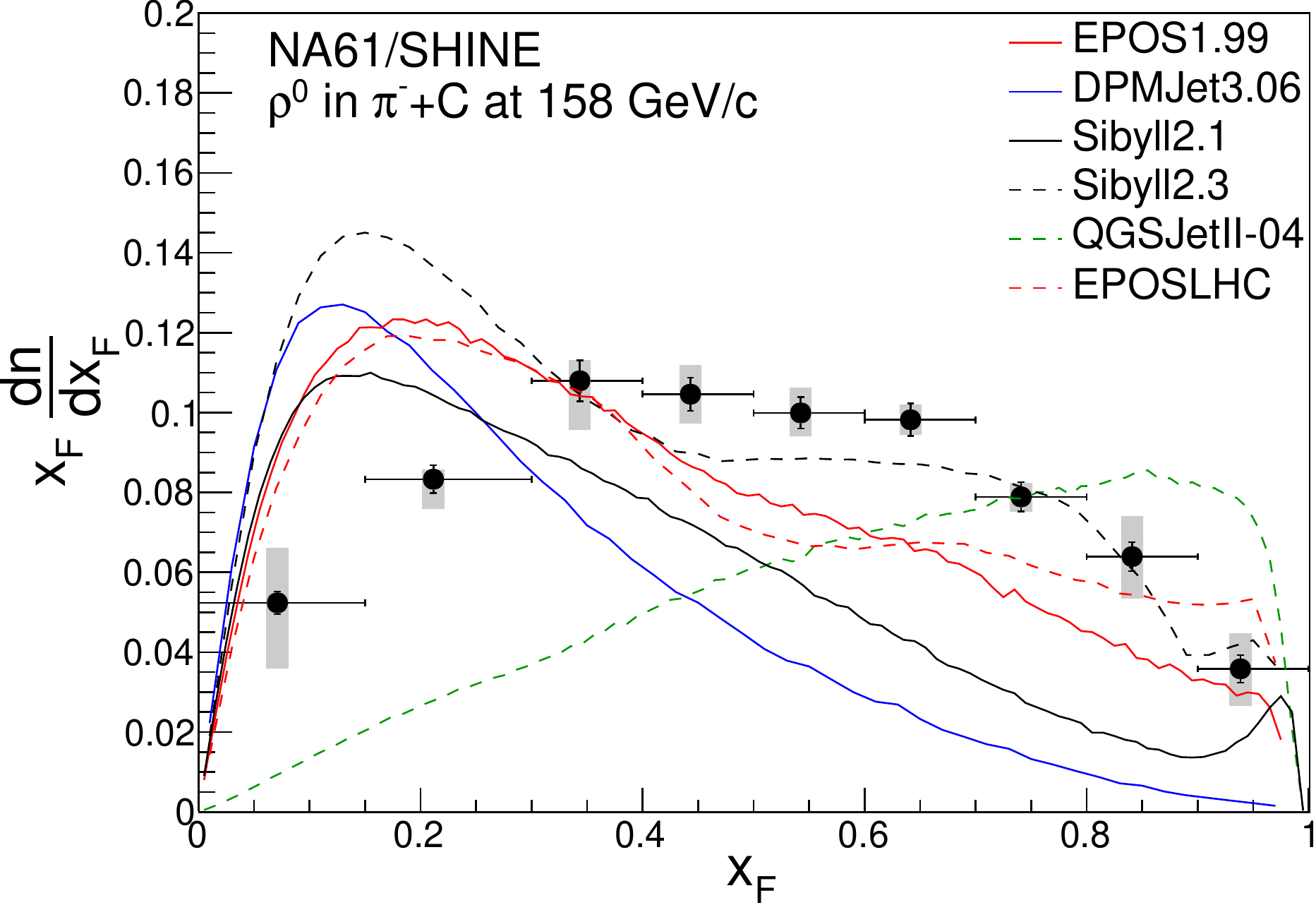}\hfill
\includegraphics[width=\figw\linewidth]{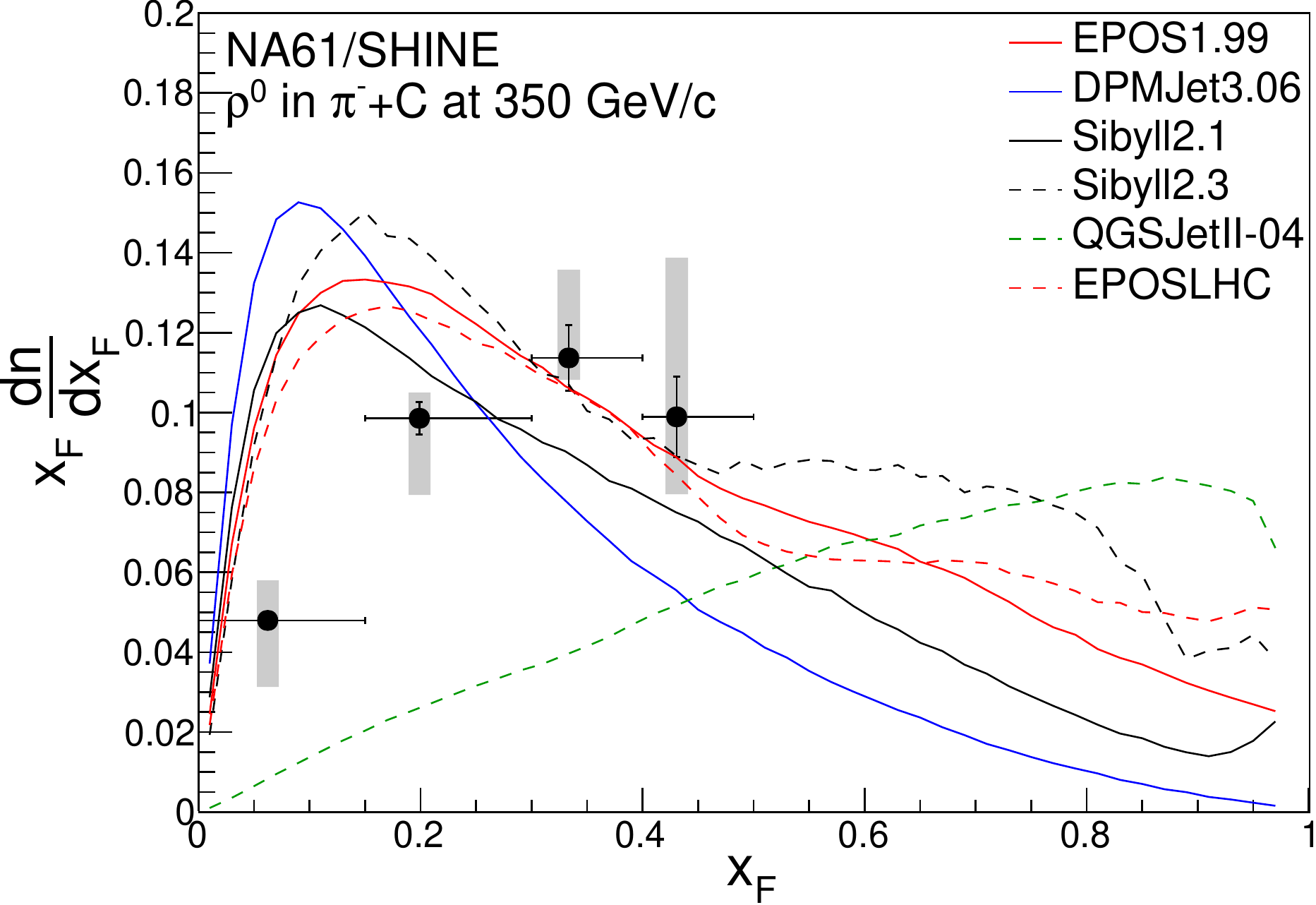}
\caption{Scaled \xF-spectra of $\rho^0$ mesons, $\xF\,\mathrm{d}n/\mathrm{d}\xF$,
in $\pi^-$+C production interactions at 158 \emph{(left)} and $350\,\GeVc$ \emph{(right)}. The error bars
show the statistical, the bands indicate systematic uncertainties.
The lines depict predictions of hadronic interaction models: red --
\EposLong, blue -- \DPMJetLong, black -- \SibyllLong, dashed green --
\QGSJetLong, dashed red -- \EposLHCLong, dashed black -- \SibyllNewLong.}
\label{fig:Rho_Result}
\end{figure}

\begin{figure}[t]
\centering\
\def\figw{0.485}
\includegraphics[width=\figw\textwidth]{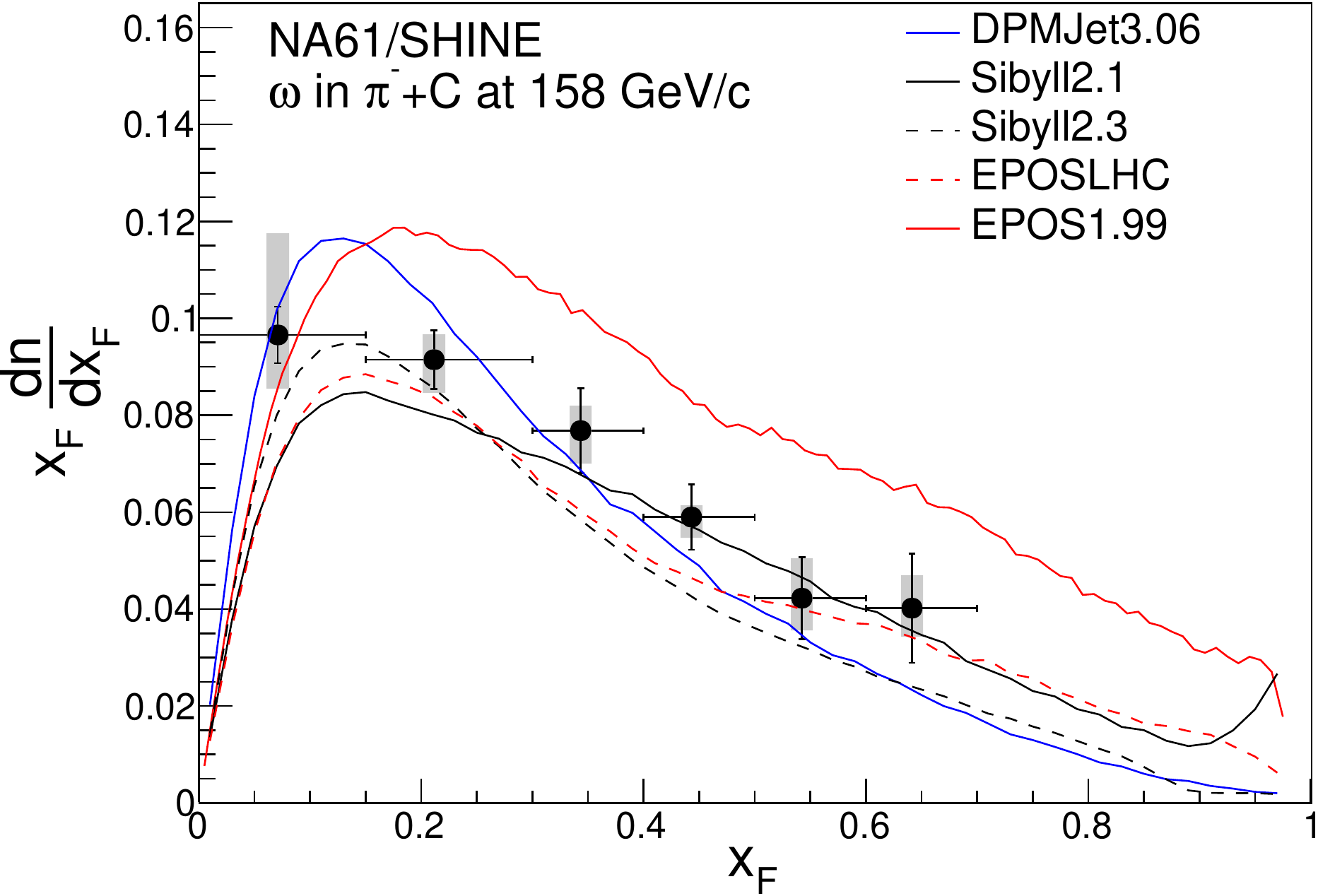}\hfill
\includegraphics[width=\figw\textwidth]{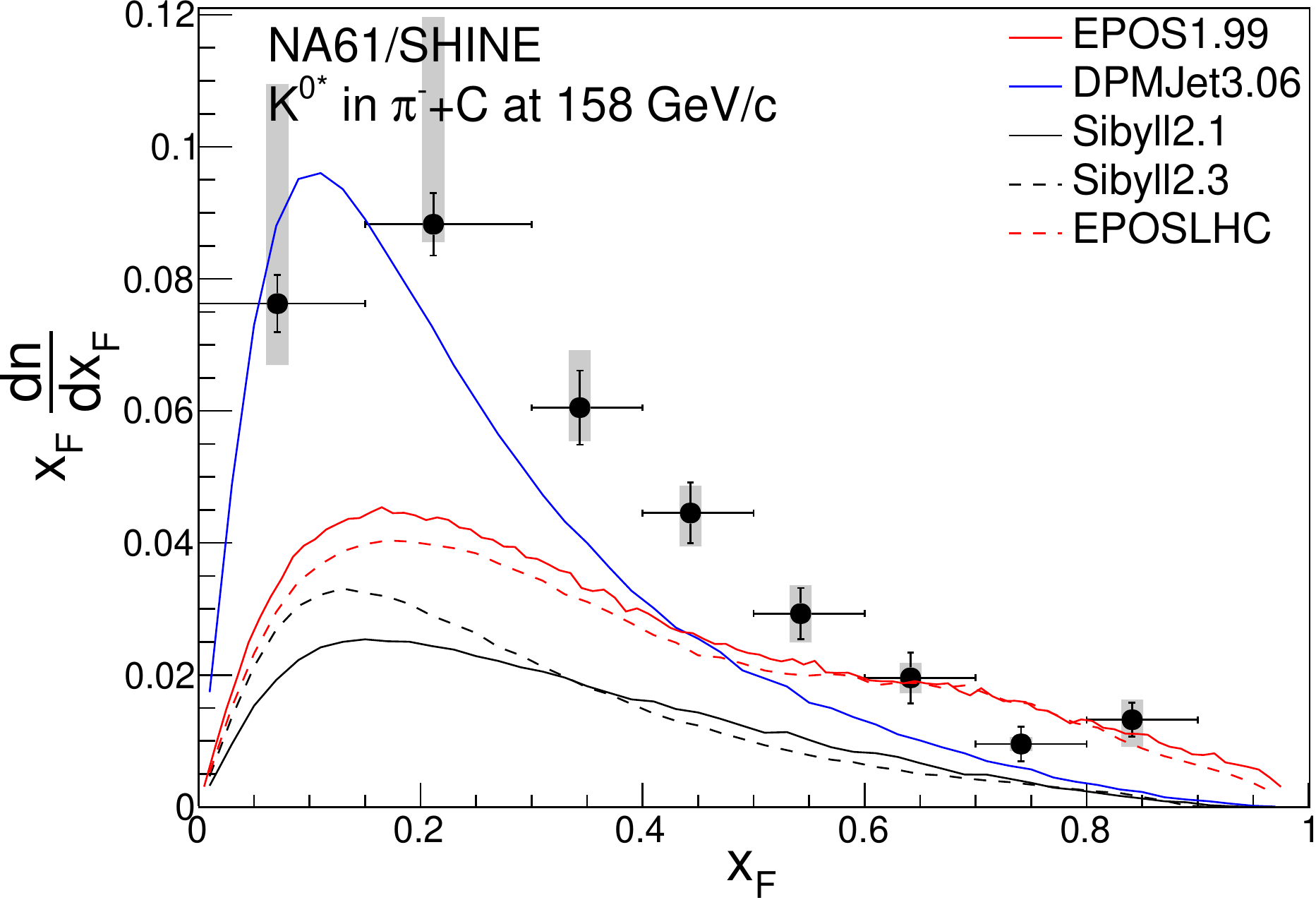}
\caption{Scaled \xF-spectra of $\omega$ \emph{(left)} and K$^{*0}$ \emph{(right)}
mesons, $\xF\,\mathrm{d}n/\mathrm{d}\xF$,
in $\pi^-$+C production interactions at $158\,\GeVc$. The error bars
show the statistical, the bands indicate systematic uncertainties.
The lines depict predictions of hadronic interaction models: red --
\EposLong, blue -- \DPMJetLong, black -- \SibyllLong, dashed red -- \EposLHCLong,
dashed black -- \SibyllNewLong.}
\label{fig:KStarOmega_Result}
\end{figure}


The measured spectra are compared to model predictions by
\QGSJetLong~\cite{Ostapchenko:2010vb}, \EposLong~\cite{Pierog:2006qv},
\DPMJetLong~\cite{Roesler:2000he}, \SibyllLong~\cite{Ahn:2009wx},
\SibyllNewLong~\cite{Riehn:2015oba} and
\EposLHCLong~\cite{Pierog:2013ria} in Figs.~\ref{fig:Rho_Result} and
\ref{fig:KStarOmega_Result}. For the purpose of display, the
multiplicities were scaled by \xF.

It can be seen that in the low \xF region ($<0.3$) all
hadronic interaction models overestimate the $\rho^0$ yield with
discrepancies of up to +80\%.  At intermediate \xF ($0.4 <
\xF < 0.7$) the $\rho^0$ production is underestimated by up to
$-60$\%. It is interesting to note that even if \QGSJetLong,
\SibyllNewLong and \EposLHCLong were tuned to $\pi^+$+p data from
NA22~\cite{Agababyan:1990df}, these models cannot reproduce the
measurement presented here. The large underestimation in \QGSJetLong
is mainly for non-forward $\rho^0$ production which is not treated
explicitly in the model.  This explains the large difference in
spectral shape compared to the other hadronic models and the large
deviations between the model and the measurement.  The best
description of our data in the forward range ($\xF>0.4$) is
given by \SibyllNewLong, which describes the data within 10\%.

The shape of the measured $\omega$ spectrum is in approximate
agreement with all of the models shown (\QGSJetLong does not include
$\omega$ mesons in the model). Also the measured normalisation is
approximately reproduced by all models but \EposLong, which produces
too many $\omega$ mesons above $\xF>0.1$.

The measured multiplicity of K$^{*0}$ mesons is not reproduced
by any of the models over the full $\xF$ range. \DPMJetLong
gives a correct description of the yields only at low $\xF$ but
underpredicts the multiplicity at large $\xF$ and the opposite
is true for \EposLHCLong and \EposLong which are in agreement with the measurement
only at $\xF\gtrsim 0.6$. \SibyllNewLong and \SibyllLong predict a too low
number of K$^{*0}$ mesons at all $\xF$ values.

The ratio between combinations of the three meson measurements are
shown in App.~\ref{app:ratio}, where it can be seen that no model can
consistently describe the results.

\begin{figure}[t]
\centering
\def\figw{0.485}
\includegraphics[width=\figw\linewidth]{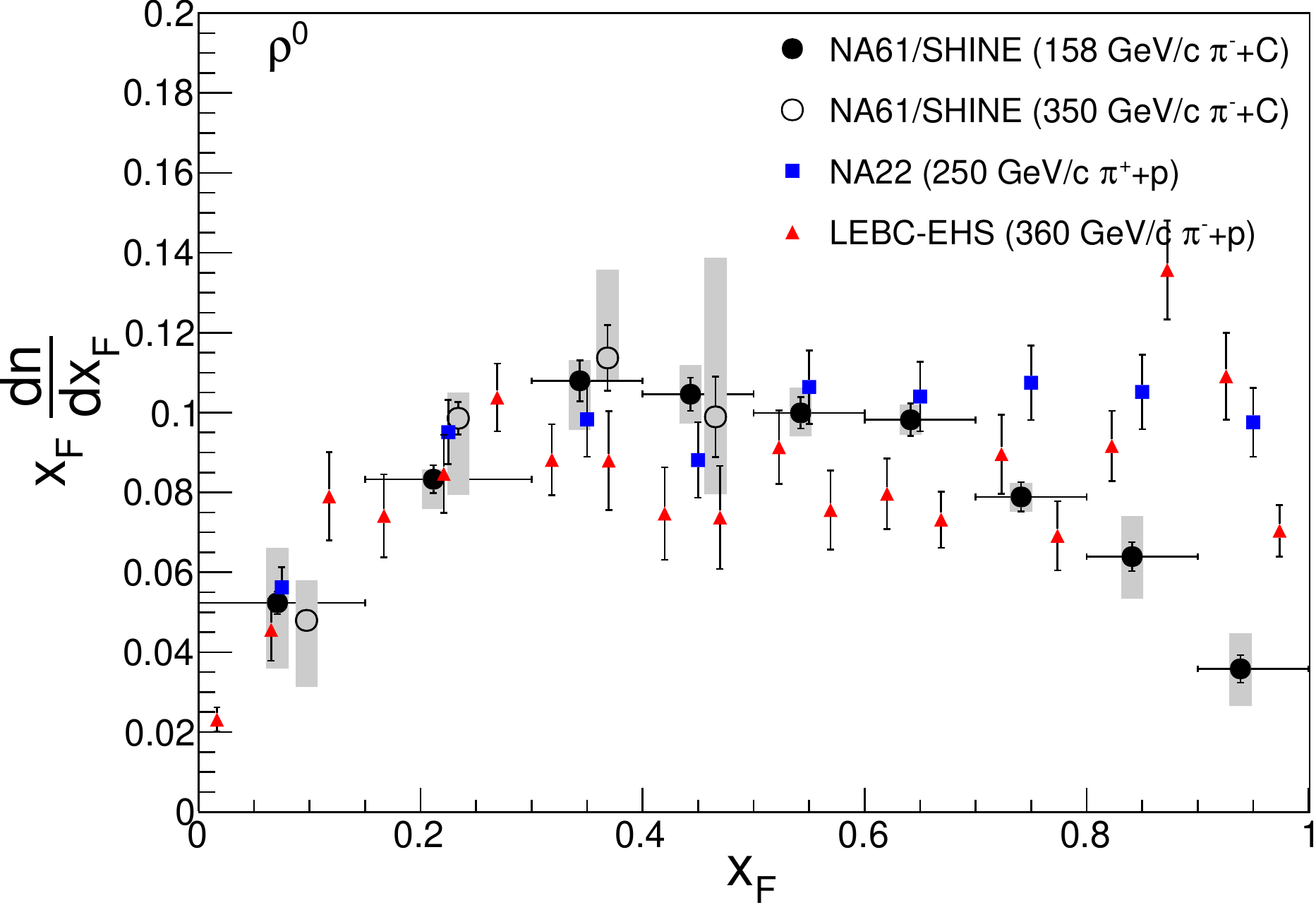}\hfill
\includegraphics[width=\figw\textwidth]{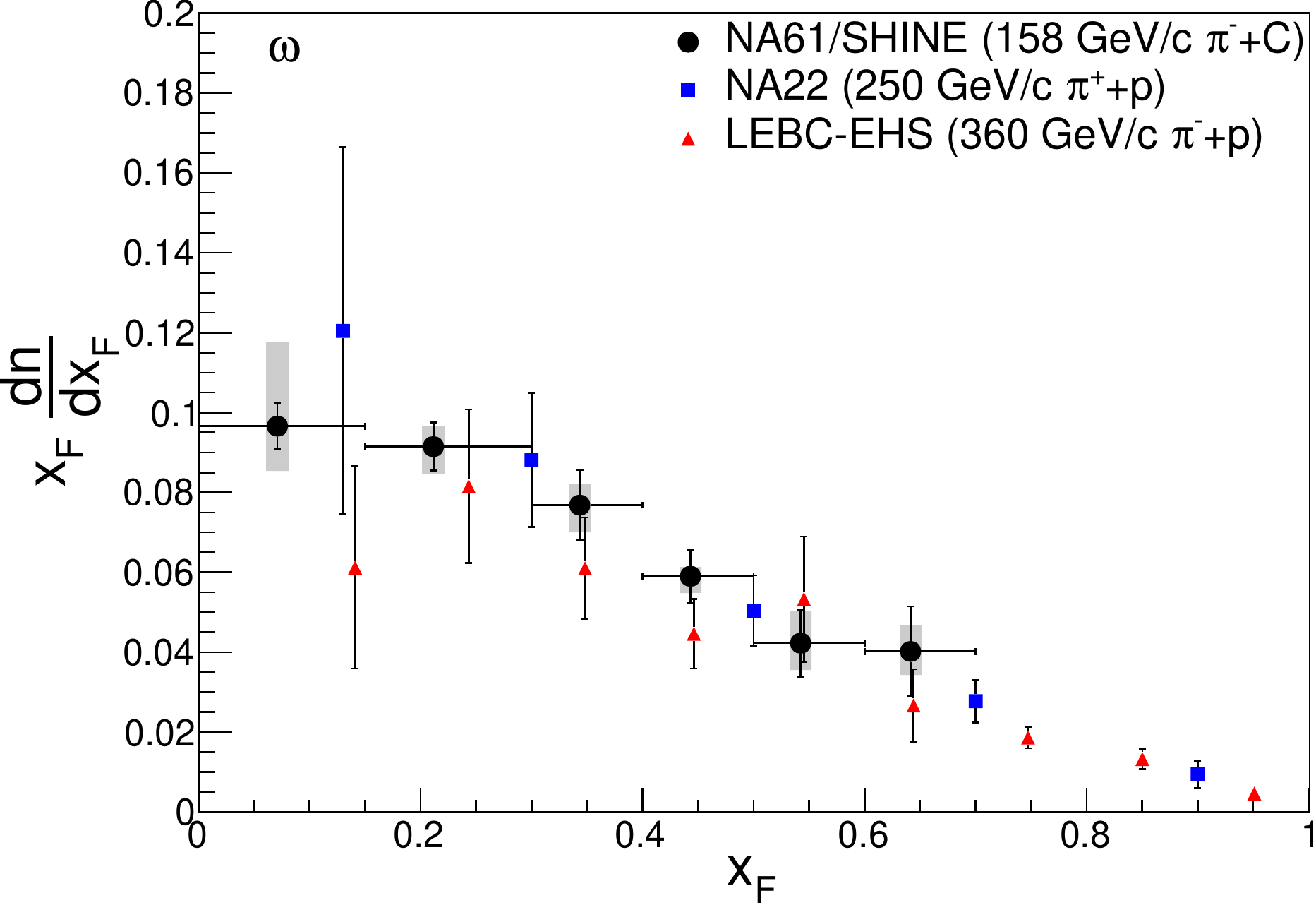}
\caption{Scaled \xF-spectra of meson production in
  $\pi^-$+C production interactions at 158 and $350\,\GeVc$
  ($350\,\GeVc$ shifted by 0.035). The error bars show the
  statistical, the bands indicate systematic uncertainties (where
  available). The black points are from this experiment, blue squares
  are from NA22~\cite{Agababyan:1990df}, red triangles are from
  LEBC-EHS (NA27)~\cite{aguilar1989vector}.  $\rho^0$ spectra are
  shown on the left and $\omega$ spectra on the right.}
\label{fig:Rho0World}
\end{figure}

The comparison between results from this analysis to measurements of
other experiments are presented in Fig.~\ref{fig:Rho0World} for
$\rho^0$ and $\omega$ mesons. The two other experiments shown are
NA22~\cite{Agababyan:1990df} and LEBC-EHS
(NA27)~\cite{aguilar1989vector}, both of which used a hydrogen
target. NA22 had a $\pi^+$ beam at 250\,\GeVc while LEBC-EHS had a
$\pi^-$ beam at 360\,\GeVc. The results from NA22 and LEBC-EHS are
scaled by their measured inelastic cross sections: $20.94\pm0.12\,\mb$
for NA22~\cite{Adamus:1986vn} and $21.6\,\mb$ for
LEBC-EHS~\cite{aguilar1989vector}.  There is good agreement between
the previous measurements with proton targets and the results from
this analysis for $\xF<0.6$. At larger \xF the
$\rho^0$ yields measured in this analysis show a decrease that is not
present in the $\pi$+p data and could thus be an effect of the nuclear
target used for the measurement presented here.  The comparison of the
measurements of the $\omega$ multiplicities shows no significant
differences between the other experiments and results from this
analysis.

  \section{Summary}
\label{sec:Summary}

This article presents experimental results on $\rho^0$, $\omega$
and K$^{*0}$ \xF-spectra
in $\pi^-$+C production interactions at $158\,\GeVc$ and the
$\rho^0$ spectra at 350\,\GeVc from the \NASixtyOne
spectrometer at the CERN SPS. These results are the first $\pi^-$+C
measurements taken in this energy range and are important to
tune hadronic interaction  models used to understand the measurements
of cosmic-ray air showers.

The comparisons of the measured spectra to predictions of hadronic
interaction models suggests that for all models further tuning is
required to reproduce the measured spectra of $\rho^0$, $\omega$ and
K$^{*0}$ mesons in the full range of \xF.  Recent re-tunes of
these models to resonance data in $\pi+p$ interactions resulted in
changes of the muon number at ground of up to
25\%~\cite{Ostapchenko:2013pia, Riehn:2015oba}. The new data provided
here for $\pi$+C interactions gives a more adequate reference for
pion-air interactions relevant for air showers and will help to
establish the effect of forward resonance production on muons in air
showers with the precision needed for using the muon number to
estimate the particle type of primary cosmic rays, as e.g.\ planned
within the upgrade of the Pierre Auger Observatory~\cite{Aab:2016vlz}.

  \nolinenumbers
  \section*{Acknowledgments}
  We would like to thank the CERN EP, BE and EN Departments for the
strong support of NA61/SHINE.

This work was supported by the Hungarian Scientific Research Fund
(grants OTKA 68506 and 71989), the J\'anos Bolyai Research Scholarship
of the Hungarian Academy of Sciences, the Polish Ministry of Science
and Higher Education (grants 667\slash N-CERN\slash2010\slash0,
NN\,202\,48\,4339 and NN\,202\,23\,1837), the Polish National Center
for Science (grants~2011\slash03\slash N\slash ST2\slash03691,
2013\slash11\slash N\slash ST2\slash03879, 2014\slash13\slash N\slash
ST2\slash02565, 2014\slash14\slash E\slash ST2\slash00018 and
2015\slash18\slash M\slash ST2\slash00125, 2015\slash 19\slash N\slash ST2 \slash01689), the Foundation for Polish
Science --- MPD program, co-financed by the European Union within the
European Regional Development Fund, the Federal Agency of Education of
the Ministry of Education and Science of the Russian Federation (SPbSU
research grant 11.38.242.2015), the Russian Academy of Science and the
Russian Foundation for Basic Research (grants 08-02-00018, 09-02-00664
and 12-02-91503-CERN), the National Research Nuclear
University MEPhI in the framework of the Russian Academic Excellence
Project (contract No. 02.a03.21.0005, 27.08.2013), the Ministry of Education, Culture, Sports,
Science and Tech\-no\-lo\-gy, Japan, Grant-in-Aid for Sci\-en\-ti\-fic
Research (grants 18071005, 19034011, 19740162, 20740160 and 20039012),
the German Research Foundation (grant GA\,1480/2-2), the EU-funded
Marie Curie Outgoing Fellowship, Grant PIOF-GA-2013-624803, the
Bulgarian Nuclear Regulatory Agency and the Joint Institute for
Nuclear Research, Dubna (bilateral contract No. 4418-1-15\slash 17),
Bulgarian National Science Fund (grant DN08/11), Ministry of Education
and Science of the Republic of Serbia (grant OI171002), Swiss
Nationalfonds Foundation (grant 200020\-117913/1), ETH Research Grant
TH-01\,07-3 and the U.S.\ Department of Energy.

  \bibliographystyle{na61Utphys}

  \bibliography{na61References}
  \newpage
  \clearpage
  \pdfbookmark[0]{Appendix}{Appendix}
  \appendix
  \section{Tables of measured resonance yields}

\begin{table}[h]
\caption{Average multiplicity of $\rho^0$ in $\pi^-$+C interactions
at 158\,\GeVc and 350\,\GeVc, binned in \xF.}
\label{tbl:rho0_multiplicity}
\begin{center}
\begin{tabular}{|c|r@{.}l@{ }r@{--}l@{ }r@{.}l|r@{.}l|r@{.}l r@{.}l r@{.}l r@{.}l|}
\hline
\multicolumn{1}{|r|}{$p_\text{beam}/(\GeVc)$} &
\multicolumn{6}{c|}{\xF} &
\multicolumn{2}{c|}{$\avg\xF$} &
\multicolumn{2}{l}{$\mathrm{d}n/\mathrm{d}\xF$} &
\multicolumn{2}{l}{$\Delta_\text{stat}$} &
\multicolumn{2}{l}{$\Delta_\text{sys}^+$} &
\multicolumn{2}{l|}{$\Delta_\text{sys}^-$}
\\
\hline
\multirow{9}{*}{158}
&0&0& && 0&15  &0&071  &0&737   &0&040    &0&194    &0&232 \\
&0&15& && 0&3  &0&212  &0&394   &0&016    &0&011    &0&035 \\
&0&3& && 0&4   &0&343  &0&314   &0&015    &0&015    &0&036 \\
&0&4& && 0&5   &0&443  &0&236   &0&009    &0&016    &0&016 \\
&0&5& && 0&6   &0&542  &0&184   &0&007    &0&012    &0&011 \\
&0&6& && 0&7   &0&641  &0&153   &0&006    &0&006    &0&006 \\
&0&7& && 0&8   &0&741  &0&106   &0&005    &0&005    &0&005 \\
&0&8& && 0&9   &0&841  &0&076   &0&004    &0&012    &0&012 \\
&0&9& && 1&0   &0&939  &0&038   &0&004    &0&010    &0&010 \\\hline
\multirow{4}{*}{350}
&0&0 && &0&15  &0&062  &0&790  &0&0419    &0&166    &0&274 \\
&0&15 && &0&3  &0&199  &0&499  &0&0202    &0&033    &0&097 \\
&0&3 && &0&4   &0&333  &0&343  &0&0246    &0&066    &0&017 \\
&0&4 && &0&5   &0&431  &0&230  &0&0235    &0&093    &0&045 \\\hline
\end{tabular}
\end{center}
\end{table}

\begin{table}[h!]
\caption{Average multiplicity of $\omega$ in $\pi^-$+C interactions
at 158\,\GeVc, binned in \xF.}
\label{tbl:omega_multiplicity}
\begin{center}
\begin{tabular}{|c|r@{.}l@{ }r@{--}l@{ }r@{.}l|r@{.}l|r@{.}l r@{.}l r@{.}l r@{.}l|}
\hline
\multicolumn{1}{|r|}{$p_\text{beam}/(\GeVc)$} &
\multicolumn{6}{c|}{\xF} &
\multicolumn{2}{c|}{$\avg\xF$} &
\multicolumn{2}{l}{$\mathrm{d}n/\mathrm{d}\xF$} &
\multicolumn{2}{l}{$\Delta_\text{stat}$} &
\multicolumn{2}{l}{$\Delta_\text{sys}^+$} &
\multicolumn{2}{l|}{$\Delta_\text{sys}^-$}
\\
\hline
\multirow{6}{*}{158}
&0&0 && &0&15  &0&071   &1&360   &0&082    &0&295    &0&156 \\
&0&15 && &0&3  &0&212   &0&432   &0&028    &0&025    &0&032 \\
&0&3& && 0&4   &0&343   &0&224   &0&026    &0&015    &0&020 \\
&0&4& && 0&5   &0&443   &0&133   &0&015    &0&005    &0&010 \\
&0&5& && 0&6   &0&542   &0&078   &0&016    &0&015    &0&012 \\
&0&6& && 0&7   &0&641   &0&063   &0&018    &0&011    &0&009 \\\hline
\end{tabular}
\end{center}
\end{table}

\begin{table}[h!]
\caption{Average multiplicity of K$^{*0}$ in $\pi^-$+C interactions
at 158\,\GeVc, binned in \xF.} 
\label{tbl:kstar_multiplicity}
\begin{center}
\begin{tabular}{|c|r@{.}l@{ }r@{--}l@{ }r@{.}l|r@{.}l|r@{.}l r@{.}l r@{.}l  r@{.}l|}
\hline
\multicolumn{1}{|r|}{$p_\text{beam}/(\GeVc)$} &
\multicolumn{6}{c|}{\xF} &
\multicolumn{2}{c|}{$\avg\xF$} &
\multicolumn{2}{l}{$\mathrm{d}n/\mathrm{d}\xF$} &
\multicolumn{2}{l}{$\Delta_\text{stat}$} &
\multicolumn{2}{l}{$\Delta_\text{sys}^+$} &
\multicolumn{2}{l|}{$\Delta_\text{sys}^-$}
\\
\hline
\multirow{9}{*}{158}
&0&0 && &0&15   &0&071   &1&073   &0&061    &0&468    &0&131 \\
&0&15 && &0&3   &0&212   &0&417   &0&022    &0&149    &0&013 \\
&0&3& && 0&4    &0&343   &0&176   &0&016    &0&025    &0&015 \\
&0&4& && 0&5    &0&443   &0&101   &0&010    &0&009    &0&011 \\
&0&5& && 0&6    &0&542   &0&054   &0&007    &0&008    &0&008 \\
&0&6& && 0&7    &0&641   &0&030   &0&006    &0&004    &0&004 \\
&0&7& && 0&8    &0&741   &0&013   &0&004    &0&001    &0&002 \\
&0&8& && 0&9    &0&841   &0&016   &0&003    &0&004    &0&005 \\\hline
\end{tabular}
\end{center}
\end{table}

  \newpage
  \section{Examples of templates of resonances and background}
\label{app:templates}

\begin{figure}[!h]
\centering
\def\figw{0.31}
\subfigure[charge mixing]{\includegraphics[width=\figw\textwidth]{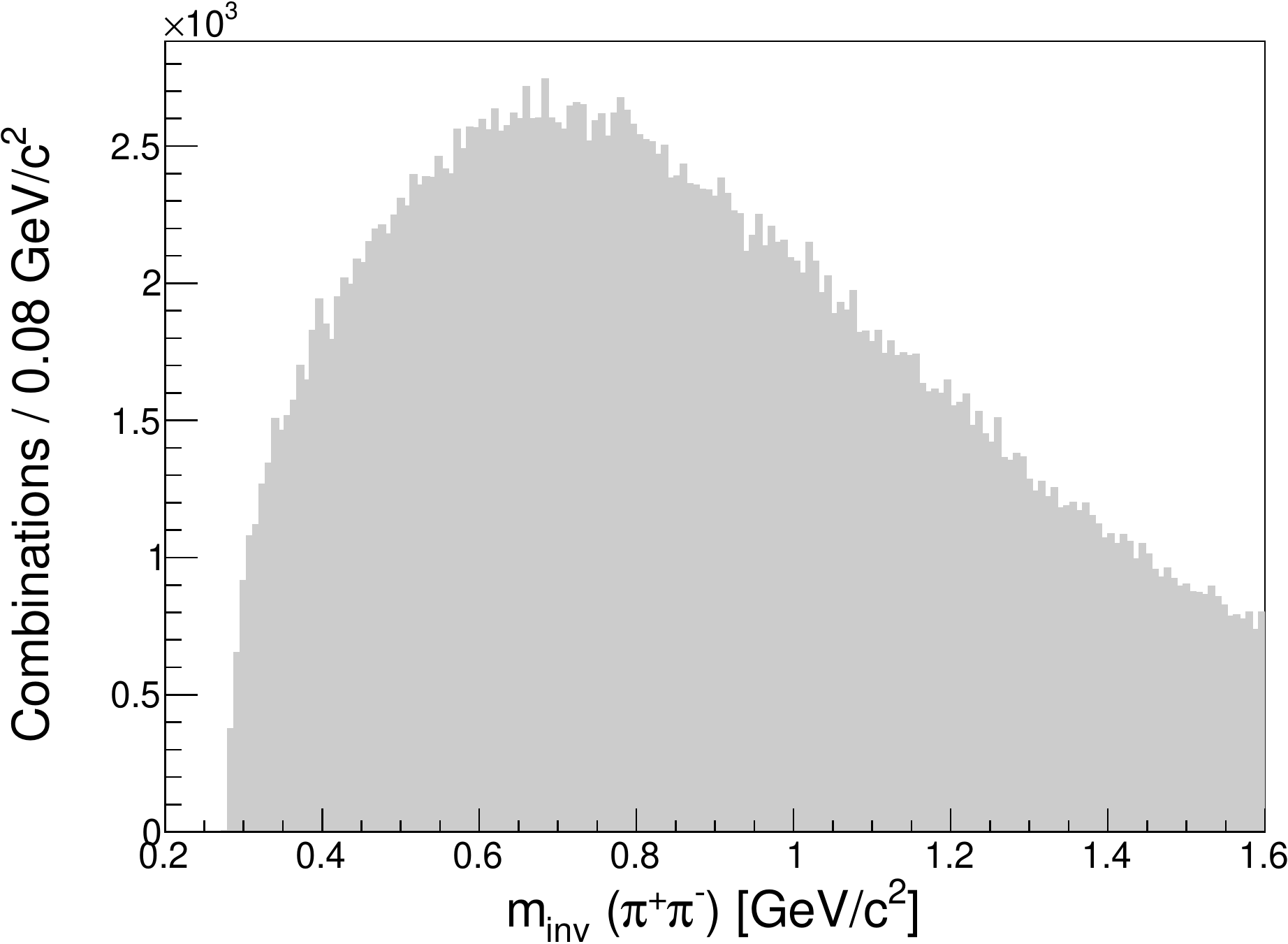}}\hfil
\subfigure[$\rho^0$]{\includegraphics[width=\figw\textwidth]{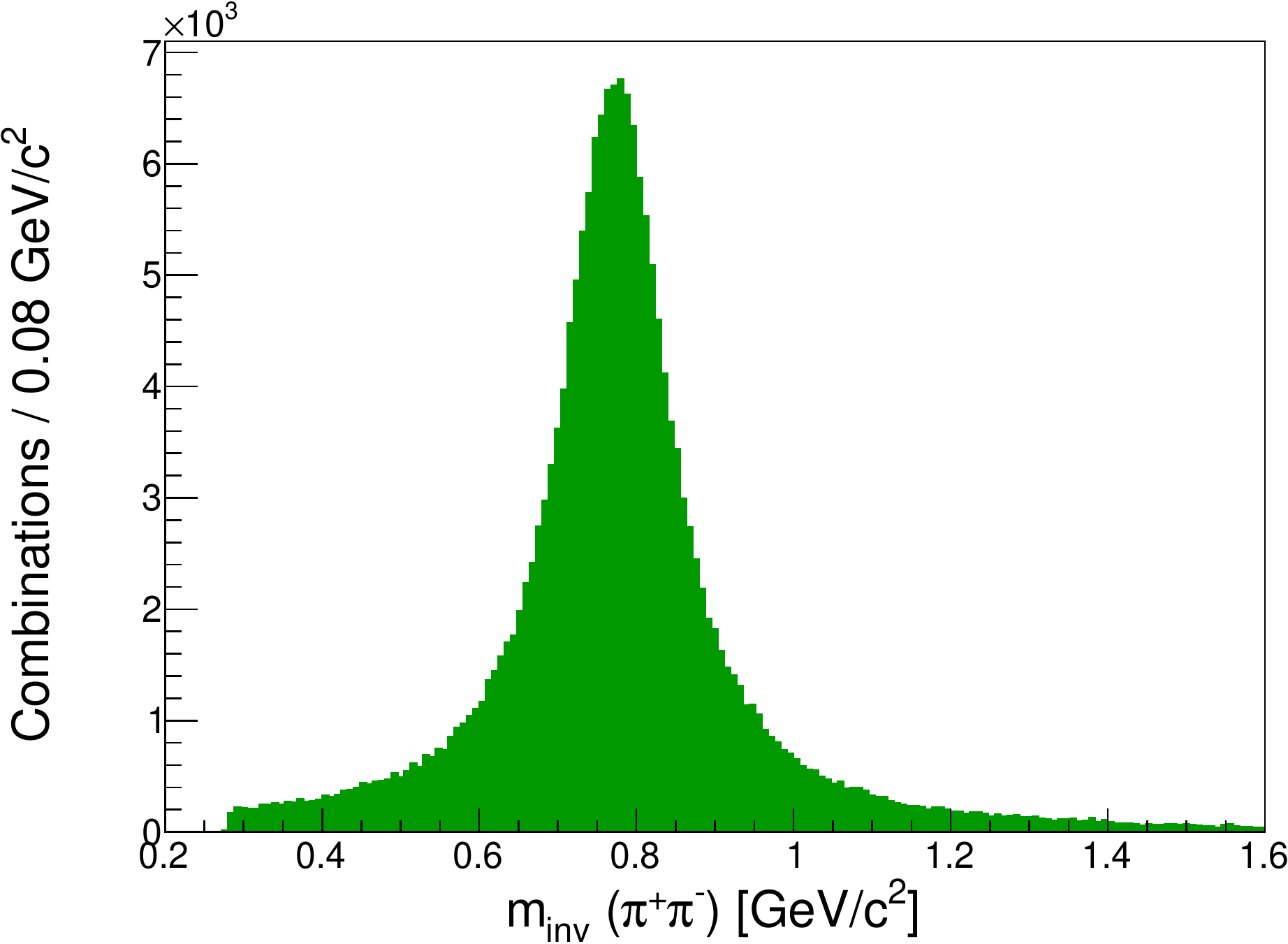}}\hfil
\subfigure[K$^{*0}$]{\includegraphics[width=\figw\textwidth]{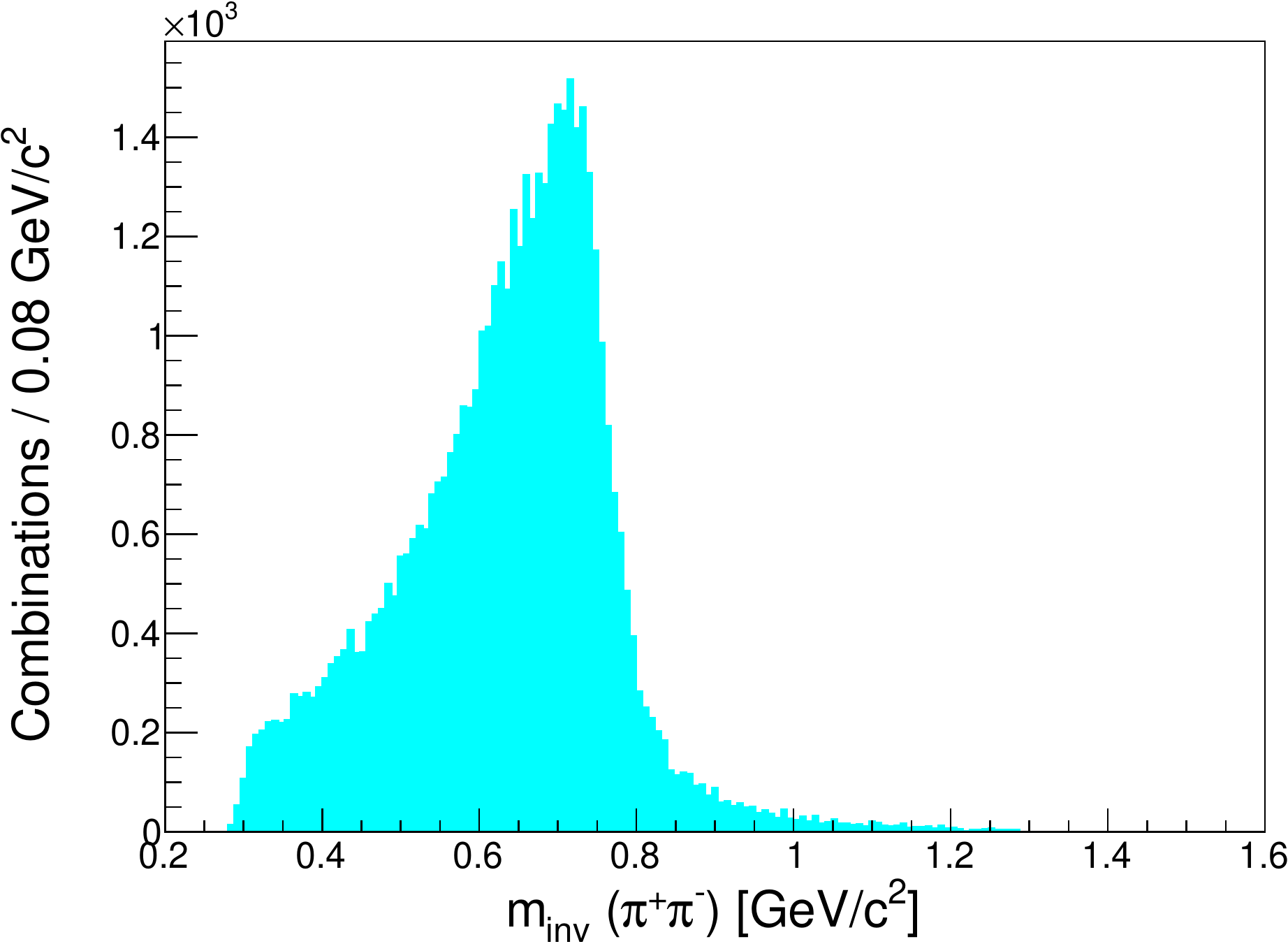}}\hfil
\subfigure[$\omega$]{\includegraphics[width=\figw\textwidth]{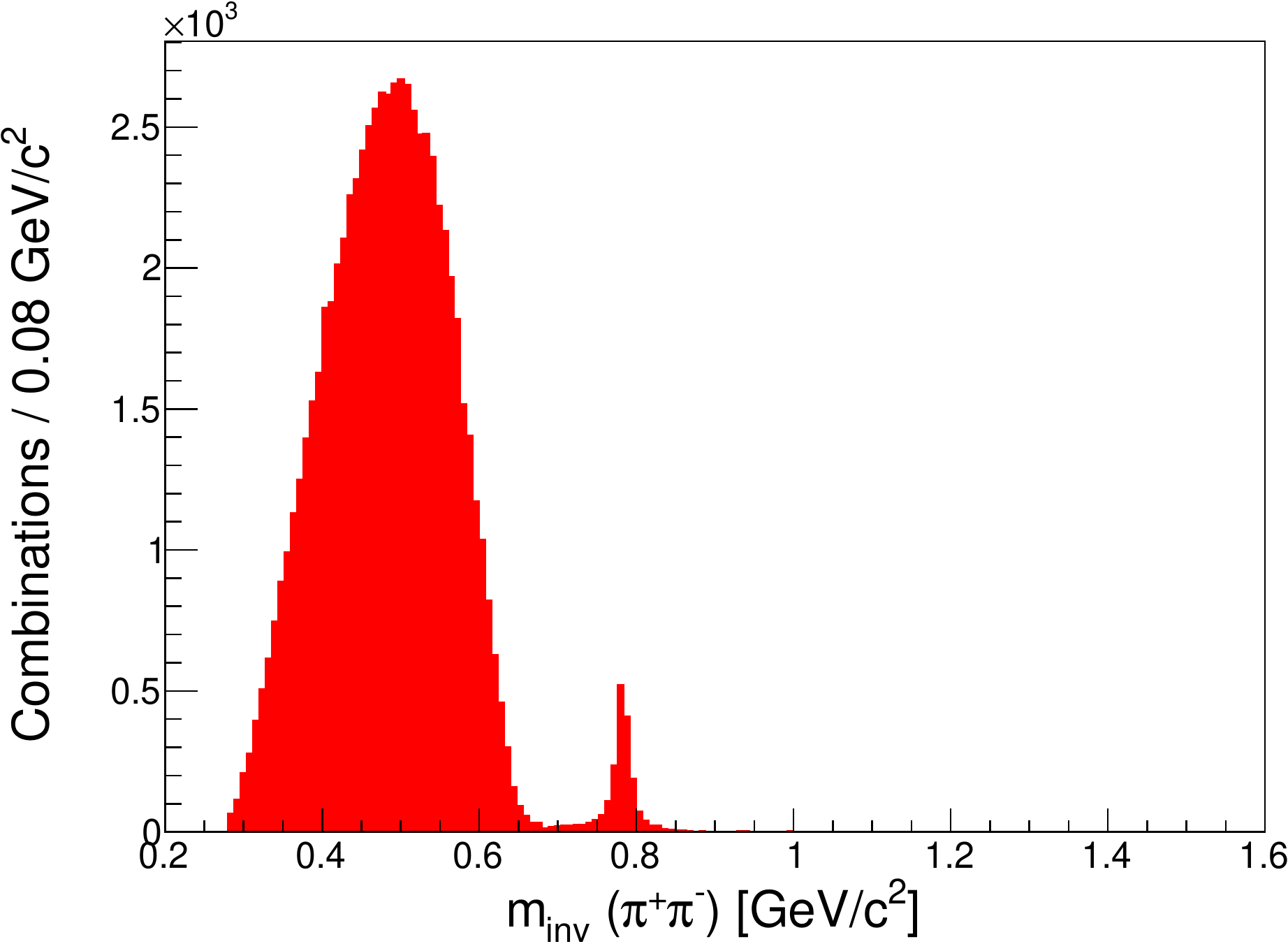}}\hfil
\subfigure[f$_2$]{\includegraphics[width=\figw\textwidth]{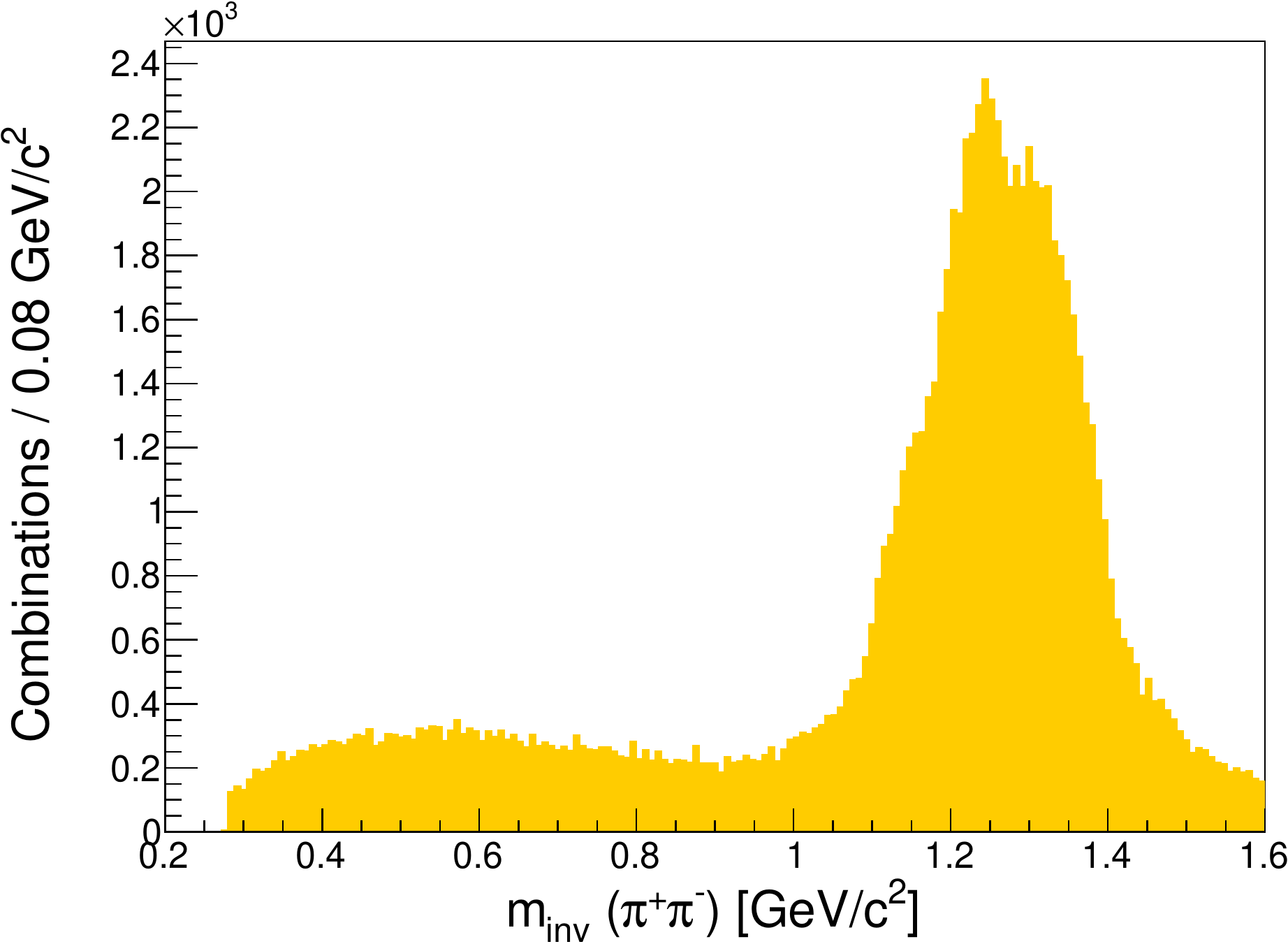}}\hfil
\subfigure[a$_2$]{\includegraphics[width=\figw\textwidth]{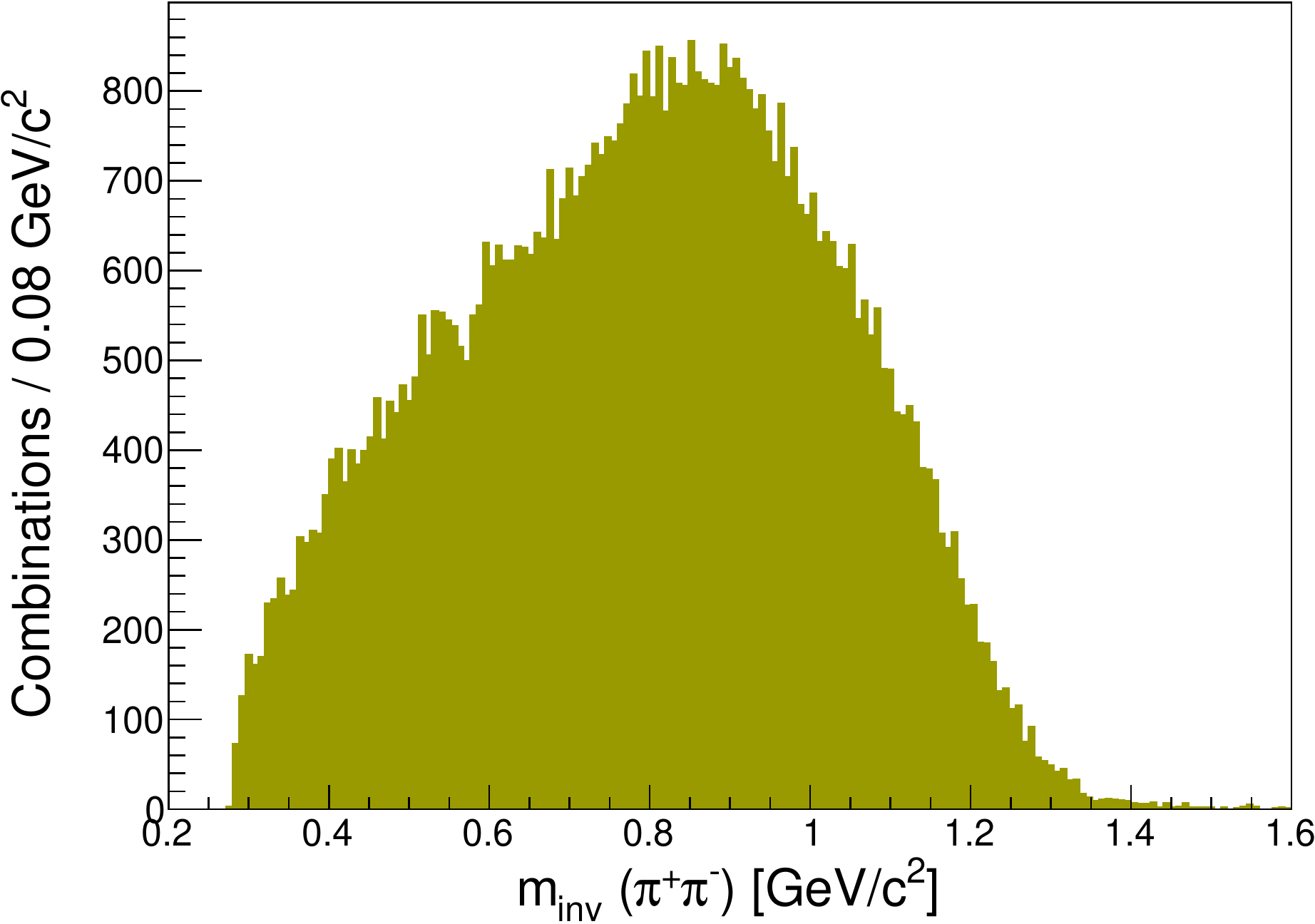}}\hfil
\subfigure[f$_0$]{\includegraphics[width=\figw\textwidth]{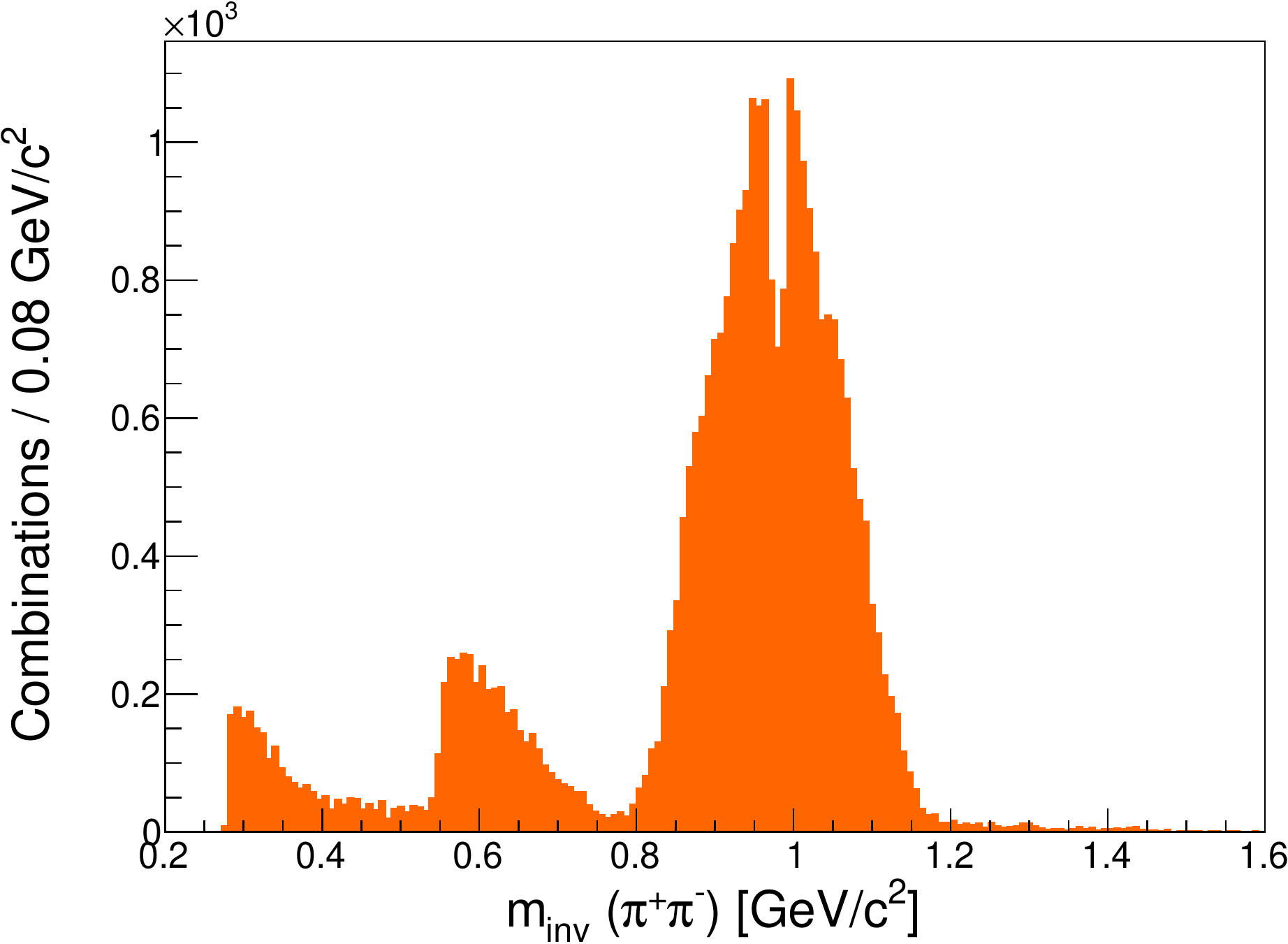}}\hfil
\subfigure[$\rho^3$]{\includegraphics[width=\figw\textwidth]{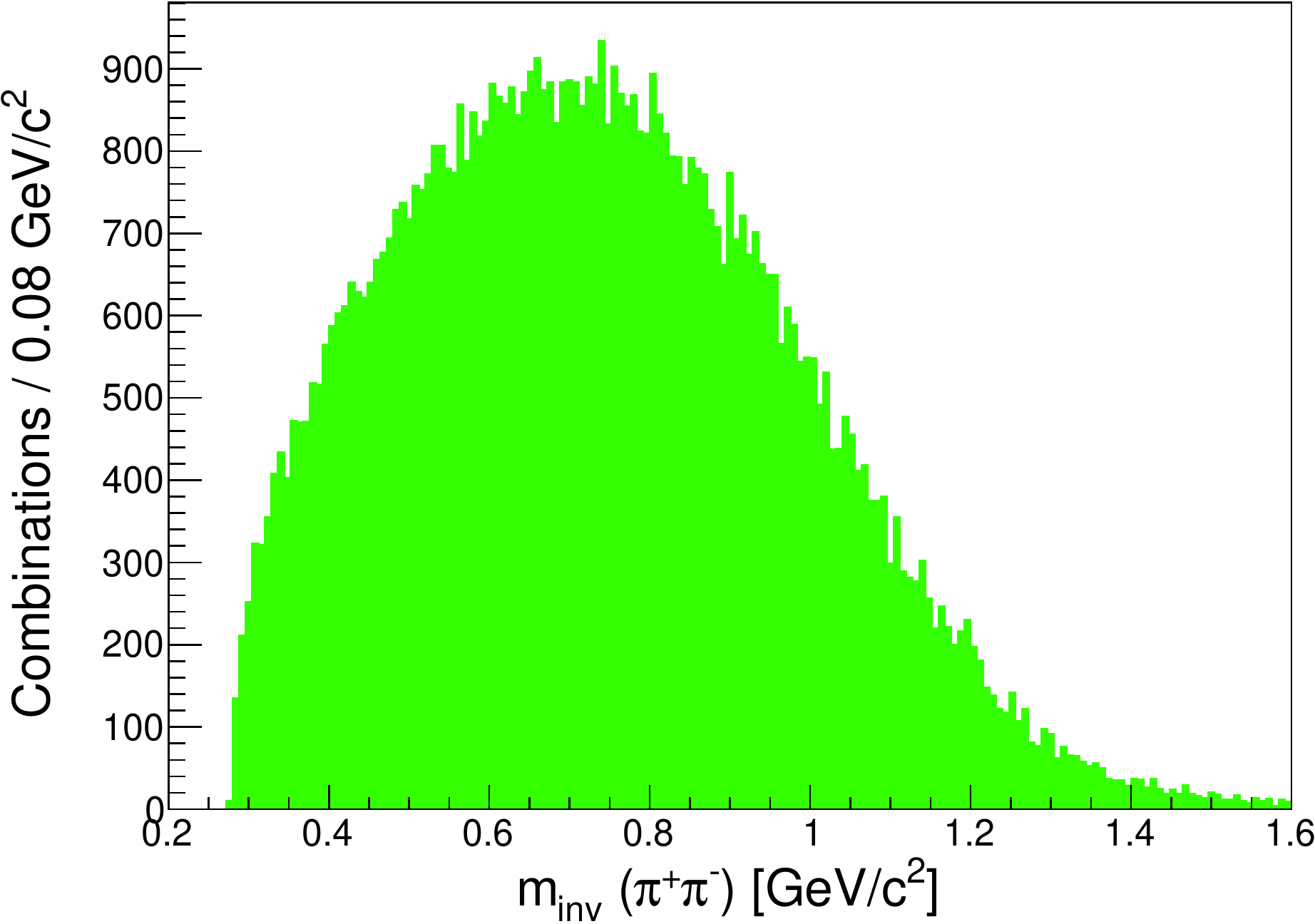}}\hfil
\subfigure[K$^0_\text{S}$]{\includegraphics[width=\figw\textwidth]{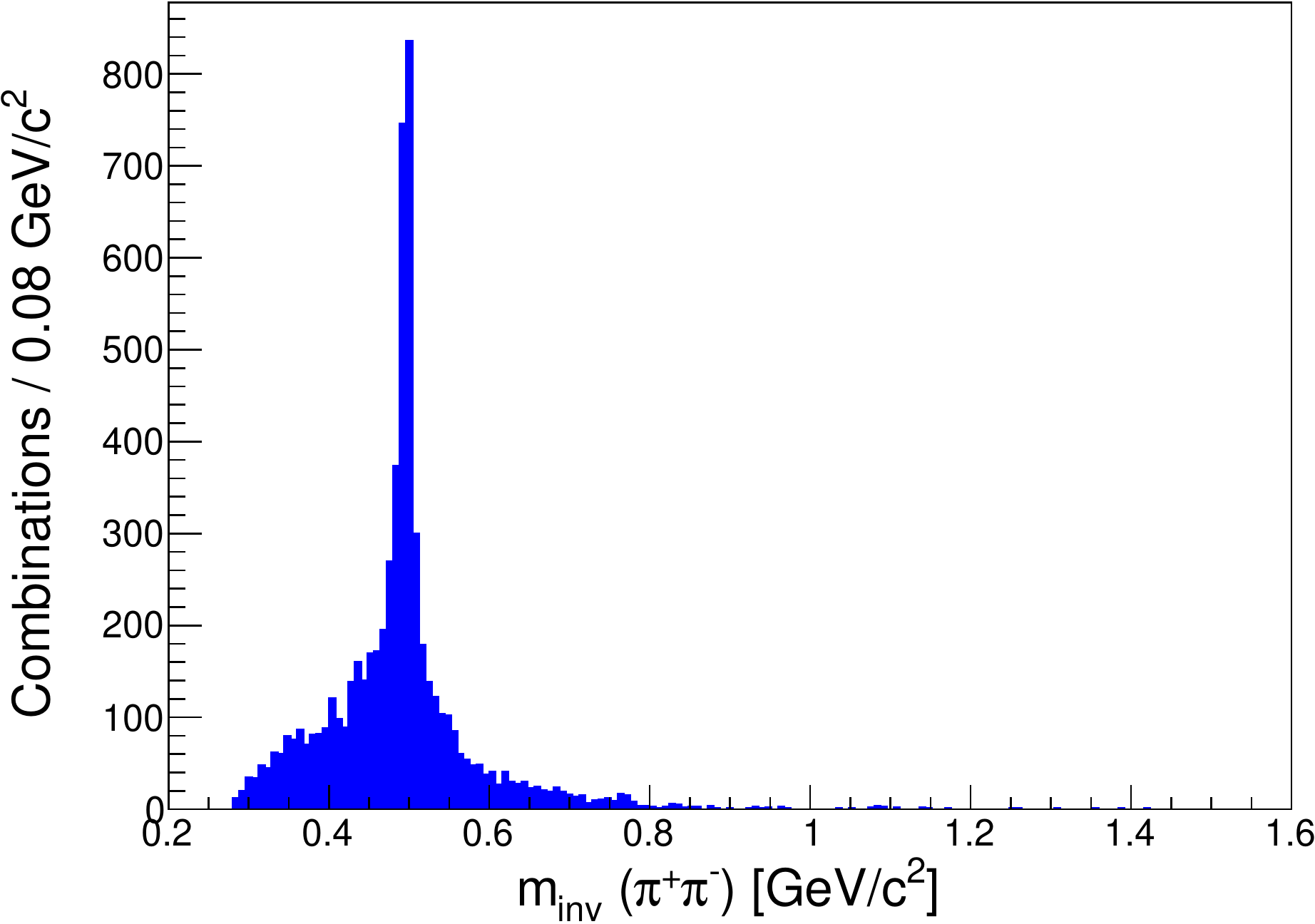}}\hfil
\subfigure[$\eta$]{\includegraphics[width=\figw\textwidth]{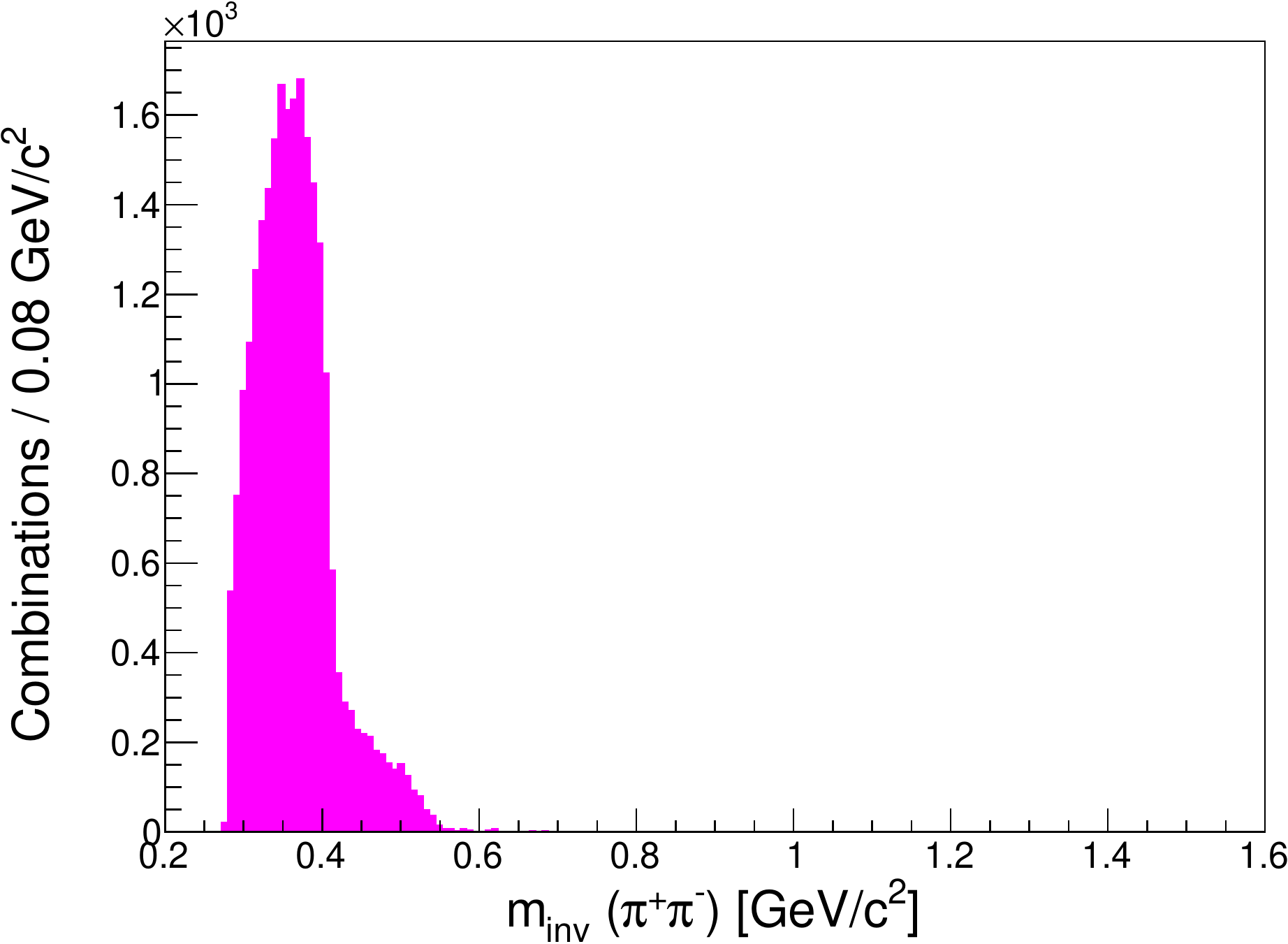}}
\caption{Templates of the invariant mass spectra of resonances and background at 158\,\GeVc in the range
$0.4<\xF<0.5$ assuming pion masses.}
\end{figure}

\clearpage

\section{Results of template fits}
\label{app:fits}

\begin{figure*}[h!]
\centering
\def\figw{0.46}
\subfigure[$0<\xF<0.15$]{
\includegraphics[width=\figw\textwidth]{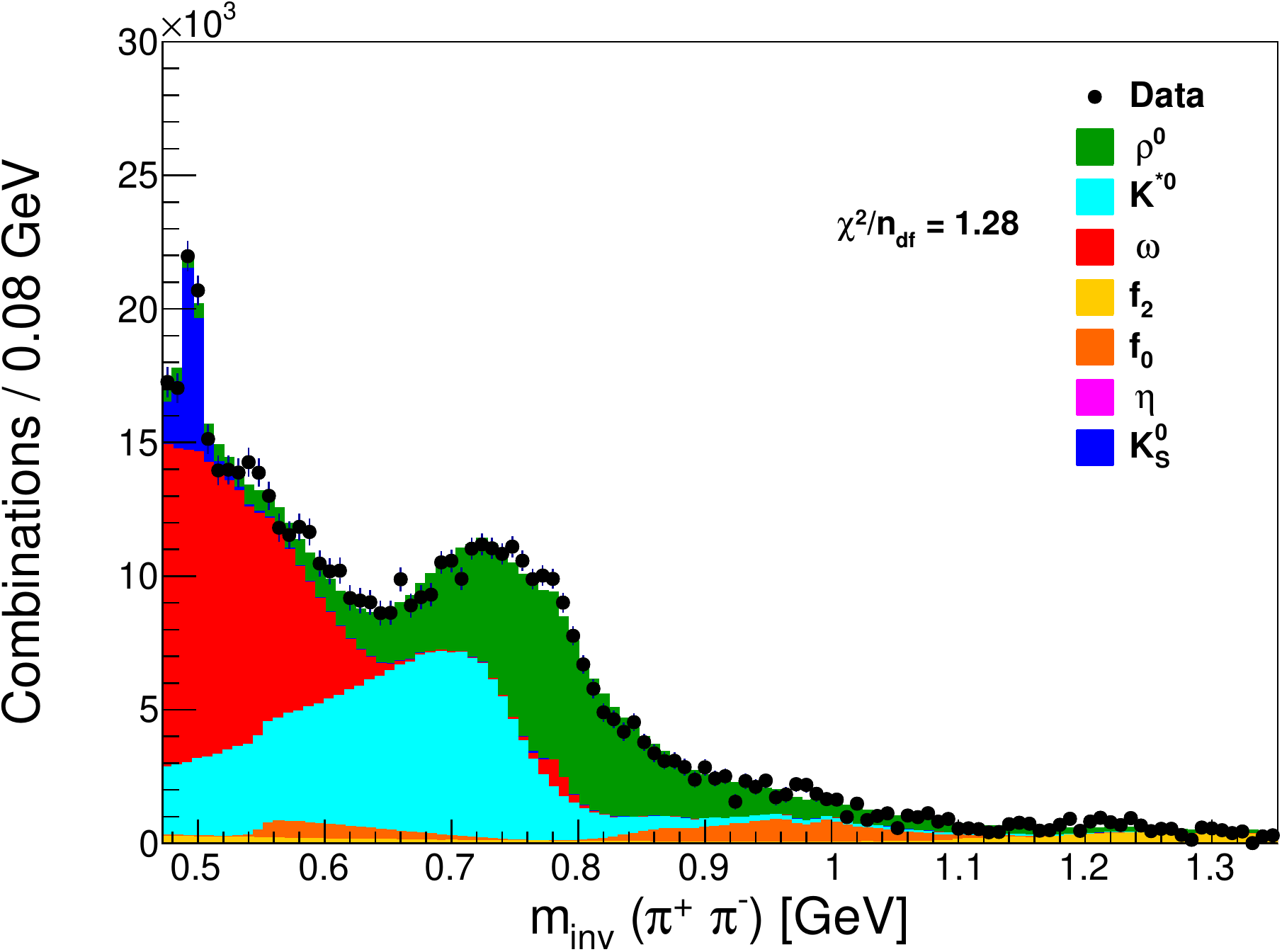}\qquad
\includegraphics[width=\figw\textwidth]{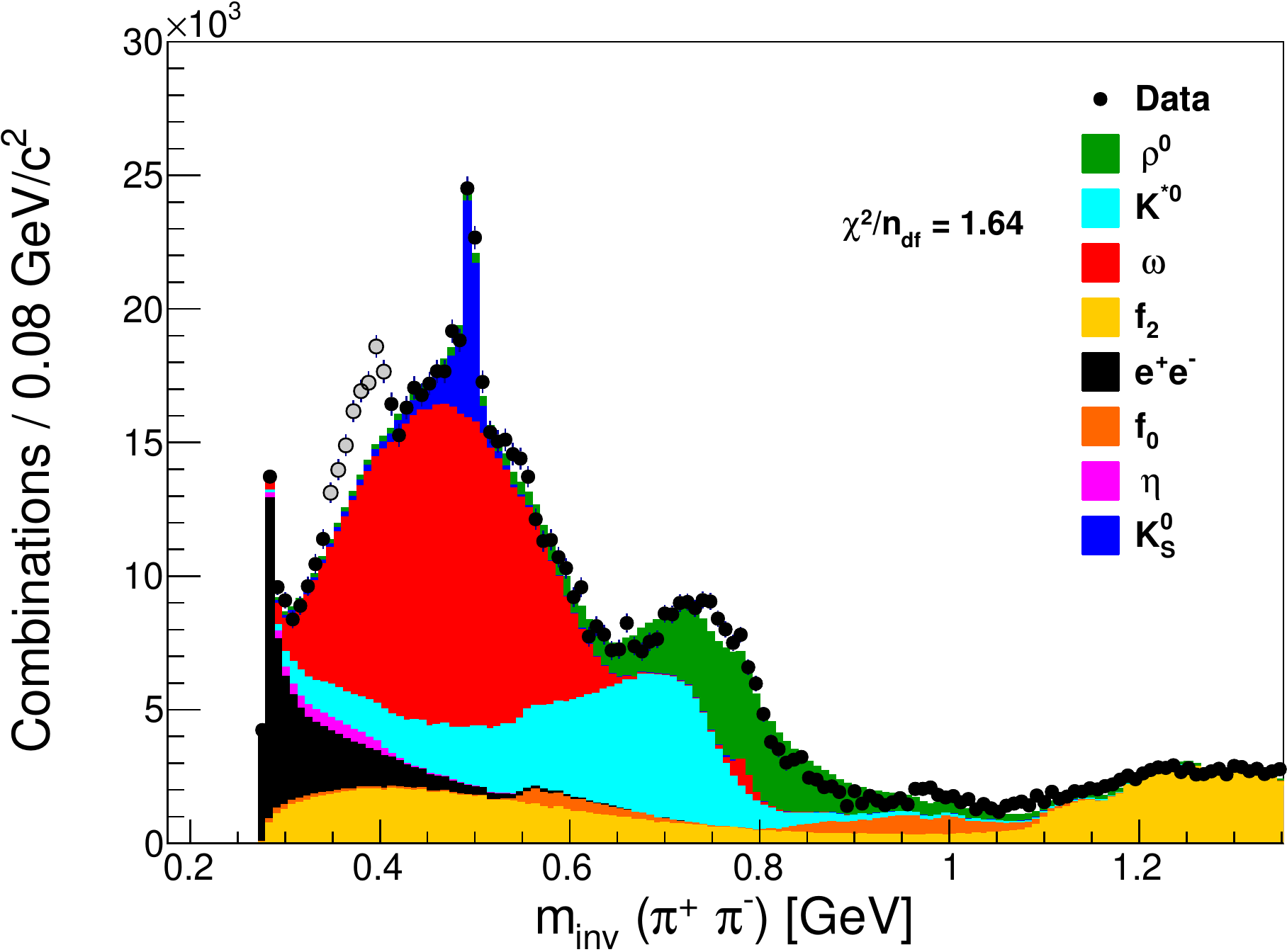}
}
\subfigure[$0.15<\xF<0.3$]{
\includegraphics[width=\figw\textwidth]{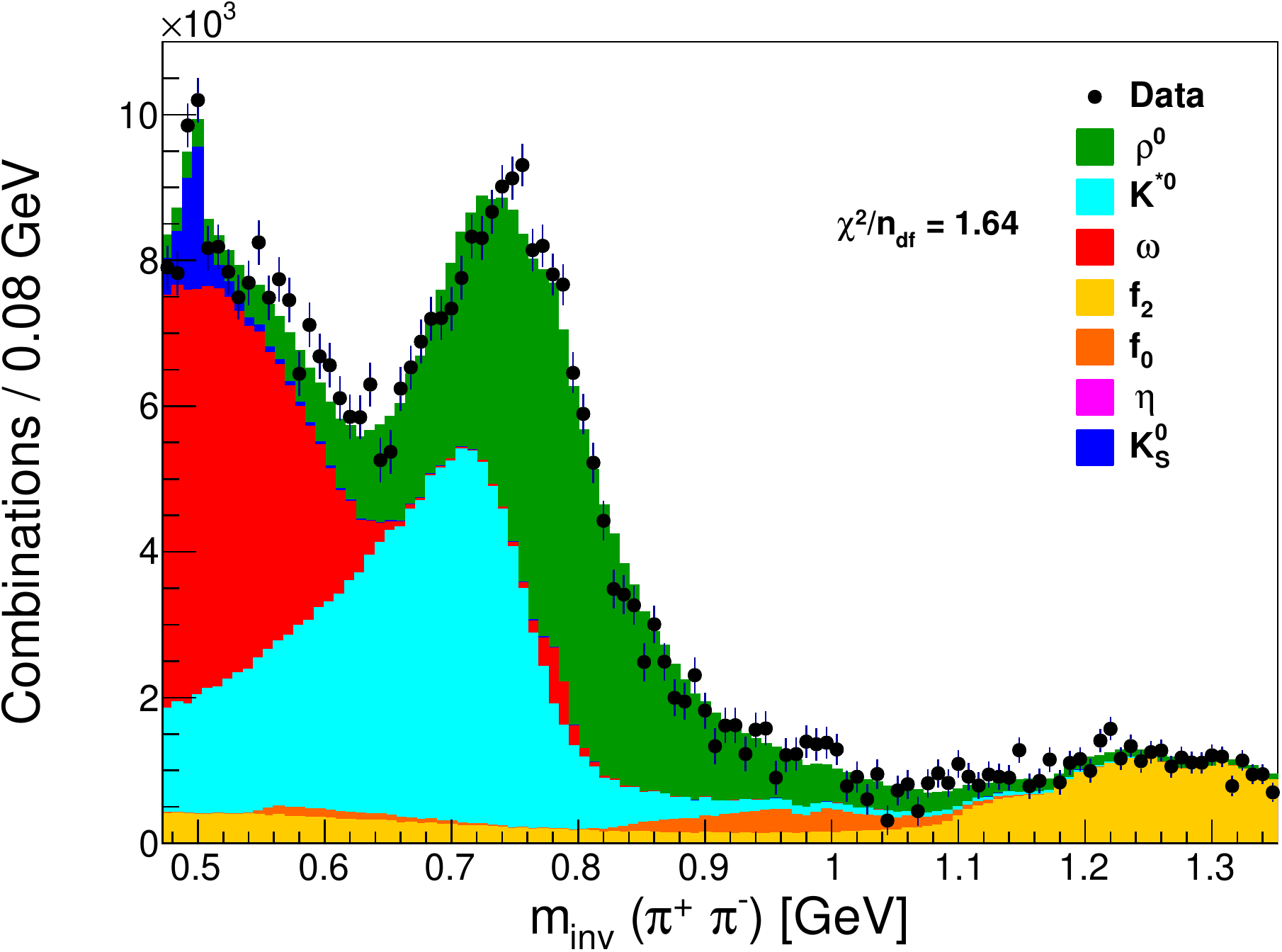}\qquad
\includegraphics[width=\figw\textwidth]{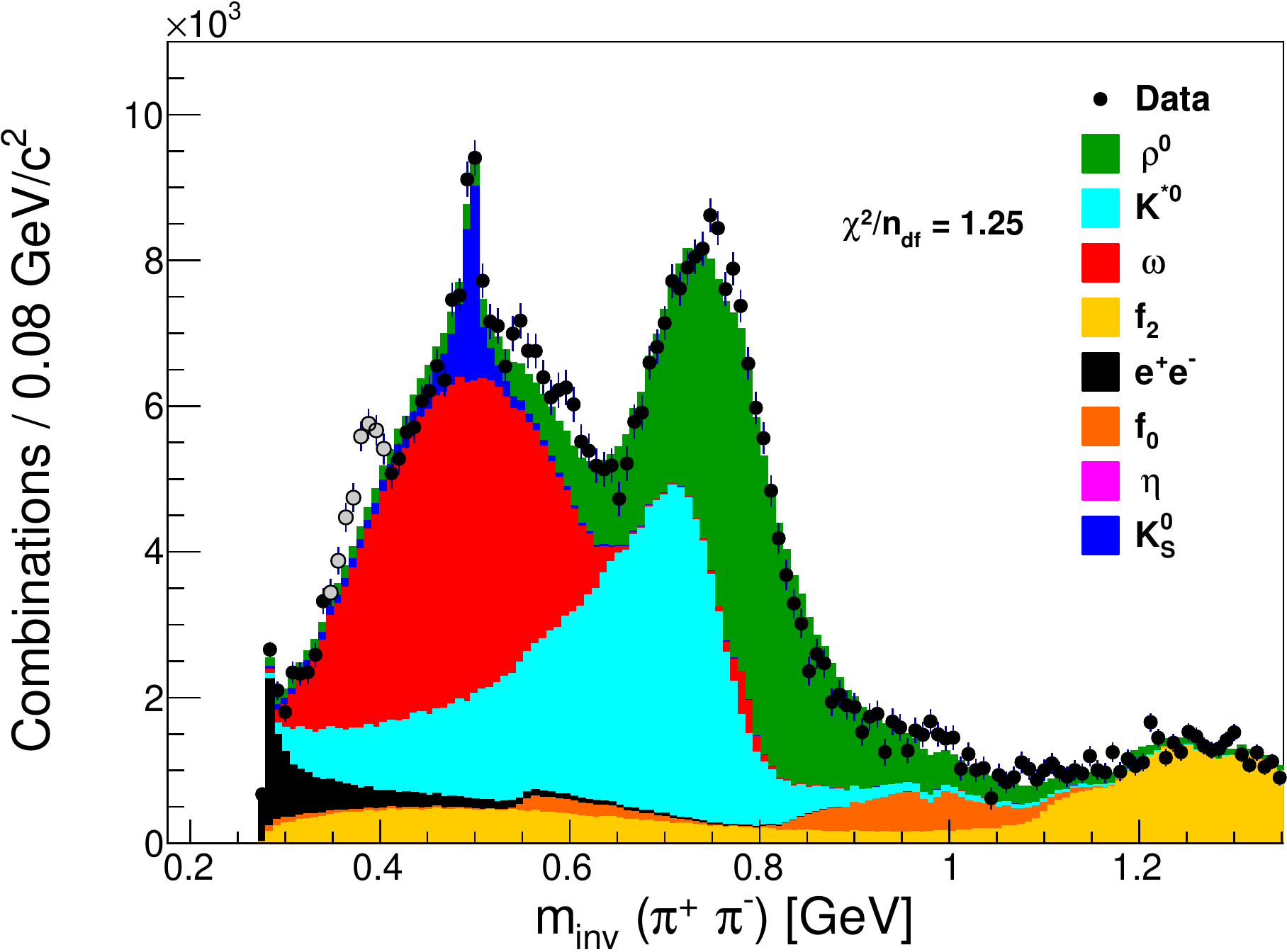}
}
\subfigure[$0.3<\xF<0.4$]{
\includegraphics[width=\figw\textwidth]{Fits_158-03}\qquad
\includegraphics[width=\figw\textwidth]{Fits_158_LongRange-03}
}
\caption{Invariant mass distribution of opposite charged particles, calculated assuming
pion masses, in $\pi^-$+C interactions
at 158\,\GeVc. Dots with error bars denote the
data and the fitted resonance templates are shown as filled histograms.
The fitted background and high mass resonances have been subtracted. Two
fits with different $\minv(\pi^+\pi^-)$ ranges are shown on the left and right
column. The fit range is equal to the displayed range, but in the extended-range fit
on the right the mass region $0.35<\minv(\pi^+\pi^-)<0.4$ is excluded
(see discussion App.~\ref{app:discussion}), as indicated by the grey points.}
\label{fig:Fits_1}
\vspace*{-4cm}
\end{figure*}

\clearpage

\begin{figure*}[h!]
\centering
\def\figw{0.46}
\subfigure[$0.4<\xF<0.5$]{
\includegraphics[width=\figw\linewidth]{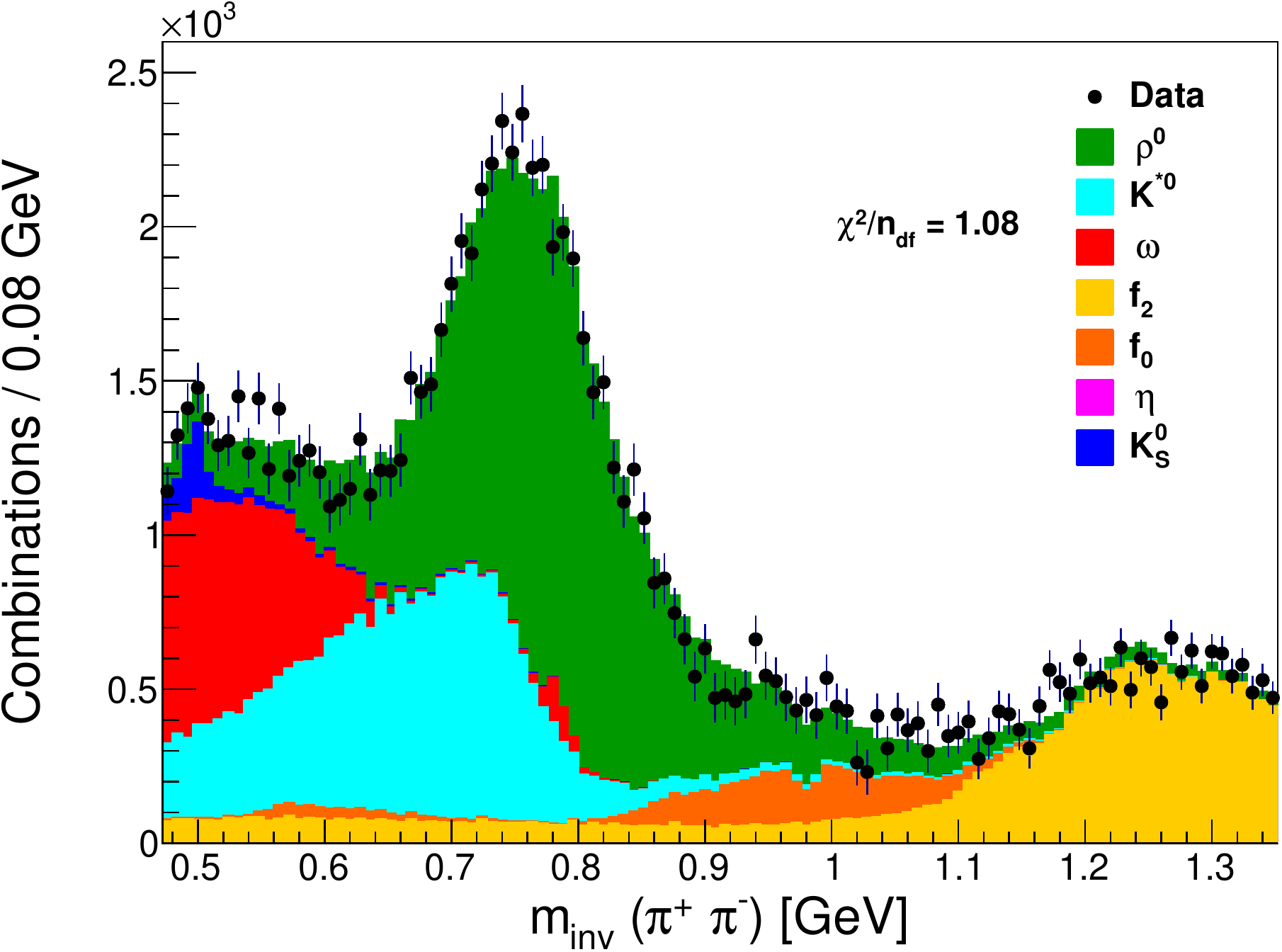}\qquad
\includegraphics[width=\figw\linewidth]{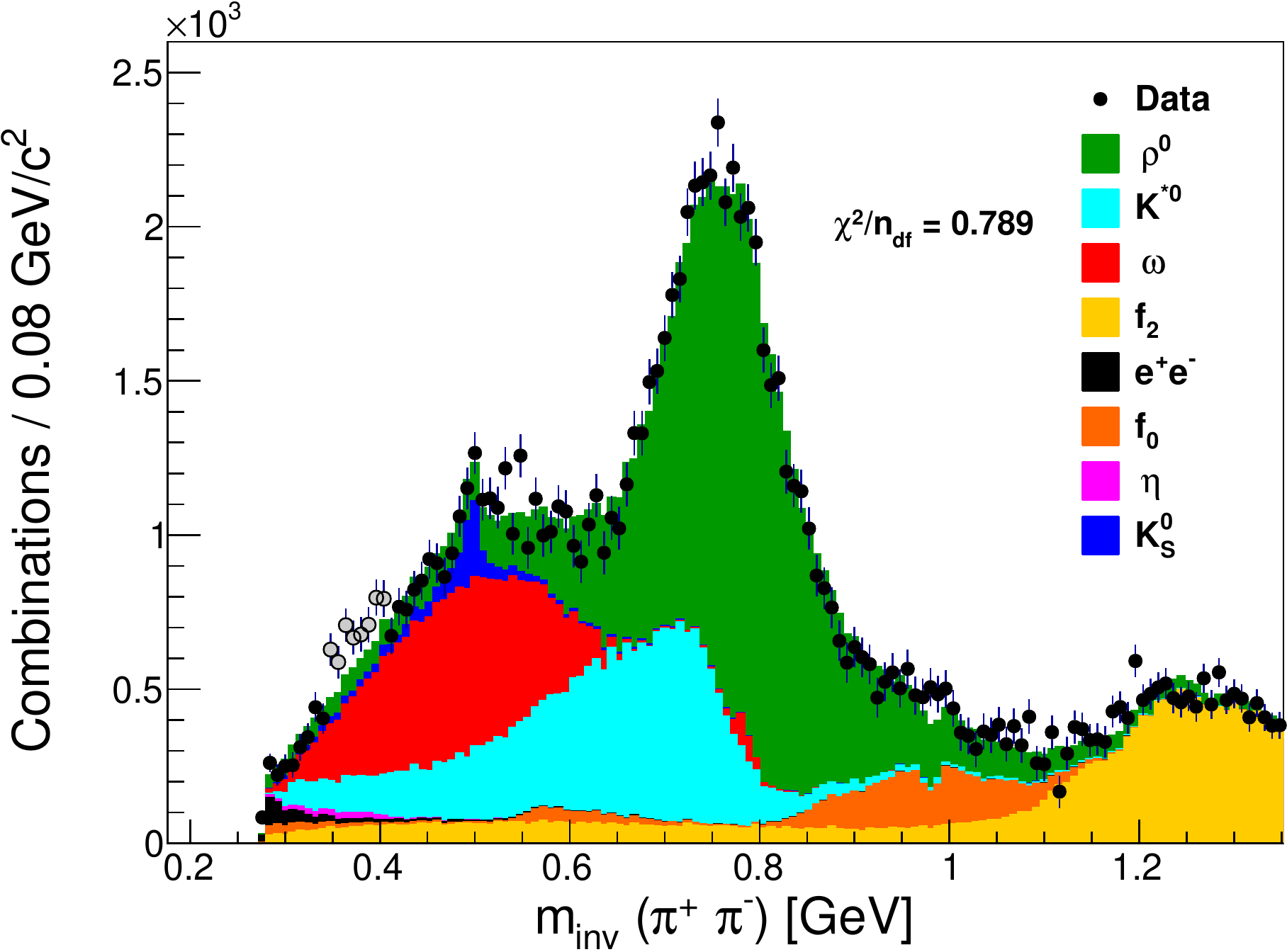}
}
\subfigure[$0.5<\xF<0.6$]{
\includegraphics[width=\figw\linewidth]{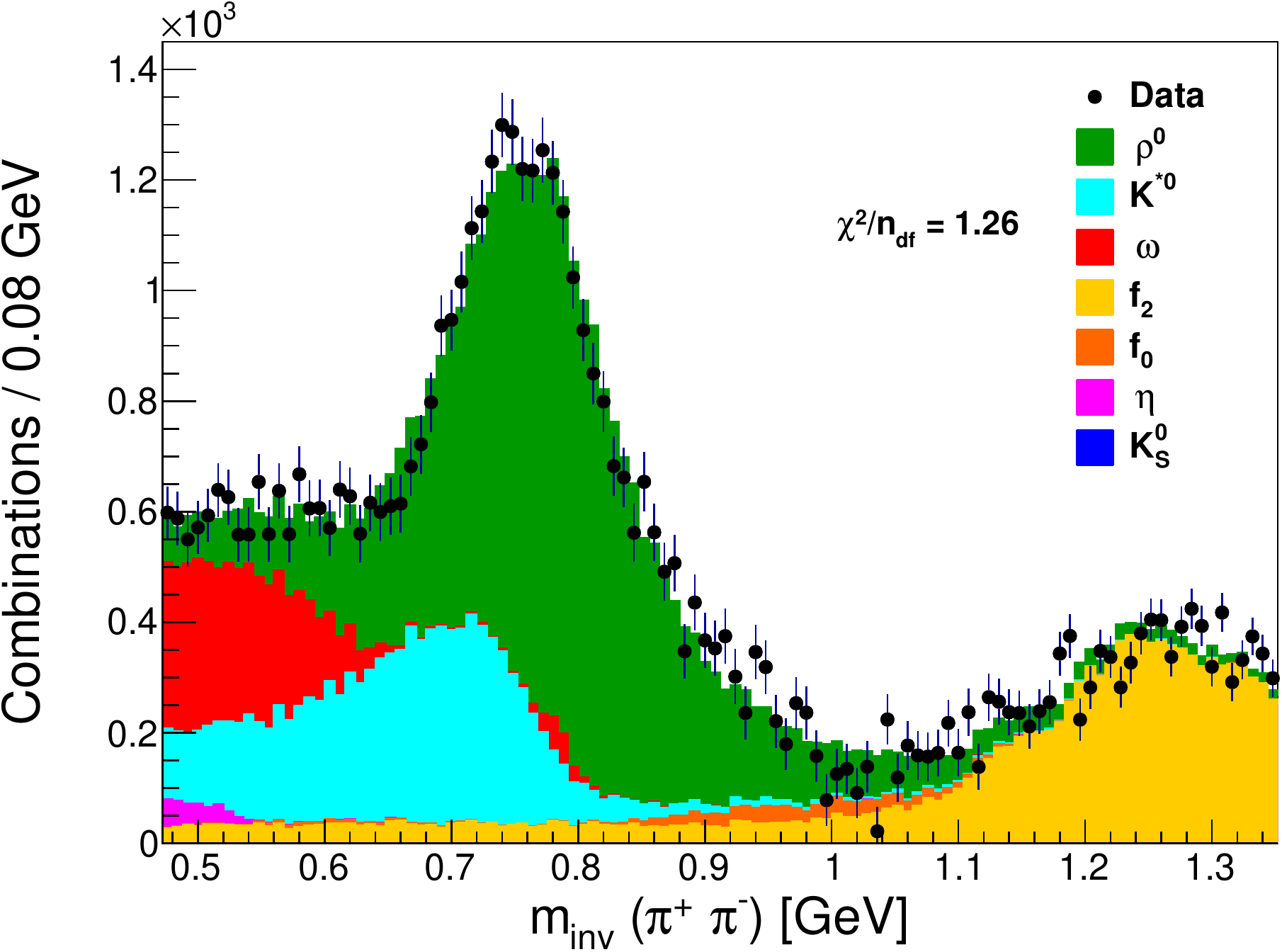}\qquad
\includegraphics[width=\figw\linewidth]{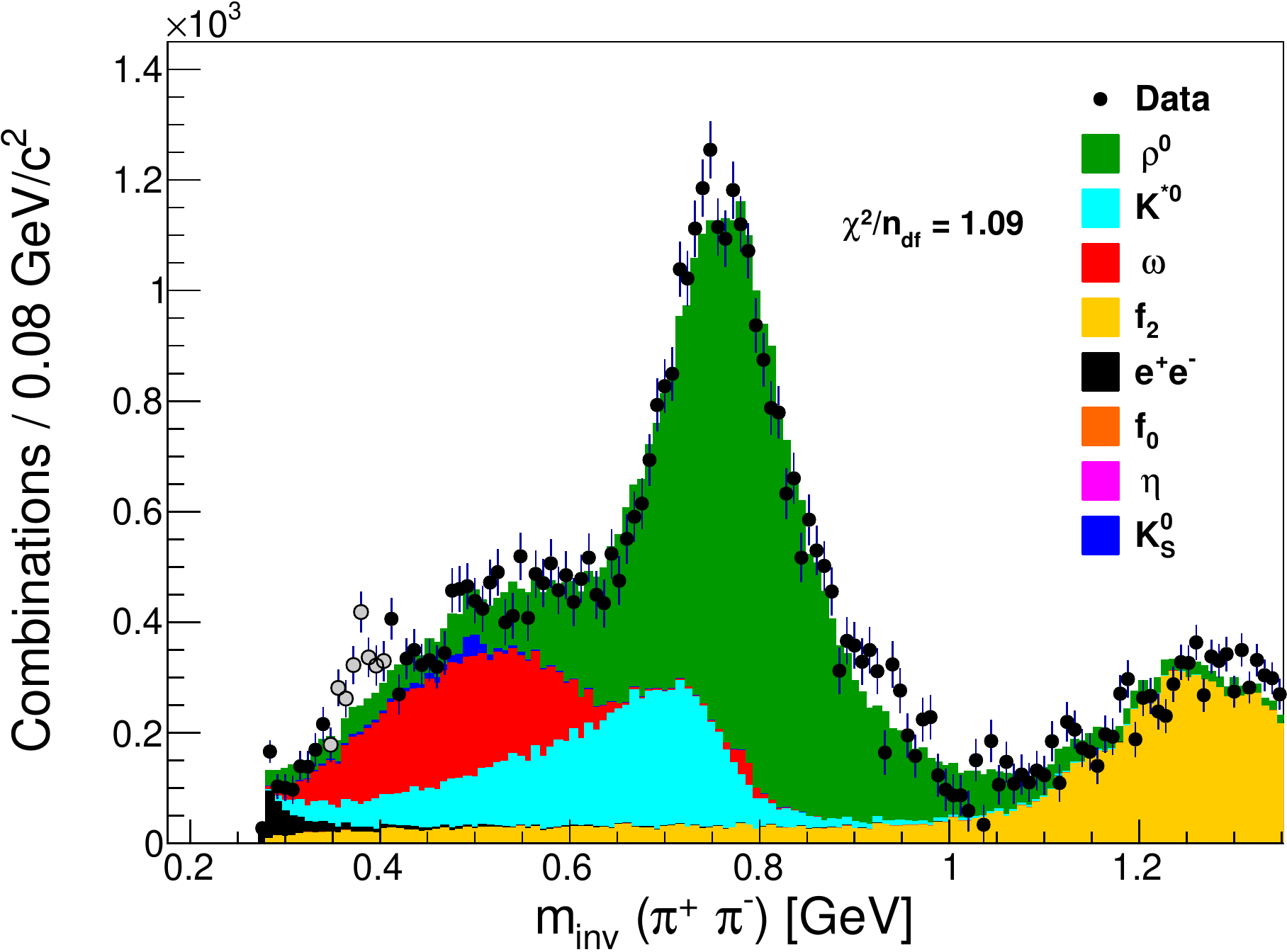}
}
\subfigure[$0.6<\xF<0.7$]{
\includegraphics[width=\figw\linewidth]{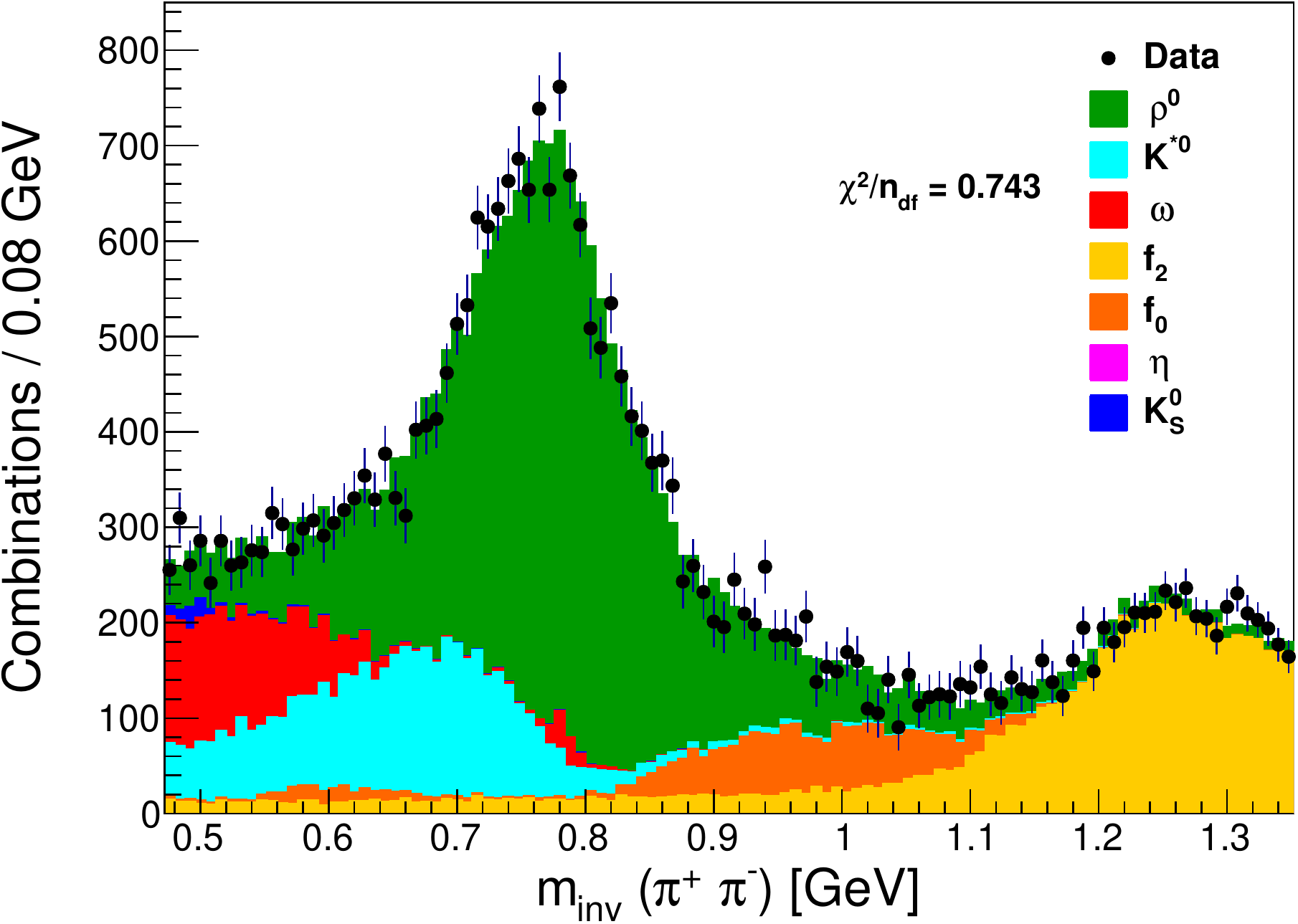}\qquad
\includegraphics[width=\figw\linewidth]{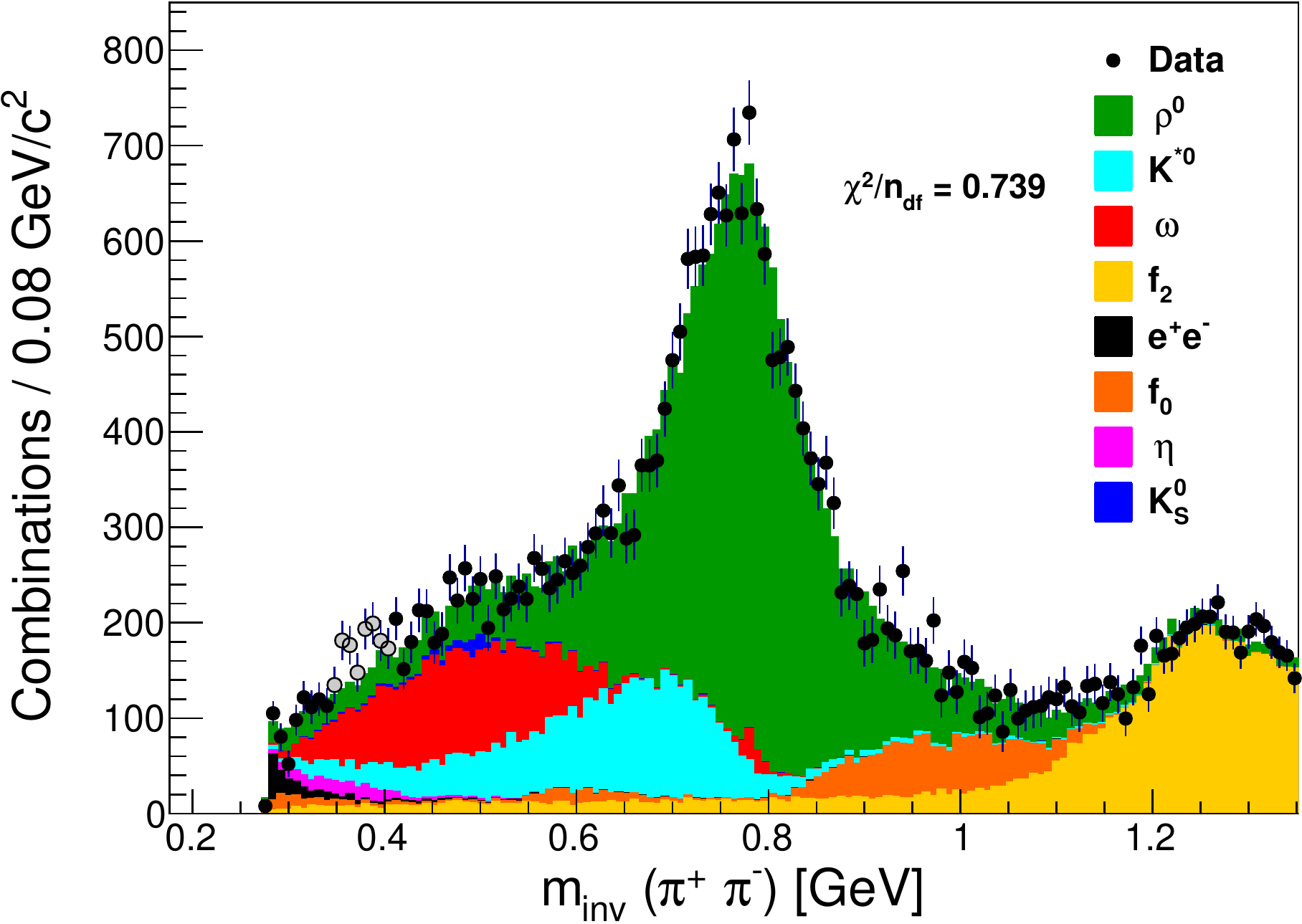}
}
\caption{Invariant mass distribution of opposite charged particles, calculated assuming
pion masses, in $\pi^-$+C interactions
at 158\,\GeVc. Dots with error bars denote the
data and the fitted resonance templates are shown as filled histograms.
The fitted background and high mass resonances have been subtracted. Two
fits with different $\minv(\pi^+\pi^-)$ ranges are shown on the left and right
column. The fit range is equal to the displayed range, but in the extended-range fit
on the right the mass region $0.35<\minv(\pi^+\pi^-)<0.4$ is excluded
(see discussion App.~\ref{app:discussion}), as indicated by the grey points.}
\label{fig:Fits_2}
\end{figure*}

\clearpage

\begin{figure*}[h!]
\def\figw{0.46}
\subfigure[$0.7<\xF<0.8$]{
\includegraphics[width=\figw\linewidth]{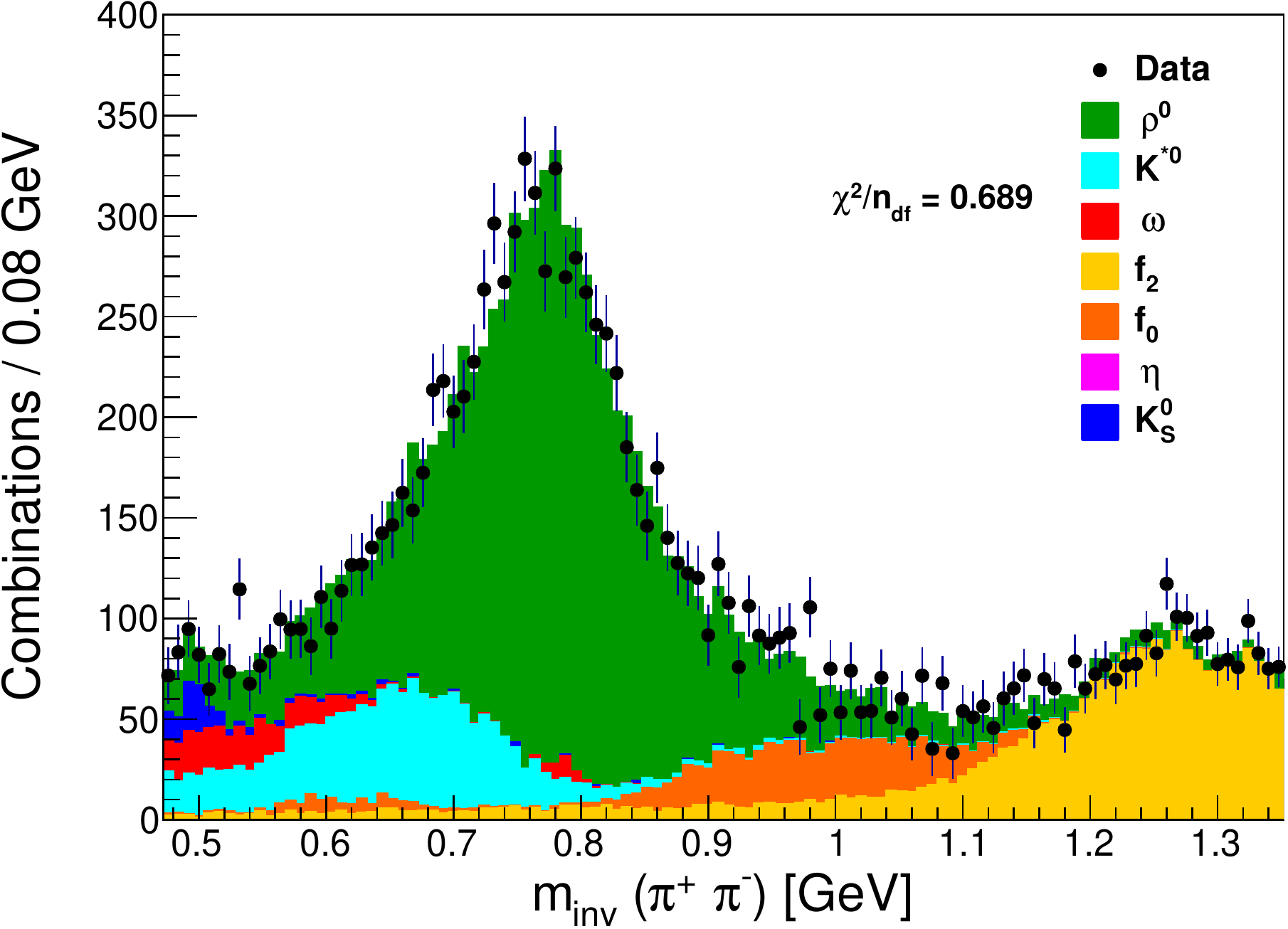}\qquad
\includegraphics[width=\figw\linewidth]{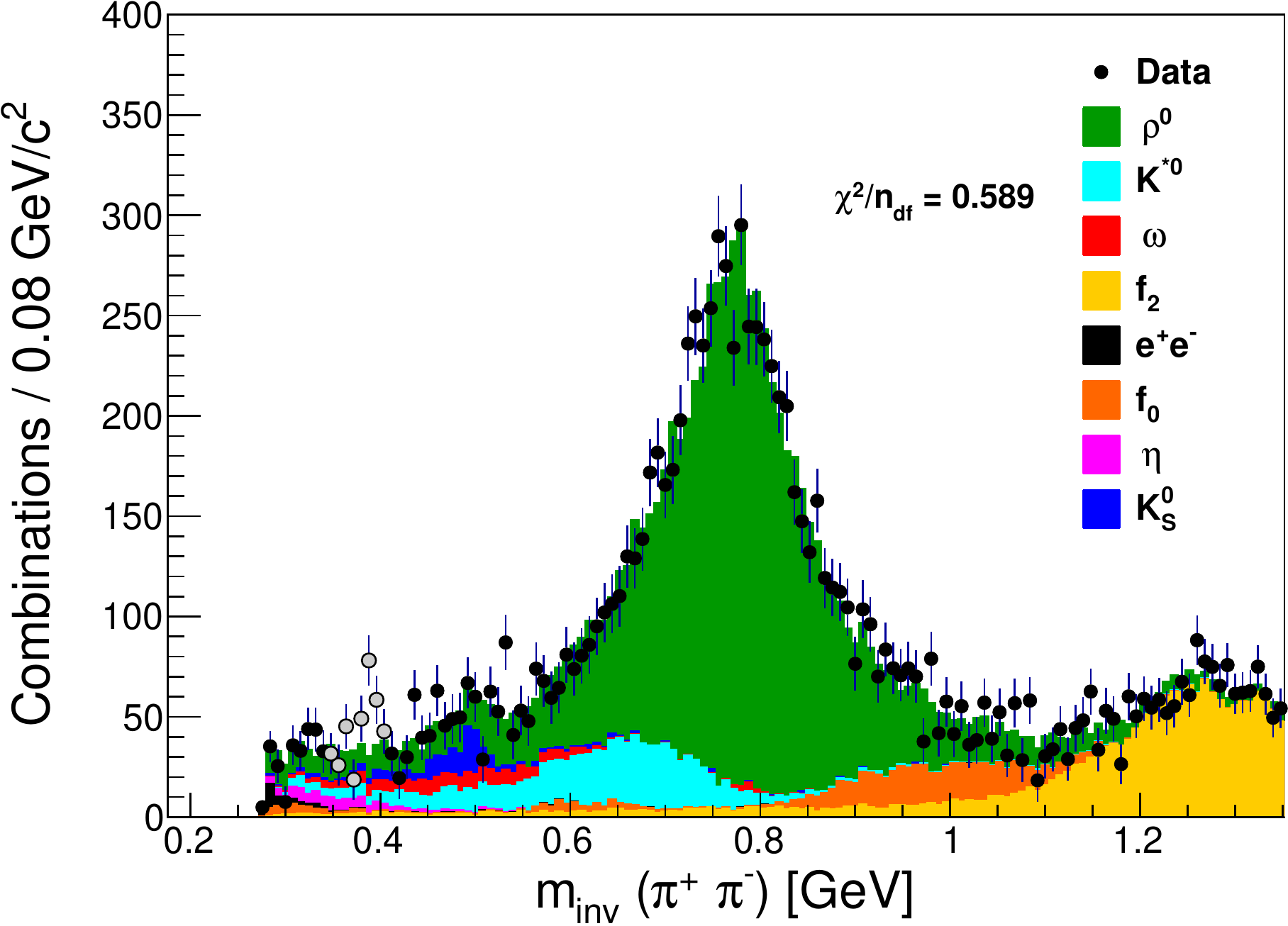}
}
\subfigure[$0.8<\xF<0.9$]{
\includegraphics[width=\figw\linewidth]{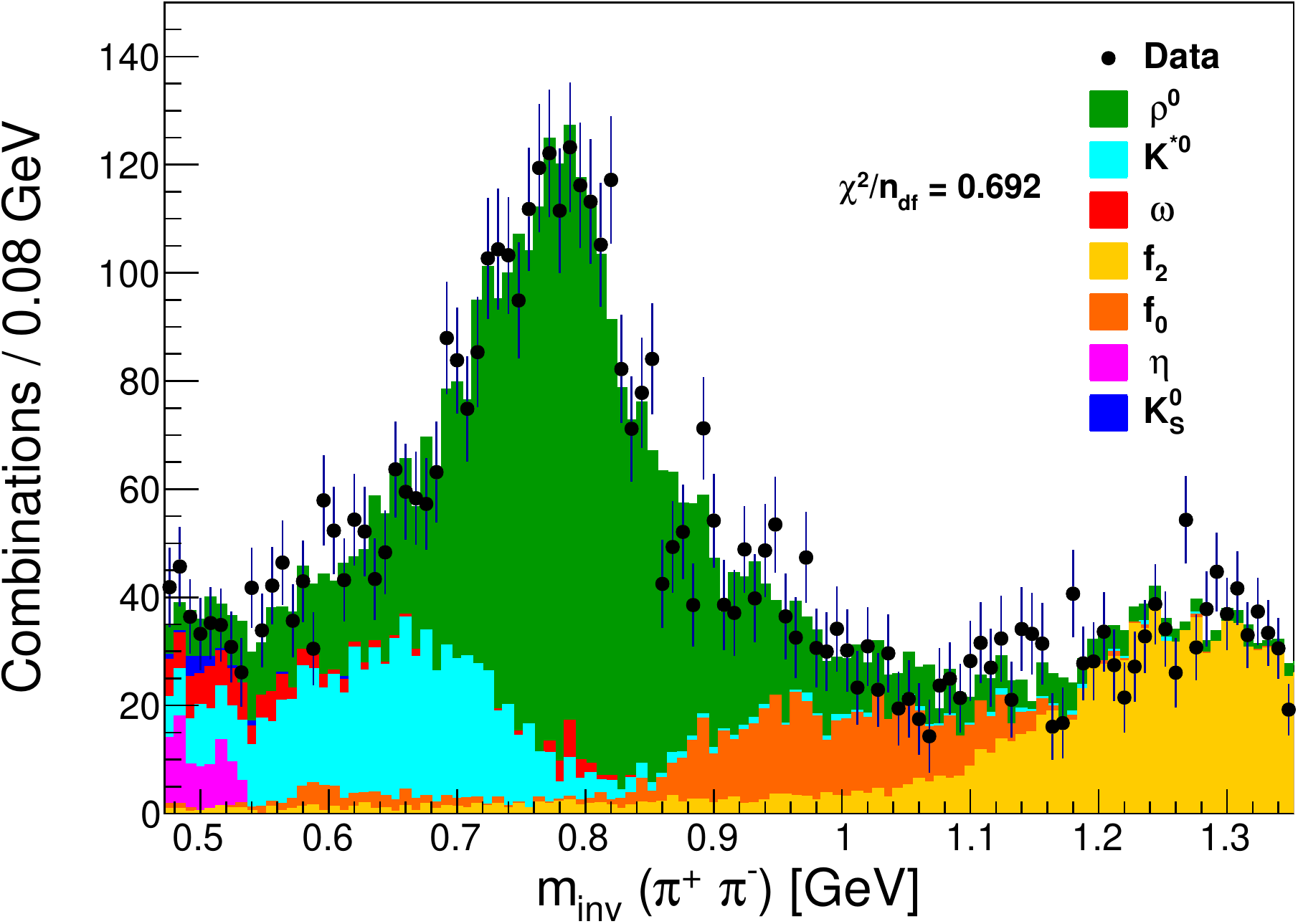}\qquad
\includegraphics[width=\figw\linewidth]{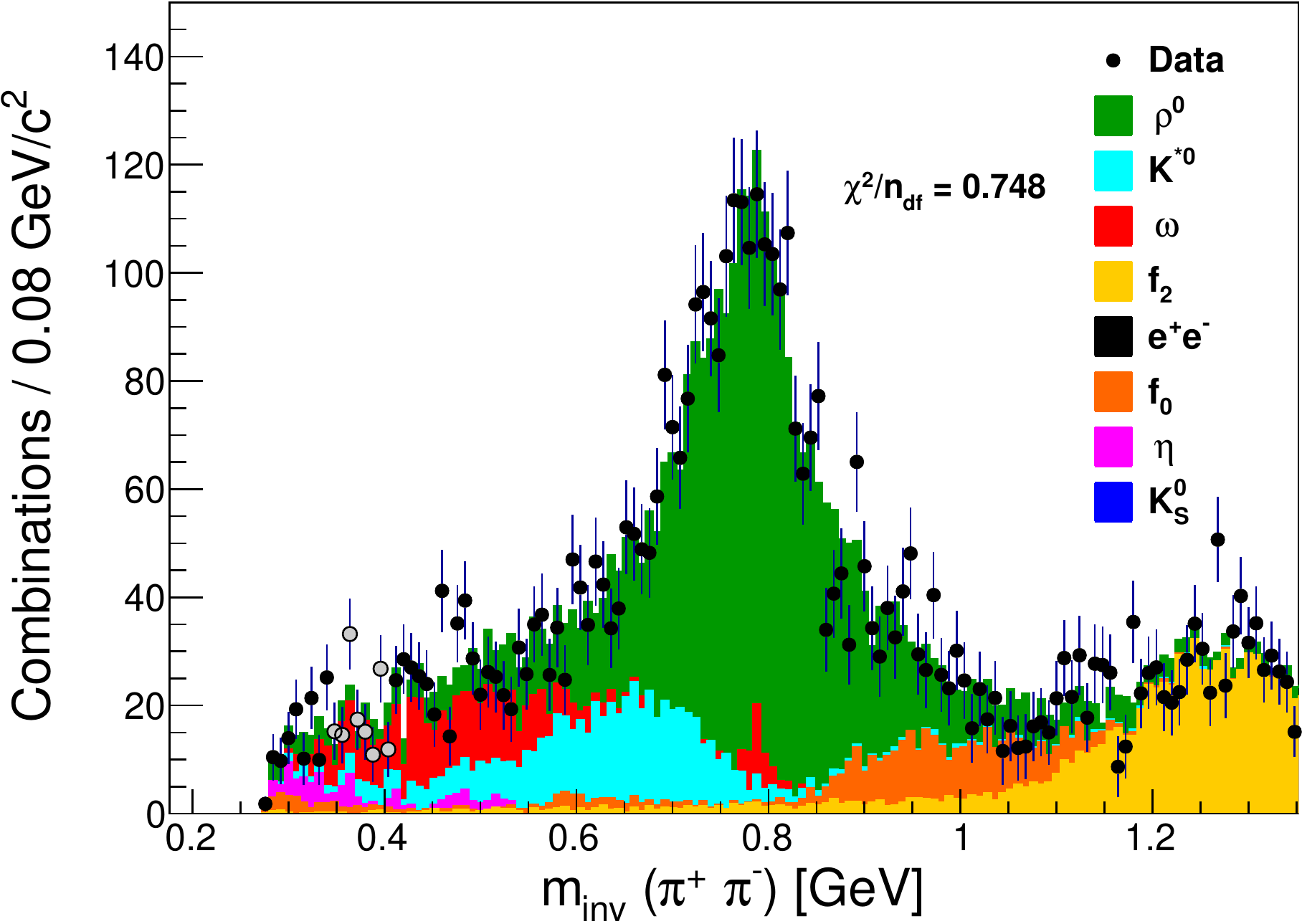}
}\\
\subfigure[$0.9<\xF<1$]{
\includegraphics[width=\figw\linewidth]{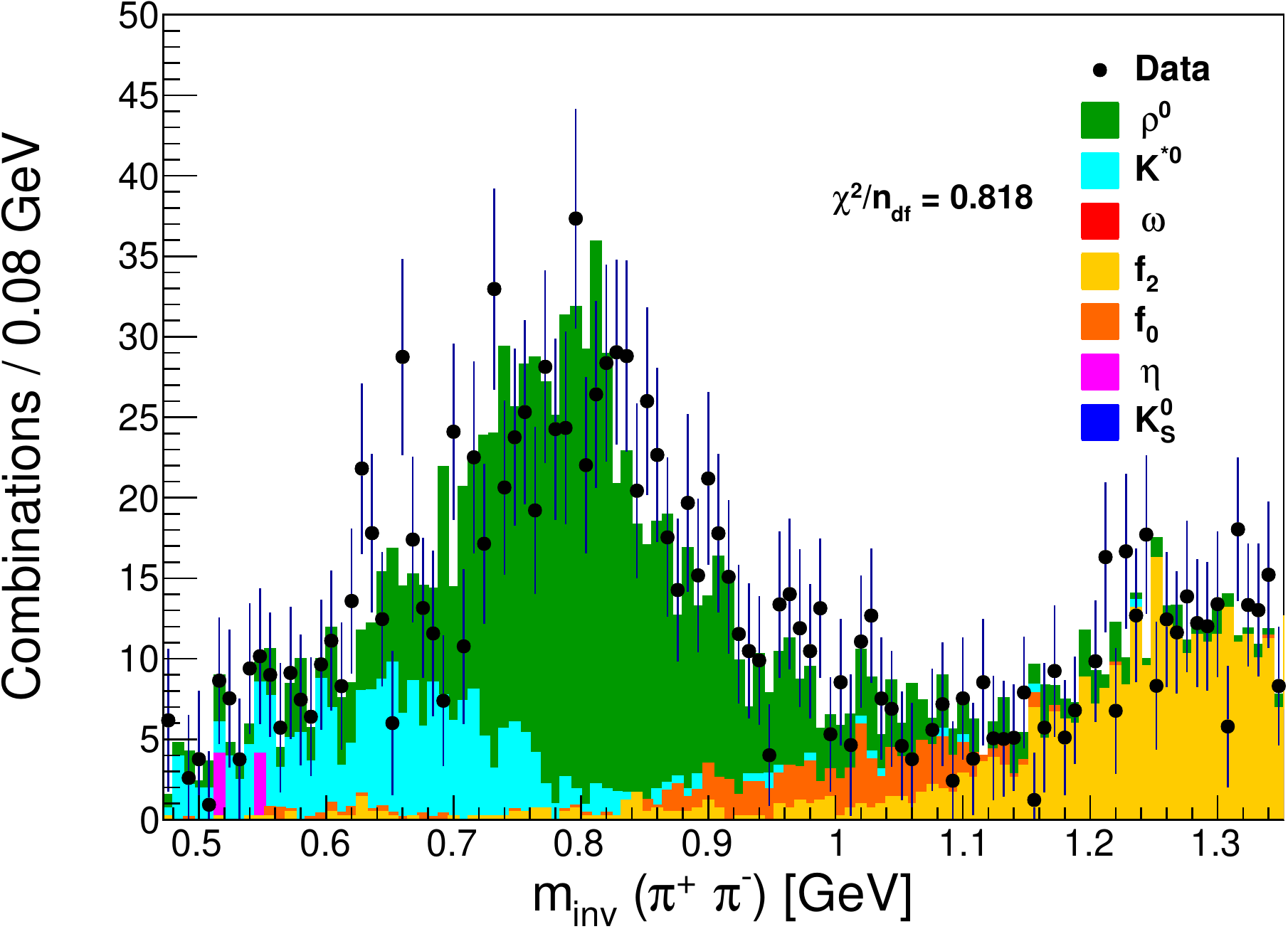}\qquad
\includegraphics[width=\figw\linewidth]{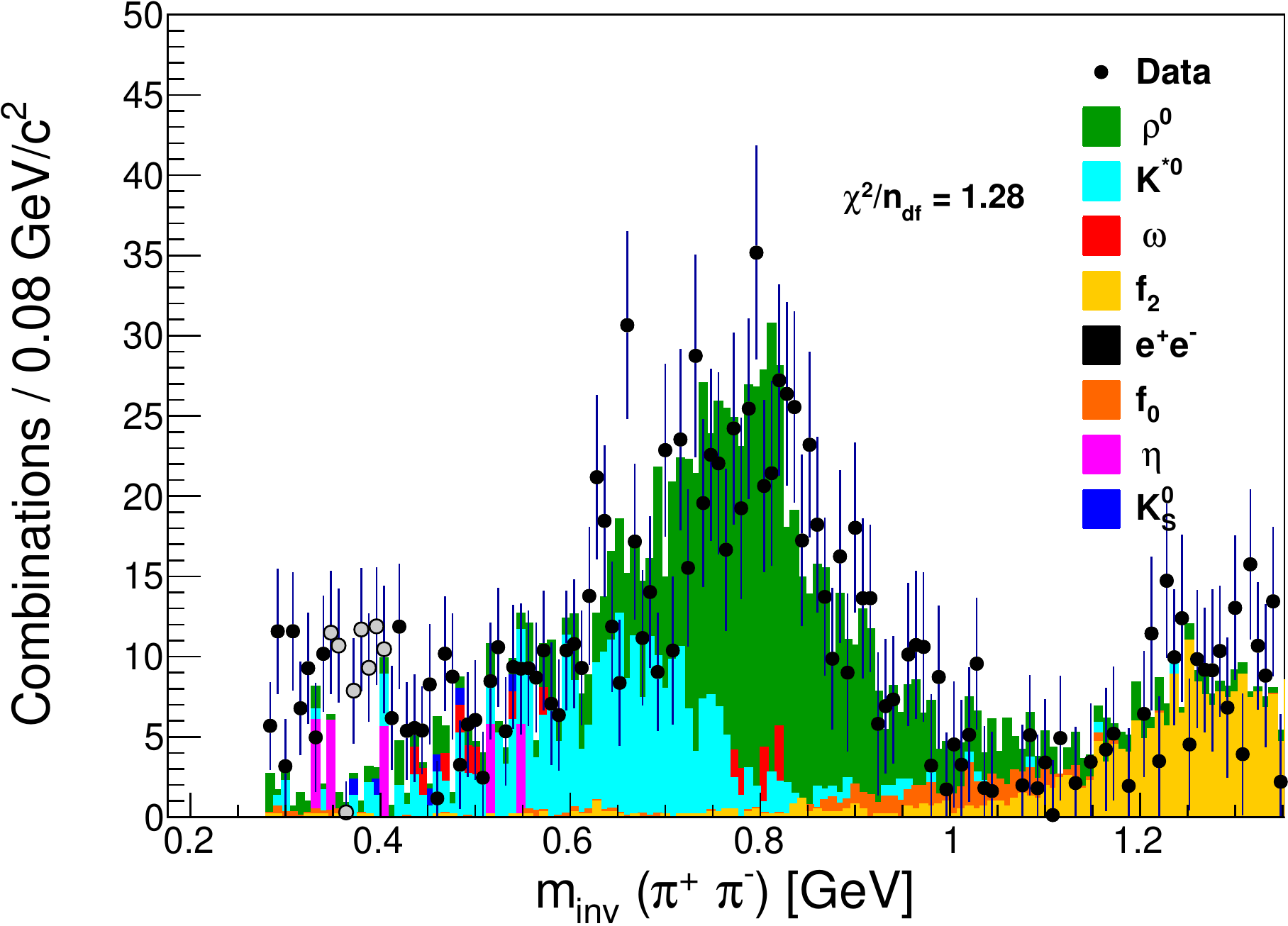}
}
\caption{Invariant mass distribution of opposite charged particles, calculated assuming
pion masses, in $\pi^-$+C interactions
at 158\,\GeVc. Dots with error bars denote the
data and the fitted resonance templates are shown as filled histograms.
The fitted background and high mass resonances have been subtracted. Two
fits with different $\minv(\pi^+\pi^-)$ ranges are shown on the left and right
column.
The fit range is equal to the displayed range, but in the extended-range fit
on the right the mass region $0.35 < \minv(\pi^+\pi^-) < 0.4$ is excluded
(see discussion App.~\ref{app:discussion}), as indicated by the grey points.}
\label{fig:Fits_3}
\end{figure*}

\clearpage

\begin{figure*}[h!]
\centering
\def\figw{0.46}
\subfigure[$0 < \xF < 0.15$]{
\includegraphics[width=\figw\linewidth]{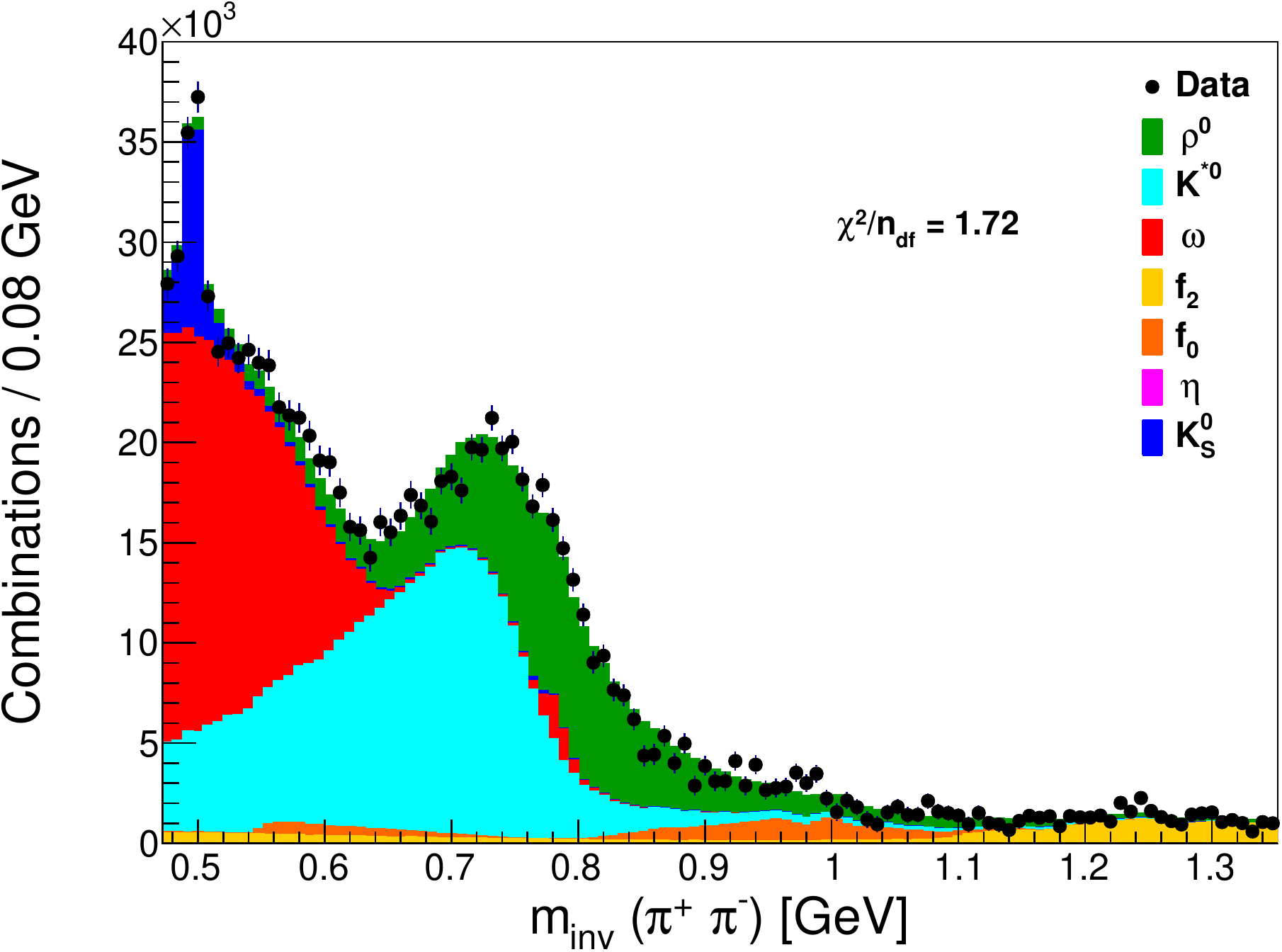}\qquad
\includegraphics[width=\figw\linewidth]{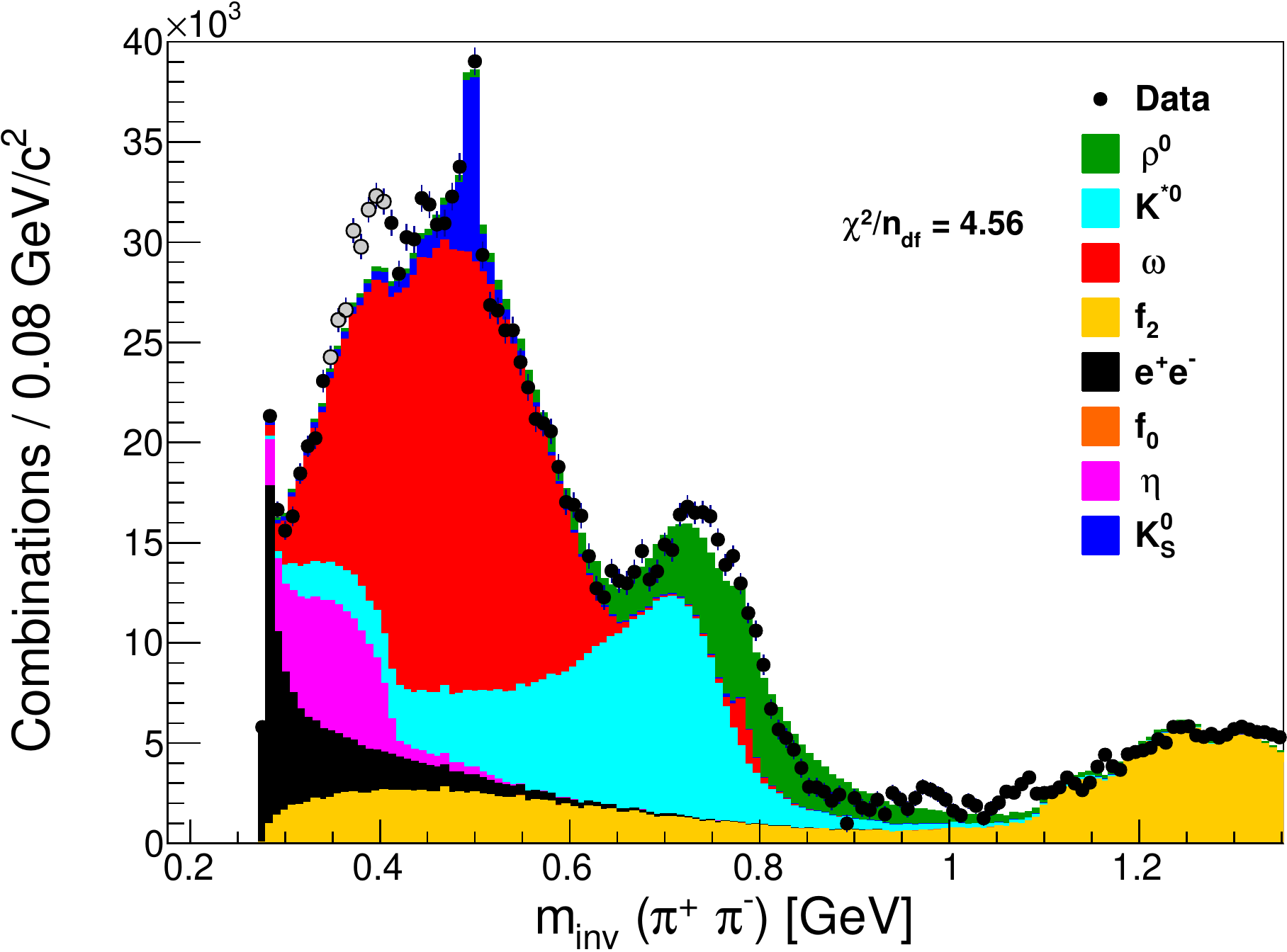}
}
\subfigure[$0.15 < x_\text{F} < 0.3$]{
\includegraphics[width=\figw\linewidth]{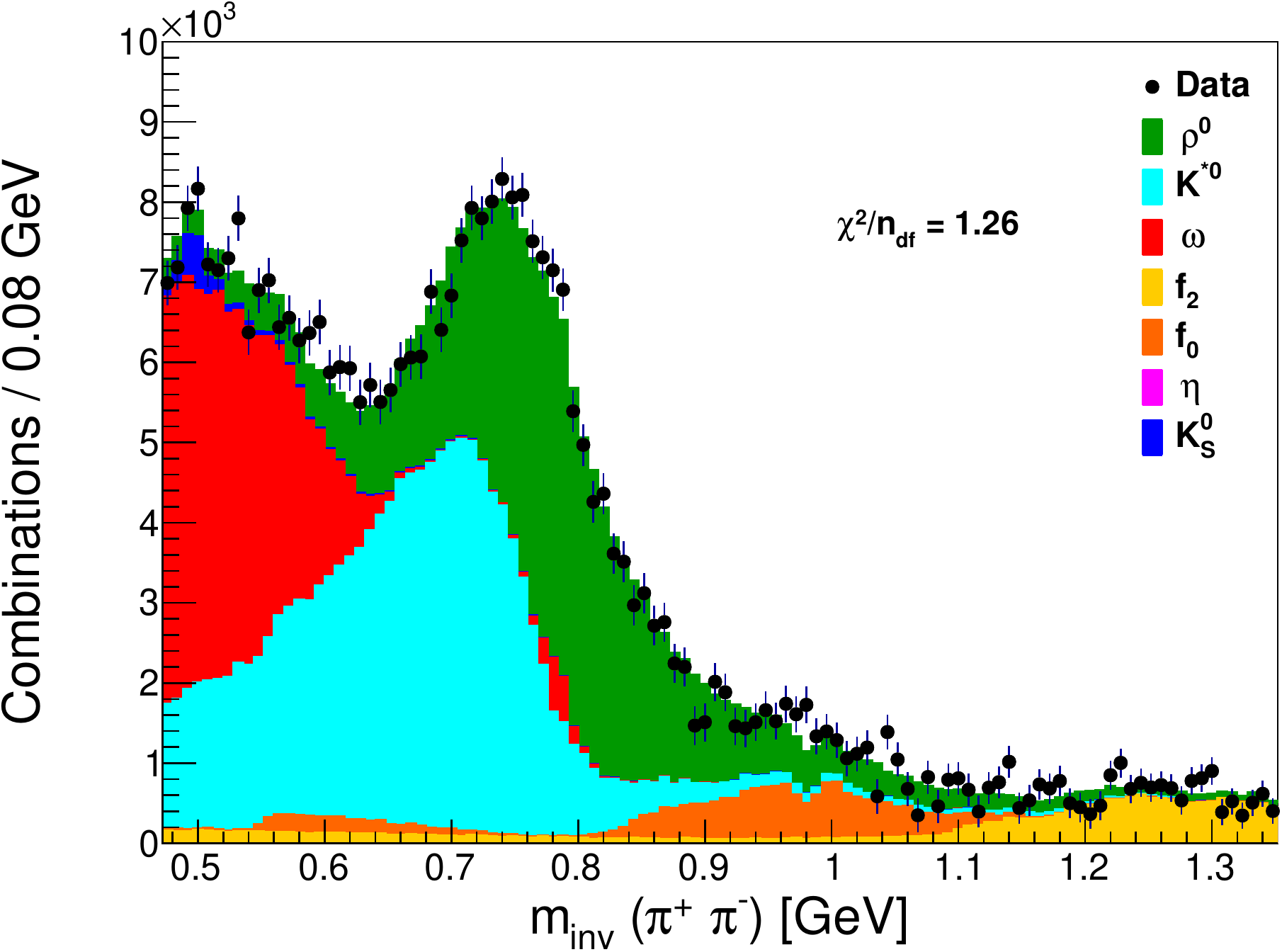}\qquad
\includegraphics[width=\figw\linewidth]{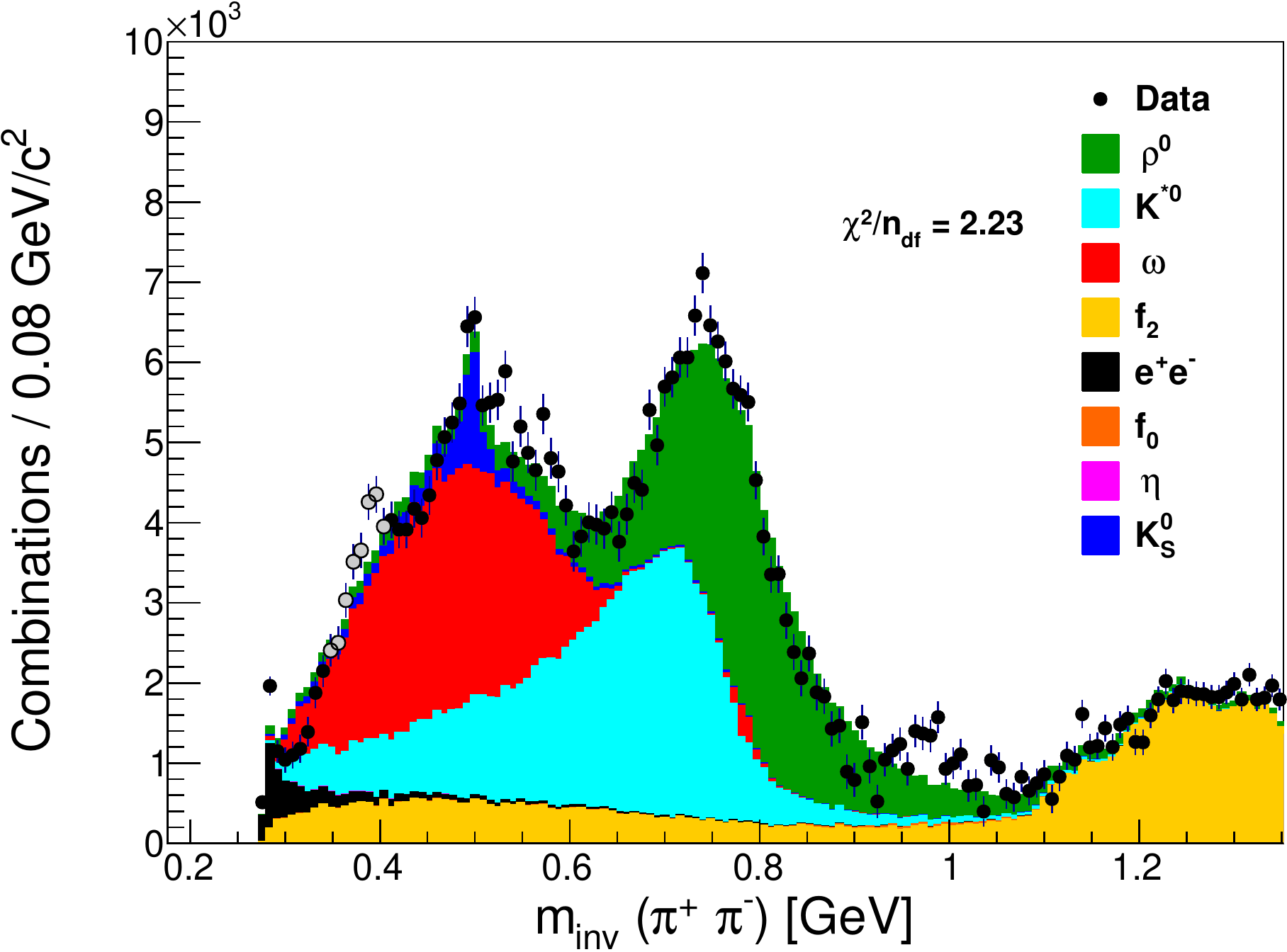}
}
\subfigure[$0.3 < x_\text{F} < 0.4$]{
\includegraphics[width=\figw\linewidth]{Fits_350-03}\qquad
\includegraphics[width=\figw\linewidth]{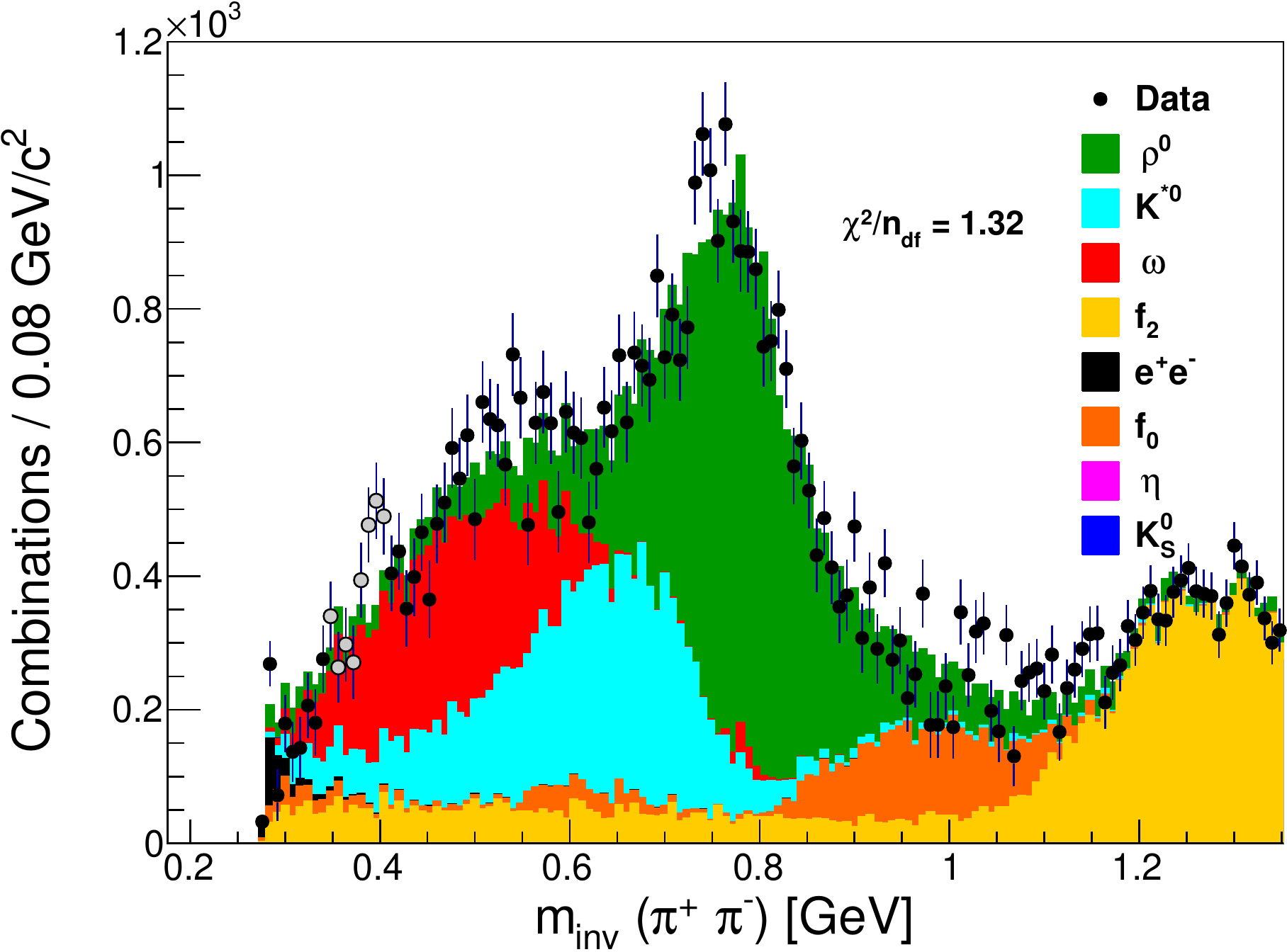}
}
\caption{Invariant mass distribution of opposite charged particles, calculated assuming
pion masses, in $\pi^-$+C interactions
at 350\,\GeVc. Dots with error bars denote the
data and the fitted resonance templates are shown as filled histograms.
The fitted background and high mass resonances have been subtracted. Two
fits with different $\minv(\pi^+\pi^-)$ ranges are shown on the left and right
column.
The fit range is equal to the displayed range, but in the extended-range fit
on the right the mass region $0.35 < \minv(\pi^+\pi^-) < 0.4$ is excluded
(see discussion App.~\ref{app:discussion}), as indicated by the grey points.}
\label{fig:Fits_4}
\end{figure*}

\clearpage

\begin{figure*}[h!]
\centering
\def\figw{0.46}
\subfigure[$0.4 < \xF < 0.5$]{
\includegraphics[width=\figw\linewidth]{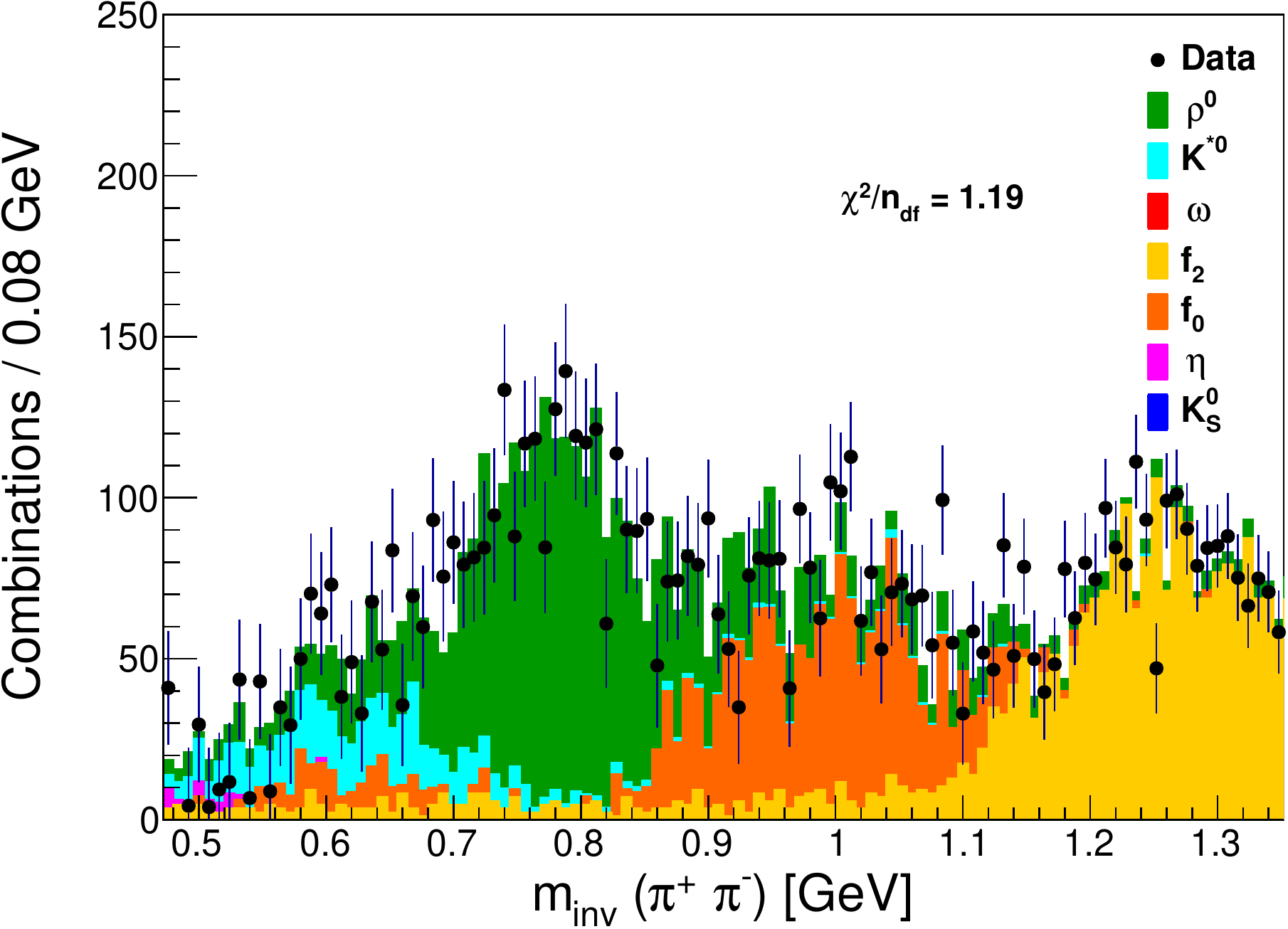}\qquad
\includegraphics[width=\figw\linewidth]{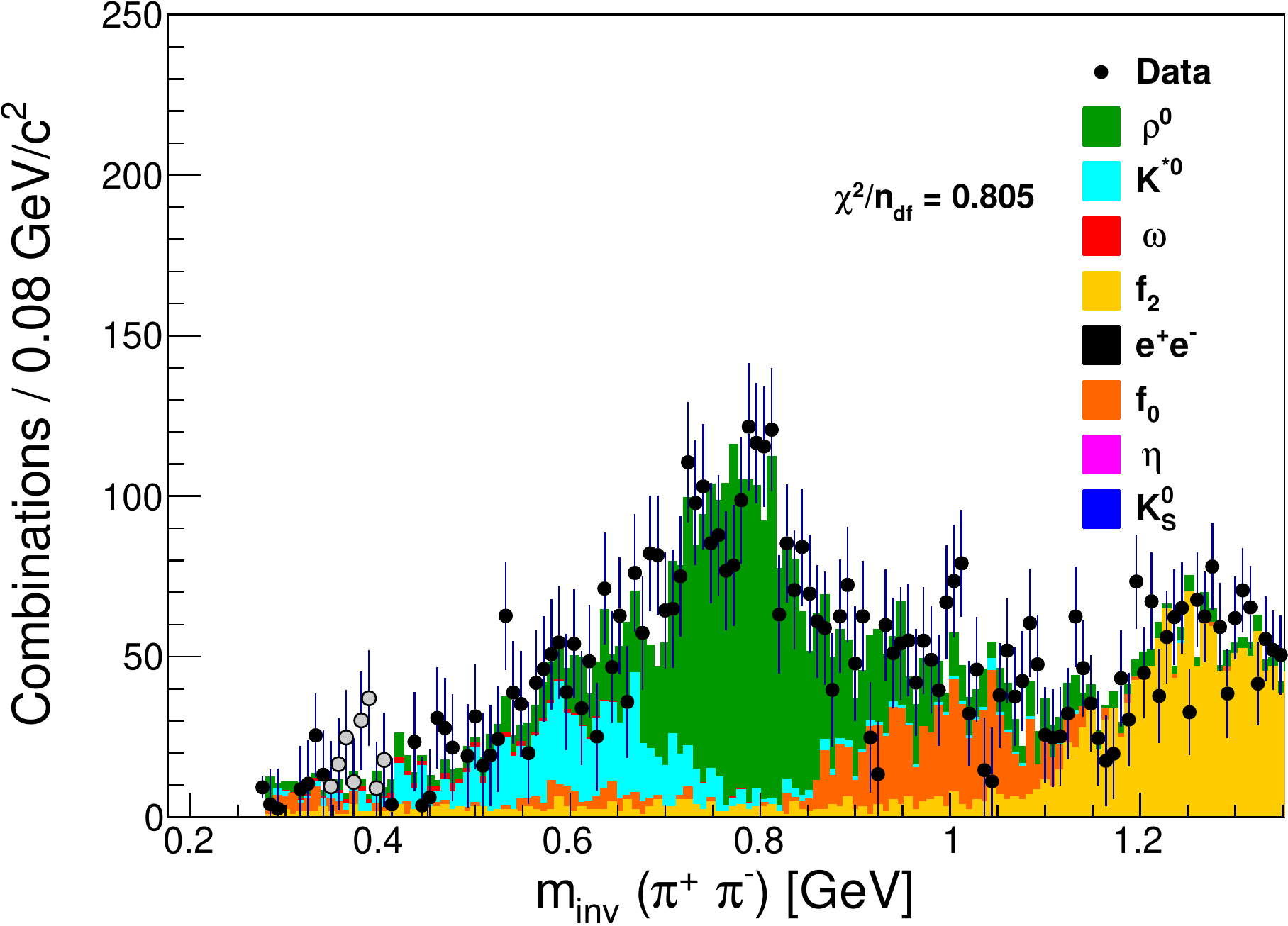}
}
\caption{Invariant mass distribution of opposite charged particles, calculated assuming
pion masses, in $\pi^-$+C interactions
at 350\,\GeVc. Dots with error bars denote the
data and the fitted resonance templates are shown as filled histograms.
The fitted background and high mass resonances have been subtracted.
The fit range is equal to the displayed range, but in the extended-range fit
on the right the mass region $0.35 < \minv(\pi^+\pi^-) < 0.4$ is excluded
(see discussion App.~\ref{app:discussion}), as indicated by the grey points.}
\label{fig:Fits_5}
\end{figure*}

\clearpage

\section{Discussion of the bump in the extended range fit}
\label{app:discussion}

The fits with the extended invariant mass range show a bump in the
data at low \xF that is not described by the template fit (see right
columns in Figs.~\ref{fig:Fits_1} to \ref{fig:Fits_5}).  A large
number of templates from different resonances were investigated to
describe the excess of combinations at low invariant masses. The
resonances were chosen from the particles with the highest yield in
the region of invariant mass where the excess was located. Most of
these resonances have a dominant decay which is either into three (or
more) particles, or two-body decays but into particles other than two
pions. As the invariant mass in this analysis is calculated assuming
the particles are pions from a two-body decay, this will shift the
calculated mass away from the true mass of the resonance. The studied
resonances are listed in the table below and they were chosen by
looking at the invariant mass distribution of particles produced in
\EposLong that produce a combination of negative and positive tracks
in the region of $\minv(\pi^+\pi^-) \approx 0.375\,\GeVcc$. Particles
not produced by this model were not considered.

\begin{table}[!htb]
\begin{center}
\small
\bgroup
\def\arraystretch{1.2}
\begin{tabular}{c|c|c|c}
\hline
resonance & mass / $(\GeVcc)$  & ${\approx}$peak in $\minv(\pi^+\pi^-)/(\GeVcc)$ & dominant decay
\\
\hline
      $\phi$ & 1.020 & 0.37 & K$^+$K$^-$ \\
      $\Lambda$ & 1.115 & 0.34 & p\,$\pi^-$ \\
      $\Delta$ & 1.230 & 0.58 & N\,$\pi\,\pi$ \\
      $N (1440)$ & 1.440 & 0.43 & N\,$\pi\,\pi$ \\
      $a_{2}^{\pm}$ & 1.320 & 0.46 & 3$\pi$\\
      $\rho_{3}^{\pm}$ & 1.690 & 0.50 & 4$\pi$, 2$\pi$\\
      $\eta^{'}$ & 0.958 & 0.35 & $\pi^+\pi^-\eta$\\
      $f_{2}^{'}$ & 1.525 & 1.15 & K\,$\bar{\text{K}}$ \\
      $f_{0} (1500)$ & 1.500 & 0.45 & 2$\pi$, 4$\pi$\\
      $f_{1}$ & 1.285 & 0.41 & 4$\pi$, $\eta\,2\pi$\\
      $f_{1} (1420)$ & 1.420 & 0.42 & K\,$\bar{\text{K}}$\,$\pi$\\
      K$_{L}^{0}$ & 0.497 & 0.39 & $\pi^+\pi^-\pi^0$ \\
      K & 0.494 & 0.44 & $\pi^+\pi^+\pi^-$\\
\hline
\end{tabular}
\egroup
\end{center}
\label{tab:Extra_Templates}
\end{table}

We found that none of these resonances can describe the bump seen at a
$\minv(\pi^+\pi^-) \approx 0.375\,\GeVcc$.  The best-fit particles
are the first two in the table: $\phi$ and $\Lambda$. However both of
these have features that are not present in the data. $\phi$ has a
peak in $\minv(\pi^+\pi^-)$ just below the bump and the
$\Lambda$-template is too broad with no peak near the bump. All other
templates were either too broad, had no peak, or their peak was too
far away from the bump. The conversion of $\gamma$ into $e^+e^-$ was
also investigated, but the corresponding templates also can not
describe the bump. Furthermore, we tried combinations of the
resonances listed above without success, though we can not exclude
that a particular combination could fit the bump since not all
possible combinations were explored.

From a study of the ionisation energy deposit of the tracks in the
TPCs we conclude that the bump is caused by pion combinations.  The
bump is not caused by re-interactions in the detector or the decay of
long lived particles as it remains present even under the tightening
of impact parameter cuts, which would remove such particles. It is
interesting to note that the mass of the bump compatible with the f$_0$
(500) meson, however the width seen here is much smaller than quoted
by the particle data group~\cite{Olive:2016xmw}.

\clearpage

\section{Yield ratios}
\label{app:ratio}
\begin{figure}[!ht]
\centering
\def\figw{0.49}
\subfigure[$\rho^0$/$\omega$]{\includegraphics[width=\figw\linewidth]{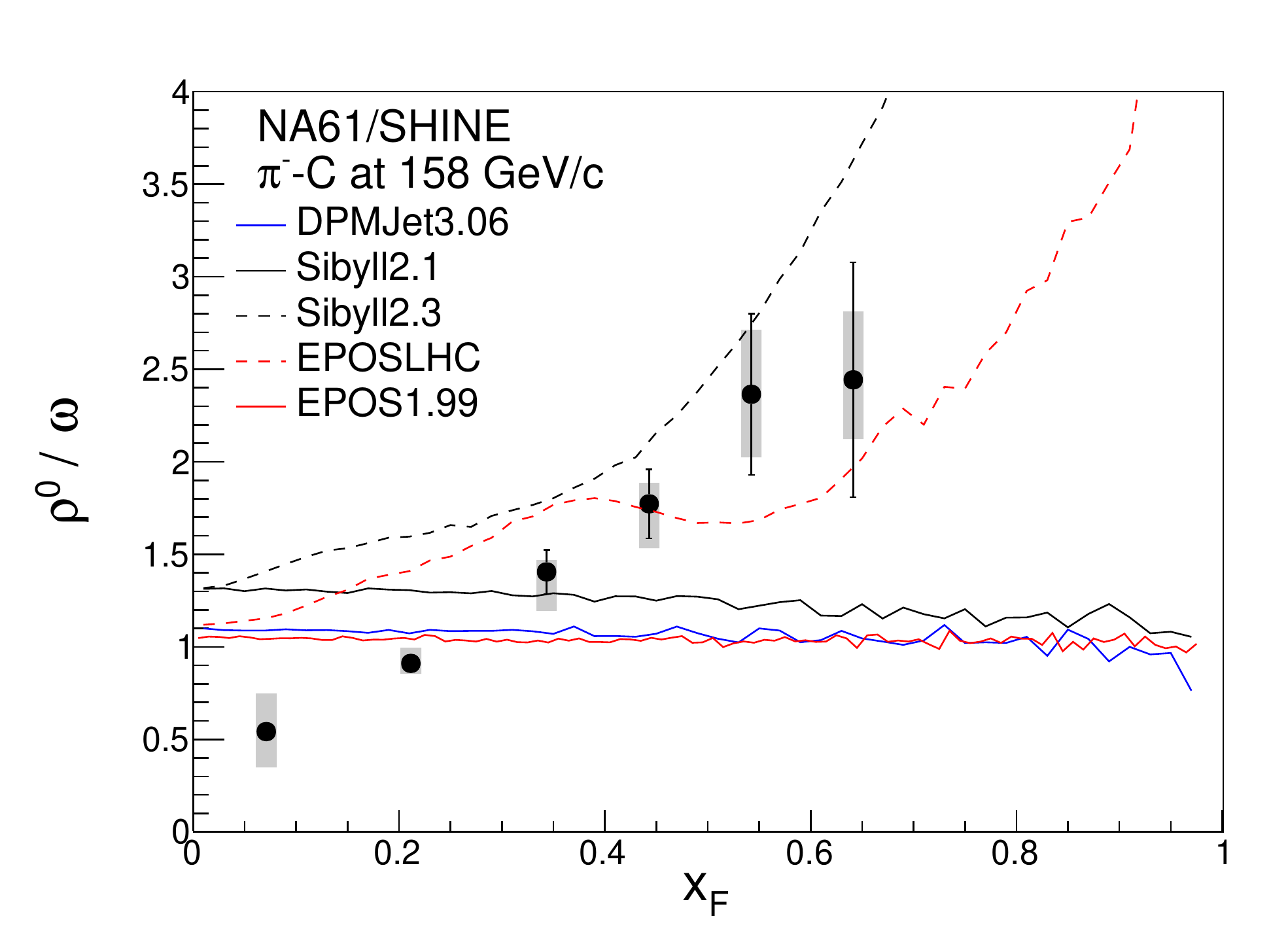}}\hfill
\subfigure[K$^{*0}$ and $\rho^0$]{\includegraphics[width=\figw\linewidth]{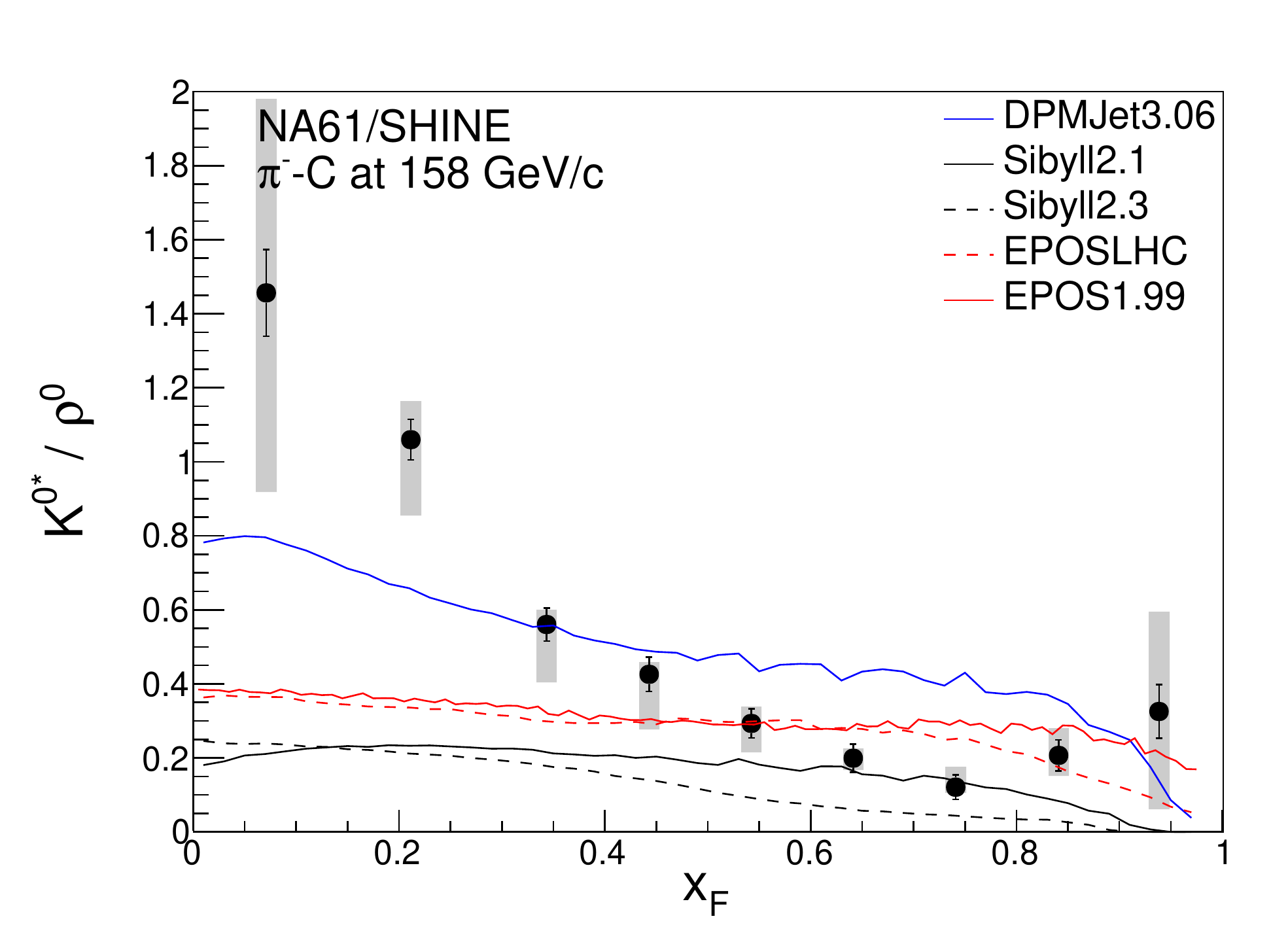}}
\\
\subfigure[K$^{*0}$/$\omega$]{\includegraphics[width=\figw\linewidth]{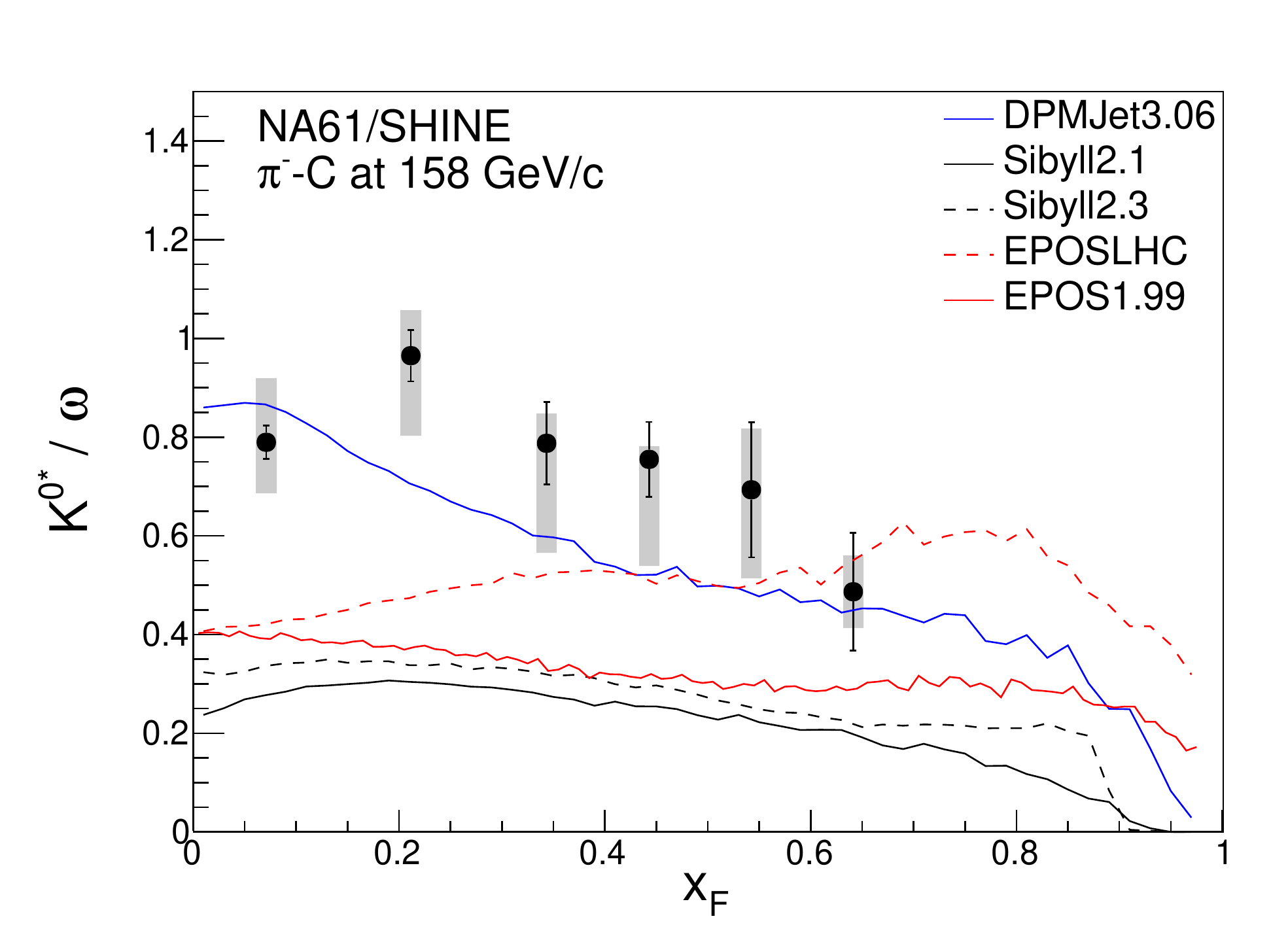}}
\caption{Ratio of meson spectra
in $\pi^-$+C production interactions at $158\,\GeVc$. The dots with error bars show the data and its statistical uncertainties. The shaded boxes denote the systematic uncertainties. The lines depict predictions of hadronic interaction models:
red -- \EposLong, blue -- \DPMJetLong, black -- \SibyllLong, dashed red --
\EposLHCLong, dashed black -- \SibyllNewLong.}.
\label{fig:Ratio_Plots}
\end{figure}

  \newpage
  \clearpage
  \pdfbookmark[0]{The NA61/SHINE Collaboration}{NA61/SHINE}
  {\Large The \NASixtyOne Collaboration}
  \bigskip
  \begin{sloppypar}

\noindent
A.~Aduszkiewicz$^{\,16}$,
Y.~Ali$^{\,13}$,
E.V.~Andronov$^{\,22}$,
T.~Anti\'ci\'c$^{\,3}$,
B.~Baatar$^{\,20}$,
M.~Baszczyk$^{\,14}$,
S.~Bhosale$^{\,11}$,
A.~Blondel$^{\,25}$,
M.~Bogomilov$^{\,2}$,
A.~Brandin$^{\,21}$
% ...

  \end{sloppypar}
  \begin{multicols}{2}
   \small

\noindent
% ...
$^{2}$Faculty of Physics, University of Sofia, Sofia, Bulgaria\\
$^{3}$Ru{\dj}er Bo\v{s}kovi\'c Institute, Zagreb, Croatia\\
% ...
$^{11}$H. Niewodnicza\'nski Institute of Nuclear Physics of the
      Polish Academy of Sciences, Krak\'ow, Poland\\
% ...
$^{13}$Jagiellonian University, Cracow, Poland\\
$^{14}$University of Science and Technology, Cracow, Poland\\
% ...
$^{16}$University of Warsaw, Warsaw, Poland\\
% ...
$^{20}$Joint Institute for Nuclear Research, Dubna, Russia\\
$^{21}$National Research Nuclear University (Moscow Engineering Physics Institute), Moscow, Russia\\
$^{22}$St. Petersburg State University, St. Petersburg, Russia\\
% ...
$^{25}$University of Geneva, Geneva, Switzerland\\
% ...

  \end{multicols}
\end{document}